\documentclass[A4paper,12pt]{article}
\usepackage[a4paper]{geometry}
\usepackage{amsmath,amsthm}
\usepackage{natbib}

\usepackage[colorlinks,citecolor=blue,urlcolor=blue]{hyperref}
\usepackage{subfigure}
\usepackage{xcolor}
\usepackage{upgreek}

\usepackage{amssymb}
\usepackage{mathtools}
\usepackage{morefloats}

\usepackage{bm}
\newcommand{\vectr}[1]{\ensuremath{\bm{\mathrm{#1}}}}
\newcommand{\matrx}[1]{\ensuremath{\bm{\mathrm{#1}}}}

\newtheorem{theorem}{Theorem}
\newtheorem{lemma}{Lemma}

\begin{document}

{
	\title{\bf Optimal Cox Regression Subsampling Procedure with Rare Events}
	\author{Nir Keret and Malka Gorfine \hspace{.2cm}\\
		Department of Statistics and Operations Research\\
		 Tel Aviv University, Israel}
	  \date{}	
	  \maketitle
}

\begin{abstract}
	Massive sized survival datasets are becoming increasingly prevalent with the development of the healthcare industry. Such datasets pose computational challenges unprecedented in traditional survival analysis use-cases. In this work we analyze the UK-biobank colorectal cancer data with genetic and environmental risk factors, including a time-dependent coefficient, which critically inflate the dataset size.
	 A popular way for coping with massive datasets is downsampling them to a more manageable size, such that the computational resources can be afforded by the researcher. Cox proportional hazards regression has remained one of the most popular statistical models for the analysis of survival data to-date.
	This work addresses the settings of right censored and possibly left-truncated data with rare events, such that the observed failure times constitute only a small portion of the overall sample. We propose Cox regression subsampling-based estimators that approximate their full-data partial-likelihood-based counterparts, by assigning optimal sampling probabilities to censored observations, and including all observed failures in the analysis. The suggested methodology is applied on the UK-biobank for building a colorectal cancer risk-predicition model, while reducing the computation time.
	Additionally, asymptotic properties of the proposed estimators are established under suitable
	regularity conditions, and simulation studies are carried out to evaluate the finite sample performance of the estimators.
\end{abstract}

{\bf keywords:} A-optimal; Big data; L-optimal; Left-truncation; Loewner ordering; Optimal sampling; Rare events; Survival analysis.

\newpage

\section{Introduction}
Survival analysis is an important, well-studied branch of statistics aimed at the analysis of time-to-event data, which frequently arise from a wide range of fields, such as medicine, sociology, engineering and economics, to name a few.
Since its introduction, the Cox proportional-hazards model \citep{cox1972regression} has consistently maintained its status as one of the most popular methods in use for time-to-event data. In order to estimate the model parameters, the second-order Newton-Raphson (NR) algorithm is typically employed, and indeed this is the default optimizer in the widely used {\tt coxph} function in the R \citep{R} Survival package \citep{survival-package}, as well as in Python's Lifelines package \citep{Davidson-Pilon2019}.  In modern applications, massive sized datasets with survival data become increasingly prevalent, with the number of observations go far beyond $10^6$ \citep{mittal2014high}. The healthcare industry has been traditionally one of the principal generators of survival data, and the amount of data accumulated in that industry has been growing very rapidly in recent years \citep{raghupathi2014big,wang2021fast}.  Massive datasets might pose a computational barrier to the analysis, due to the large number of observations and covariates. Despite the large number of observations, many times the event of interest constitutes only a very small portion of the overall dataset, so that these failure times are referred to as ``rare events''. In the UK-Biobank (UKB) and in the China Kadoorie Biobank, there are about 2800 and 3000 incident cases of colorectal cancer (CRC), out of some 500,000 and 509,500 observations, respectively, yielding a failure rate of only about 0.6\% \citep{UKBnum,pang2018adiposity}. 

We focus in this work on building a CRC risk-prediction model, based upon the UKB data. The UKB is a large-scale health resource containing rich information of about 500,000 participants in the UK, recruited at ages 40--69. All participants provided consent for follow-up through linkage to their health-related records. The collected data include genetic, medical and environmental information measured in various ways, such as blood, urine and saliva samples as well as personal questionnaires. Out of 484,918 UKB participants with available genetic and environmental data, there is a total number of 2,792 incident CRC cases during the follow-up time. As a result of the ascertainment procedure, the UKB data are characterized by right censoring with delayed entry, and our analyses are conducted accordingly. After including 72 established CRC-related single-nucleotide polymorphisms (SNPs) in the model, in addition to enviromental risk factors, diagnostics suggested that a time-dependent coefficient may be needed, since there were hints for violations of the proportional-hazards assupmtion. Incorporating a time-dependent coefficient critically inflated the size of the dataset to more than 33 million rows, instead of just 500,000 to begin with. However, the number of rows with observed failure times remains unchanged.

In this work, we propose a two-step subsampling procedure to deal with the computational challenge of massive data with rare events. In the context of rare events, our goal is to select a subsample from the censored observations in the most efficient way, while retaining all observed failure times in the analyzed dataset. The main rationale for this type of subsampling design is that the failure times contribute more information than the censored times. In addition, instability issues may potentially be brought about by a subsample with too few (even 0) failure times had the sampling been performed from the entire dataset irrespectively. The suggested procedure is comprised of two steps, because the expressions for the optimal sampling probabilities involve the true, unknown, parameters. Therefore, a first step with uniform subsampling from the censored is necessary for approximating the optimal sampling probabilities. Due to the wide range of use-cases that fall under the ``rare events'' framework, we focus in this work on these settings. When the number of observed failure times is large, such that a subsampling routine is required also for the failure times, similar methodology can be developed, but the asymptotic theory is not covered by the current work. We defer such work for future research. 

Lastly, it is noteworthy, that the {\tt coxph} function in R is very comprehensive, and includes many appealing features. Our methods use {\tt coxph} directly, thus enabling access to all of its features. If one tried to reduce computations using other optimization routines, such as quasi-Newton or gradient descent, the programming will need to be carried out from scratch, instead of using {\tt coxph} in the {\tt R} Survival package.

\subsection{Related Work}

Subsampling methods on the premise of massive data have been developed for generalized linear and quantile regression models, to alleviate the computational burden. \cite{dhillon2013new} and \cite{ma2015statistical} proposed a subsampling algorithm for least squares regression; \cite{wang2018optimal} derived the optimal sampling probabilities for logistic regression; \cite{ai2018optimal}, \cite{wang2020optimal} and \cite{yu2020optimal} then extended it to generalized linear models, quantile regression and quasi-likelihood estimators, respectively. In the context of survival analysis, \cite{zuo2020sampling} derived the optimal sampling probabilities for the additive hazards model, under the non-rare-events setting. Practical usage of subsampling methods with survival data for reducing the computational burden, was observed in \cite{johansson2015family} 
for Poisson regression, and in \cite{gorfine2020marginalized} for a new semi-competing risks model. In these two works, all failure times were included in the subsample, while a subset of the censored observations was drawn using the inefficient uniform sampling probabilities.

Two other well-known designs in the epidemiological literature that are related to our approach are the case-cohort (CC) \citep{prentice1986case} and the nested-case-control (NCC) \citep{liddell1977methods}. These designs are put to practice when some covariates are too costly to procure for the entire cohort, and therefore only a subset of the full data undergo expensive measuring. Since these methods are geared at a different use-case, they cannot be fully efficient for our purposes, as will be demonstrated in our simulation study. In the CC design, only the observations that failed, termed ``cases'' and a random subset, uniformly sampled from the censored observations, termed ``subcohort'' are measured for their expensive covariates. Our subsampling proecedure with uniform sampling probabilities, coincides with the CC (up to the fact that we sample with replacement, in contrast to CC, as will be discussed later). In the NCC design, all cases are kept, and for each case a small number of ``controls'', typically ranging from 1 to 4, are uniformly sampled from its corresponding riskset. An improvement upon the classic NCC was suggested by \cite{samuelsen1997psudolikelihood}, by using each sampled control for more than one failure time.
Some other works for improving the efficiency of classic CC designs \citep*{chen1999case,kulich2004improving,kim2013more,zheng2017improving} and of NCC designs \citep{langholz1995counter,keogh2013using}, were published, however these works are established on the grounds of missing covariates, and most importantly, they do not consider computational complexity as a key aspect. 

\subsection{Main Contributions}
In the models listed above with a derived optimal subsampling procedure (for example, logistic or quantile regression), the estimating equations take the simple form of a sum of independent terms, over all observations. Derivation of the optimal sampling probabilities for the Cox regression is substantially more complicated, as the estimating equation  is a sum of ratios, such that the observations appear both in the numerator and in the denominator of more than one ratio term. 
In addition,  incorporating all failure times in the analysis, as we do under the rare events settings, produces unpredictable stochastic processes and standard martingale theory does not suffice for deriving all asymptotic properties.

Due to the different nature of the estimating equations, and the different treatment of failure and censored observations, previous results derived under other models cannot be directly applied to the Cox model. While this work is inspired by the ideas in \cite{wang2018optimal} for logistic regression and \cite{wang2020optimal} for quantile regression, the optimal probabilities for Cox regression under rare events must be derived and justified from scratch, and the asymptotic properties are also to be developed separately. The contribution of this work is threefold:

$\bullet$ This work presents a novel subsampling approach addressing the above mentioned computational challenge for time-to-event data with rare events. Our method is shown to be both highly efficient and computationally fast, by means of both simulations and theoretical rigor. We prove the consistency and asymptotic normality of the estimated regression coefficients and the estimated cumulative hazard function of the Cox model. 

$\bullet$ Our method and asymptotic theory naturally accommodate delayed entry (left truncation), stratified analysis, time-dependent covariates and time-dependent coefficients, as discussed in Section 3. 

$\bullet$ One of the elegant computational devices in the {\tt coxph} function, is the transformation of a dataset into time-dependent form, thus enabling easy accommodation of time-dependent covariates or time-dependent coefficients. Each observation is broken into several pseudo-observations, as elaborated in \cite{therneau2017using} and in Section 3.2. One of our novel findings is that consistent and efficient estimators can be achieved by means of sampling from the pseudo-observations, and not necessarily from the original observations. The implication is that it is efficient to use only informative parts of the individual's observed follow-up trajectory.

\subsection{Paper Outline}
The rest of the paper is organized as follows. 	 	
Section 2 presents the proposed approach and the main methodological details of this paper. Section 3 elaborates on the utility of our subsampling procedures also to popular refinements of the Cox model.
Section 4.1 demonstrates the performance of our methods using simulations, including the settings of delayed entry, time-dependent covariates and time-dependent coefficients. In Section 4.2 we apply our subsampling methods on the CRC incidence cases of the UKB data, with common environmental risk factors and 72 SNPs that have been identified by genome-wide association studies (GWAS) to be associated with CRC \citep{jeon2018determining}.
Section 5 is dedicated to additional research directions, and concluding discussion.

\section{Methodology}\label{sec:Methodology}

\subsection{Notation and Model Formulation}

Let $V$ denote a failure time, $C$ a censoring time and $T$ the observed time $V \wedge C$. Denote $\Delta = I(V<C)$ and let $\vectr{X}$ be a vector of possibly time-dependent covariates of size $r\times1$, and for notational convenience $\vectr{X}$ is used instead of $\vectr{X}(t)$. Suppose there is a fixed number of independent and identically distributed (iid) observations, denoted $n$, then the observed data are $\mathcal{D}_n=\left\{T_i,\Delta_i,\vectr{X}_i; i=1,\ldots,n\right\}$. Out of the $n$ observations, a random number, denoted $n_c$, are censored and $n_e=n-n_c=\sum_{i=1}^n\Delta_i$ have their failure times observed. It is assumed that $n_e/n$ converges to a small positive constant as $n\rightarrow\infty$, and that $r$ remains fixed. Let $q$ be the number of censored observations sampled out the full data. The value of $q$ is decided by the researchers, according to their computational limitations, and it is assumed that $q$ is substantially smaller than $n$, but $q/n$ converges to some small positive constant as $q$ and $n$ go to infinity. For instance, a design that follows such assumptions will have $q=d\times n_e$, where $d$ is some positive integer. Out of computational considerations, the sampling will be performed with replacement, and practically, since $n_c / q$ is assumed very large, there should be little difference between sampling with and without replacement. However, when sampling without replacement, after each drawn observation, the remaining sampling probabilities should be updated. This sequential updating procedure greatly increases the runtime and is thus avoided.
Denote $\mathcal{C}$ as the set of all censored observations in the full data ($|\mathcal{C}|=n_c$), $\mathcal{E}$  as the set of all observations whose failure time was observed ($|\mathcal{E}|=n_e$) and
$\mathcal{Q}$ as the set containing $\mathcal{E}$, in addition to all censored observations included in the subsample ($|\mathcal{Q}|=n_e+q$). 

We assume throughout this work the Cox proportional-hazards model. Extensions of the Cox model will be addressed in Section 3. Let $\vectr{\beta}$ be the vector of coefficients corresponding to $\vectr{X}$, then the instantaneous hazard of observation $i$ at time $t$ takes the form
\[
\lambda(t|\vectr{X}_i) = \lambda_0(t)e^{\vectr{\beta}^T \vectr{\vectr{X}}_i},
\]
where $\lambda_0(t)$ is an unspecified positive function, and $\Lambda_0(t)$ is the cumulative baseline hazard function $\int_{0}^{t}\lambda_0(u)du$. Denote $\vectr{\beta}^o$,$\lambda_0^o$ $\Lambda_0^o$ as the true unknown values of $\vectr{\beta}$, $\lambda_0$ and $\Lambda_0$, respectively.
Let us now introduce the following common notation $\vectr{S}^{(k)}(\vectr{\beta},t) = \sum_{i=1}^ne^{\vectr{\beta}^T\vectr{X}_i}Y_i(t)\vectr{X}^{\otimes k}_i$, $k=0,1,2$,
where  $\vectr{X}^{\otimes0} = 1$,  $\vectr{X}^{\otimes1} = \vectr{X}$ and $\vectr{X}^{\otimes2} = \vectr{X}\vectr{X}^T$, and $Y_i(t) = I(T_i \ge t)$ is the indicator that observation $i$ is at risk at time $t$.
Denote $\hat{\vectr{\beta}}_{PL}$ the full-sample partial-likelihood (PL) estimator for $\vectr{\beta}$ and $\tau$ the maximal follow-up time. So, $\hat{\vectr{\beta}}_{PL}$ is the vector that solves the following vectorial equation
\[
\frac{\partial l(\vectr{\beta})}{\partial \vectr{\beta}^T} = \sum_{i=1}^{n} \Delta_i \left\{ \vectr{X}_i - \frac{\vectr{S}^{(1)}(\vectr{\beta},T_i)}{S^{(0)}(\vectr{\beta},T_i)}\right\} = \vectr{0} \, .
\]
Let $\vectr{p}=(p_1,\ldots,p_{n_c})^T$ be the vector of sampling probabilities for the censored observations, where $\sum_{i=1}^{n_c}p_i = 1$, and let us set $w_i = (p_i q)^{-1}I(\Delta_i=0, p_i>0) + I(\Delta_i=1)$.
The subsample-based counterpart of the previously introduced notation, is  $\vectr{S}_w^{(k)}(\vectr{\beta},t) = \sum_{i\in\mathcal{Q}}w_ie^{\vectr{\beta}^T\vectr{X}_i}Y_i(t)\vectr{X}^{\otimes k}_i$.
Now, suppose that a subsample of size $q$ was drawn from the censored observations, then the estimator based on $\mathcal{Q}$, denoted $\tilde{\vectr{\beta}}$, is the vector that solves
\[
\frac{\partial l^*(\vectr{\beta})}{\partial \vectr{\beta}^T} =  \sum_{i\in\mathcal{Q}} \Delta_i  \left\{\vectr{X}_i - \frac{\vectr{S}_w^{(1)}(\vectr{\beta},T_i)}{S_w^{(0)}(\vectr{\beta},T_i)} \right\} = \vectr{0} \, .
\]
For a given vector of regression coefficients $\vectr{\beta}$, let us define the function $$\hat{\Lambda}_0(t,\vectr{\beta}) = \sum_{i=1}^n\frac{\Delta_iI(T_i\le t)}{S^{(0)}(\vectr{\beta},T_i)}\, ,$$
and the Breslow estimator \citep{breslow1972contribution} for the cumulative baseline hazard function is thus produced by $\hat{\Lambda}_0(t,\hat{\vectr{\beta}}_{PL})$. In this work we propose Breslow-type estimators, such that different consistent estimators for $\vectr{\beta}^o$ are plugged in, based on our optimal sampling procedure.

Denote $R_i$ as the random variable which counts the number of times observation $i$ was drawn into the subsample $\mathcal{Q}$, and $\vectr{R}=(R_1,\ldots,R_n)^T$. Conditionally on $\mathcal{D}_n$, $R_i = 1$ if $\Delta_i=1$, whereas $\vectr{R}_c|\mathcal{D}_n \sim Multinomial(q,\vectr{p})$,  where $\vectr{R}_c$ is the $n_c\times1$ sub-vector of $\vectr{R}$, corresponding to the censored observations. Using this notation one can replace, for example, $\sum_{i\in\mathcal{Q}}w_ie^{\vectr{\beta}^T\vectr{X}_i}Y_i(t)\vectr{X}_i $ with $\sum_{i=1}^nw_ie^{\vectr{\beta}^T\vectr{X}_i}Y_i(t)\vectr{X}_iR_i $, and both forms will be used interchangeably, depending on convenience. The two forms of presentation illuminate different aspects of the subsampling procedure in the conditional space, namely given $\mathcal{D}_n$. The first presentation is a sum of $n_e$ events and $q$ iid sampled censored observations (similarly to bootstrap sampling), while the second presentation, which spans the full data, emphasizes the dependency induced by sampling a fixed and predetermined number of observations. Additionally, let us introduce the usual counting process notation $N_i(t) = \Delta_iI(T_i\le t)$, $N_.(t) = \sum_{i=1}^nN_i(t)$, and finally, $\Vert.\Vert_2$ denotes the $l_2$ Euclidean norm.

\subsection{Asymptotic Properties}
It is important to clarify, that this section begins with the derivation of the asymptotic properties of a general subsampling-based estimator, for any vector of conditionally deterministic probabilities, given the data, that conforms with Assumption A.8, as presented below. It will be shown that in the conditional space, i.e. given the observed data, the subsampling-based estimator converges to the full-sample PL estimator, while in the unconditional space, it converges to the true parameter. Based on these results, the optimal sampling probabilities will be derived, motivated by different optimality criteria, as elaborated in Section 2.3. However, the optimal sampling probabilities involve the true unknown parameter, so Section 2.4 presents approximated optimal sampling probabilities, which are built upon the results of the preceding chapters, and do not require the true parameter. The convergence properties of the subsampling-based estimators using the probabilities in Section 2.4, will be studied in the unconditional space. 

Following is a list of all required assumptions, some of them standard for Cox regression, others unique to our case. The latter are further discussed after listing all the assumptions.

\begin{itemize}
	\item[\bf A.1] 
	The cumulative hazard function is bounded, namely, $\int_{0}^{\tau}\lambda^o_0(t)dt<\infty$, and the regression parameter space $\mathcal{B}$ is compact. 
	\item[\bf A.2] 
	There exists a neighbourhood $\mathcal{B}$ of $\vectr{\beta}^o$ and functions $\vectr{s}^{(j)}(\vectr{\beta},t), j=0,1,2$, defined on $\mathcal{B} \times [0,\tau]$ such that for $j=0,1,2$ as $n \rightarrow \infty$, 
	$$\sup_{t\in[0,\tau],\vectr{\beta}\in\mathcal{B}}\frac{1}{n}\left\| \vectr{S}^{(j)}(\vectr{\beta},t) - \vectr{s}^{(j)}(\vectr{\beta},t)\right\|_2\xrightarrow{P}0 \, .$$
	\item[\bf A.3]  
	The covariates $\vectr{X}(t)$ are uniformly bounded over $t\in(0,\tau)$.
	\item[\bf A.4]  
	For all $\vectr{\beta}\in\mathcal{B}$ and $t\in[0,\tau]$, 
	$$\partial s^{(0)}(\vectr{\beta},t)/(\partial\vectr{\beta}) = \vectr{s}^{(1)}(\vectr{\beta},t) \, ,$$ and
	$$\partial^2s^{(0)}(\vectr{\beta},t)/(\partial\vectr{\beta}^T\partial\vectr{\beta}) = \matrx{s}^{(2)}(\vectr{\beta},t) \, .$$
	For $j=0,1,2$, $\vectr{s}^{(j)}(\vectr{\beta},t)$ is a continuous function of $\vectr{\beta}$, uniformly in $t\in[0,\tau]$. The functions $\vectr{s}^{(j)}(\vectr{\beta},t)$, $j=0,1,2$, are bounded, and $s^{(0)}$ is bounded away from $0$ on $\mathcal{B} \times [0,\tau]$.
	\item[\bf A.5] The matrix
	$$\matrx{\Sigma}(\vectr{\beta}^o)=\int_{0}^{\tau}\left[\matrx{s}^{(2)}(\vectr{\beta}^o,t) -  \frac{\left\{\vectr{s}^{(1)}(\vectr{\beta}^o,t)\right\}^{\otimes2}}{s^{(0)}(\vectr{\beta}^o,t)}\right]\lambda_0(t)dt$$ 
	is positive definite.
	\item[\bf A.6] The failure time $V$ and censoring time $C$ are conditionally independent given the covariates $\vectr{X}$. Additionally, when the data also include delayed entry, the entry and failure times are conditionally quasi-independent \citep{tsai1990testing} given the covariates $\vectr{X}$.
	
	\item[\bf A.7] 
	Non-emptyness of the risk set. Namely,  $E\{ Y_i (\tau) \} > 0$ for all $i=1,\ldots,n$.
	\\
	
	A.1--A.7 are the standard assumptions for the consistency and asymptotic normality of Cox regression in the unconditional space, see for instance \cite{fleming2011counting}.
	
	\item[\bf A.8] 
	As $n\rightarrow\infty$, $p_in$ is conditionally bounded away from 0 for all $i\in\mathcal{C}$, where $\vectr{p}=(p_1,\ldots,p_{n_c})^T$ is a vector of conditionally deterministic probabilties, given the data $\mathcal{D}_n$.
	\item[\bf A.9]
	The Hessian matrices $\partial^2 l(\vectr{\beta}) / (\partial\vectr{\beta}^T \partial \vectr{\beta})$ and $\partial^2 l^*(\vectr{\beta}) / (\partial\vectr{\beta}^T \partial \vectr{\beta})$ are non-singular with probability going to 1, as $n\rightarrow\infty$ and $q\rightarrow\infty$.
	
	\item[\bf A.10]
	$n_e/n$ and $q/n$ converge to some small positive constants as $q, n\rightarrow\infty$.
\end{itemize}
Assumptions A.8-A.10 are specific to our case. A.8 guarantees that the sampling probabilities do not approach 0 too fast as the sample size increases. For instance, when the sampling probabilities are uniform, $p_in=n/n_c$, and the condition is indeed satisfied. We do, however, allow for a special set of censored observations be assigned a sampling probability of 0, as discussed in the next paragraph. A.9 ensures that as the sample size increases, the subsample-based and full-sample-based information matrices are invertible. A.10 indicates that the full data is a simple random sample from the target population, so the proportion of observed failure times tends to its frequency in the population.

Generally speaking, for the subsampling estimator to be consistent, each observation should be assigned a sampling probability strictly greater than 0. This condition is a version of the so-called ``positivity'' assumption in the theory of inverse probability weighting (IPW) estimators.  However, some observations contribute no information whatsoever to the PL-based estimating equations, and no harm will be done if they are assigned a sampling probability of 0. For instance, observations censored before the first observed failure time, contribute no information at all. Other cases of censored observations that contribute no information will be discussed where relevant in Section 3. Assigning sampling probabilities of 0 is thus allowed exactly for that subset of observations, and the theory developed below holds under this relaxation of the positivity condition. For those methods allowing 0 sampling probabilities, it is as if this subset of observations was removed in advance. This way, the PL remains unaffected, but for the theoretical results, it may be assumed that no observation has a sampling probability of 0. 

Theorem 1 establishes the consistency of the subsample-based estimators to their full-sample counterparts, conditionally on the observed data.  
\begin{theorem}
	Conditionally on $\mathcal{D}_n$ and given that A.3 and A.7--A.9 hold, as $q\rightarrow\infty$ and $n\rightarrow\infty$, 
	\begin{equation} \label{Theorem1:part1}
	\left\Vert\tilde{\vectr{\beta}} - \hat{\vectr{\beta}}_{PL}\right\Vert_2 = O_{P|\mathcal{D}_n}(q^{-1/2}) 
	\end{equation}
	and, for each $t\in[0,\tau]$,
	\begin{equation} \label{Theorem1:part2}
	\hat{\Lambda}_0(t,\tilde{\vectr{\beta}}) - \hat{\Lambda}_0(t,\hat{\vectr{\beta}}_{PL}) = O_{P|\mathcal{D}_n}(q^{-1/2}) \, ,
	\end{equation}
	where $P|\mathcal{D}_n$ stands for the conditional probability measure given $\mathcal{D}_n$.
\end{theorem}

The proof of Theorem 1 requires the following three lemmas.

\begin{lemma}
	Based on A.3 and A.8, and for a fixed vector of coefficients $\vectr{\beta}$, it holds that 
	\begin{equation} \label{lemma1.1}
	\sup_{t\in[0,\tau], \vectr{\beta}\in\mathcal{B}}\frac{1}{n}\left\lvert\vectr{S}^{(k)}_w(\vectr{\beta},t) - \vectr{S}^{(k)}(\vectr{\beta},t)\right\rvert = O_{P|\mathcal{D}_n}(q^{-1/2})
	\end{equation}
	for $k=0,1,2$.
\end{lemma}
For $k=1,2$, the lemma is in an element-wise sense in the respective vector/matrix. 
\begin{proof}[Proof of Lemma 1]
	The proof will be provided for $k=1$, but similar steps can be repeated for $k=0,2$, where in $k=2$ it is done by treating a general element in the matrix. Let us first rewrite the expression in Eq.(\ref{lemma1.1}) in the following manner
	\begin{equation} \label{lemma1.2}
	\frac{1}{n}\left\{\vectr{S}^{(1)}_w(\vectr{\beta},t) - \vectr{S}^{(1)}(\vectr{\beta},t)\right\} = \frac{1}{n}\sum_{i=1}^ne^{\vectr{\beta}^T\vectr{X}_i}Y_i(t)\vectr{X}_i\left(w_iR_i - 1\right) \, ,
	\end{equation}
	and observe that its conditional expectation is 0, since $E(R_i|\mathcal{D}_n) = w_i^{-1}$. Examining the conditional variance, it holds that,
	\begin{eqnarray}
	&& Var\left[\frac{1}{n}\left\{\vectr{S}^{(1)}_w(\vectr{\beta},t) - \vectr{S}^{(1)}(\vectr{\beta},t)\right\} |\mathcal{D}_n\right] \nonumber \\ 
	&& \hspace{0.4cm} = \frac{1}{n^2}\left\{\sum_{i\in\mathcal{C}}\frac{1}{p_iq}e^{2\vectr{\beta}^T\vectr{X}_i}Y_i(t)\vectr{X}_i^{\otimes2}  - \sum_{i,j\in\mathcal{C}}\frac{1}{q}e^{\vectr{\beta}^T(\vectr{X}_i + \vectr{X}_j)}Y_i(t)Y_j(t)\vectr{X}_i\vectr{X}_j^T \right\} \nonumber \\
	&& \hspace{0.4cm} = O_{|\mathcal{D}_n}(q^{-1})\nonumber \, ,
	\end{eqnarray}
	where $O_{|\mathcal{D}_n}$ stands for the standard big-O notation (not in the probabilistic sense), in the conditional space. The last equality stems from A.3 and A.8, and based on Chebyshev's inequality, $n^{-1}|\vectr{S}^{(1)}_w(\vectr{\beta},t) - \vectr{S}^{(1)}(\vectr{\beta},t)| = O_{P|\mathcal{D}_n}(q^{-1/2})$. Finally, the time $t$ affects only the conditionally-deterministic part in Eq.(\ref{lemma1.2}), $e^{\vectr{\beta}^T\vectr{X}_i(t)}Y_i(t)\vectr{X}_i(t)$, which is bounded due to A.1 and A.3, so the result holds also for the supremum over $t$, and the proof of Lemma 1 is complete.
\end{proof}
\begin{lemma} If A.3 and A.7--A.8 are satisfied, then conditionally on $\mathcal{D}_n$ it holds that
	
	$$\frac{1}{n} \frac{\partial l^*(\hat{\vectr{\beta}}_{PL})}{\partial \vectr{\beta}^T} = O_{P|\mathcal{D}_n}(q^{-1/2}) \, ,$$		
in an element-wise sense. 
\end{lemma}
For clarification, $\partial l^*(\hat{\vectr{\beta}}_{PL})/\partial \vectr{\beta}^T = \partial l^*(\vectr{\beta})/\partial \vectr{\beta}^T|_{\vectr{\beta}=\hat{\vectr{\beta}}_{PL}}$, and this style is adopted throughout this work for notational convenience. Lemma 2 shows that the subsample-based pseudo-score function tends to 0 at the value $\hat{\vectr{\beta}}_{PL}$, the full-sample PL estimator.
\begin{proof}[Proof of Lemma 2] First, let us write
	\begin{equation} \label{lemma2}
	\frac{1}{n} \frac{\partial l^*(\hat{\vectr{\beta}}_{PL})}{\partial \vectr{\beta}^T} = \frac{1}{n}\int_{0}^\tau\left\{\vectr{X}_i - \frac{n^{-1}\vectr{S}_w^{(1)}(\hat{\vectr{\beta}}_{PL},t)}{n^{-1}S_w^{(0)}(\hat{\vectr{\beta}}_{PL},t)}\right\}dN_.(t) \, .
	\end{equation}
	In the conditional space, $N_.(t)$ is deterministic for all $t$, and so is $\hat{\vectr{\beta}}_{PL}$. For the integrand of Eq.(\ref{lemma2}), a first order Taylor expansion for the bivariate function $f(x,y)=x/y$, where $x= n^{-1}\vectr{S}_w^{(1)}(\hat{\vectr{\beta}}_{PL},t)$, $y = n^{-1}S_w^{(0)}(\hat{\vectr{\beta}}_{PL},t)$, about $\left(n^{-1}S^{(0)}(\hat{\vectr{\beta}}_{PL},t),n^{-1}\vectr{S}^{(1)T}(\hat{\vectr{\beta}}_{PL},t)\right)^T$ yields
	\begin{eqnarray}
	&&\frac{1}{n}\int_{0}^\tau\left[\vectr{X}_i - \frac{n^{-1}\vectr{S}^{(1)}(\hat{\vectr{\beta}}_{PL},t)}{n^{-1}S^{(0)}(\hat{\vectr{\beta}}_{PL},t)} + \frac{1}{\eta_t}n^{-1}\left\{\vectr{S}_w^{(1)}(\hat{\vectr{\beta}}_{PL},t) - \vectr{S}^{(1)}(\hat{\vectr{\beta}}_{PL},t)\right\} \nonumber \right. \\
	&& \hspace{0.7cm} \left. - \frac{\vectr{\xi}_t}{\eta_t^2}n^{-1}\left\{S_w^{(0)}(\hat{\vectr{\beta}}_{PL},t) - S^{(0)}(\hat{\vectr{\beta}}_{PL},t)\right\}\right]dN_.(t) \nonumber \\
	&& \hspace{0.4cm} = \frac{1}{n} \frac{\partial l(\hat{\vectr{\beta}}_{PL})}{\partial \vectr{\beta}^T} + \frac{1}{n}\int_{0}^\tau\left[ \frac{1}{\eta_t}n^{-1}\left\{\vectr{S}_w^{(1)}(\hat{\vectr{\beta}}_{PL},t) - \vectr{S}^{(1)}(\hat{\vectr{\beta}}_{PL},t)\right\} \nonumber \right. \\
	&& \hspace{0.7cm} \left. - \frac{\vectr{\xi}_t}{\eta_t^2}n^{-1}\left\{S_w^{(0)}(\hat{\vectr{\beta}}_{PL},t) - S^{(0)}(\hat{\vectr{\beta}}_{PL},t)\right\}\right]dN_.(t) \nonumber \, ,
	\end{eqnarray}
	where $(\eta_t,\vectr{\xi}^T_t)^T$ is on the line segment between $\left(n^{-1}S^{(0)}_w(\hat{\vectr{\beta}}_{PL},t),n^{-1}\vectr{S}^{(1)T}_w(\hat{\vectr{\beta}}_{PL},t)\right)^T$ and $\left(n^{-1}S^{(0)}(\hat{\vectr{\beta}}_{PL},t),n^{-1}\vectr{S}^{(1)T}(\hat{\vectr{\beta}}_{PL},t)\right)^T$. Thanks to continuity, $\eta_t$ and $\vectr{\xi}_t$ converge to $n^{-1}\vectr{S}^{(k)}(\hat{\vectr{\beta}}_{PL},t)$, $k=0,1$, respectively, and therefore $\eta^{-1}_t$ and $\vectr{\xi}_t\eta_t^{-2}$ are conditionally bounded in probability due to the continuous mapping theorem, A.3 and A.7. Therefore, based on Lemma 1, we have proven Lemma 2.
\end{proof}

For Lemma 3 denote
\begin{eqnarray}
\matrx{\mathcal{I}}(\vectr{\beta}) &=& \frac{1}{n}\frac{\partial^2l(\vectr{\beta})}{\partial\vectr{\beta}^T\partial\vectr{\beta}} = 
-\frac{1}{n}\int_0^{\tau}\left\{\frac{\matrx{S}^{(2)}(\vectr{\beta},t)}{S^{(0)}(\vectr{\beta},t)} - \left(\frac{\vectr{S}^{(1)}(\vectr{\beta},t)}{S^{(0)}(\vectr{\beta},t)}\right)^{\otimes2}\right\}dN_.(t) \nonumber
\end{eqnarray}
and,
\begin{eqnarray}
\tilde{\matrx{\mathcal{I}}}(\vectr{\beta}) &=& \frac{1}{n}\frac{\partial^2l^*(\vectr{\beta})}{\partial\vectr{\beta}^T\partial\vectr{\beta}} = -\frac{1}{n}\int_0^{\tau}\left\{\frac{\matrx{S}_w^{(2)}(\vectr{\beta},t)}{S_w^{(0)}(\vectr{\beta},t)} - \left(\frac{\vectr{S}_w^{(1)}(\vectr{\beta},t)}{S_w^{(0)}(\vectr{\beta},t)}\right)^{\otimes2} \right\}dN_.(t) \nonumber \, .
\end{eqnarray}

Additionally, denote, 
$$S_{k}^{(1)}(\vectr{\beta},t) = \sum_{i=1}^{n} e^{\vectr{\beta}^T\vectr{X}_i}Y_i(t)X_{ik} \, ,$$
$$S_{wk}^{(1)}(\vectr{\beta},t) = \sum_{i=1}^{n} w_i e^{\vectr{\beta}^T\vectr{X}_i}Y_i(t)X_{ik}R_i \, ,$$

where $X_{ik}$ stands for the $k$'th element in the vector $\vectr{X}_i$, $k=1,\ldots,r$.
\begin{lemma}
	If A.3 and A.7--A.8 are satisfied, then conditionally on $\mathcal{D}_n$, for a vector of fixed coefficients $\vectr{\beta}$,
	\begin{equation} \label{lemma3}
	\tilde{\matrx{\mathcal{I}}}(\vectr{\beta}) - \matrx{\mathcal{I}}(\vectr{\beta}) = O_{P|\mathcal{D}_n}(q^{-1/2}) \, ,
	\end{equation}
	in the sense that it holds for each element in the matrix. 
\end{lemma}
Lemma 3 shows that the subsample-based observed information matrix for $\vectr{\beta}$ converges to the corresponding observed information matrix based on the full sample. 

\begin{proof}[Proof of Lemma 3]
	\begin{eqnarray}
	\tilde{\matrx{\mathcal{I}}}(\vectr{\beta}) - \matrx{\mathcal{I}}(\vectr{\beta}) &=& \frac{1}{n}\int_0^{\tau}\left\{\frac{\matrx{S}^{(2)}(\vectr{\beta},t)}{S^{(0)}(\vectr{\beta},t)} - \frac{\matrx{S}_w^{(2)}(\vectr{\beta},t)}{S_w^{(0)}(\vectr{\beta},t)}\right\}dN_.(t) \nonumber \\
	&& -  \frac{1}{n}\int_0^{\tau}\left[\left\{\frac{\vectr{S}^{(1)}(\vectr{\beta},t)}{S^{(0)}(\vectr{\beta},t)}\right\}^{\otimes2} -	\left\{\frac{\vectr{S}_w^{(1)}(\vectr{\beta},t)}{S_w^{(0)}(\vectr{\beta},t)}\right\}^{\otimes2}\right]dN_.(t) \, . \nonumber
	\end{eqnarray}
	The first addend can be shown to be $O_{P|\mathcal{D}_n}(q^{-1/2})$ in an identical manner to Lemma 1, by examining a general element in the matrix. We now turn to showing that the second addend is also $O_{P|\mathcal{D}_n}(q^{-1/2})$. Let us examine a general element in the matrix, in the $m$'th row and the $l$'th column and use a Taylor expansion for the trivariate function $f(x,y,z) = xy/z^2$, where $x=n^{-1}S_{wm}^{(1)}(\vectr{\beta},t) $, $y=n^{-1}S_{wl}^{(1)}(\vectr{\beta},t)$ and $z=n^{-1}S_w^{(0)}(\vectr{\beta},t)$, about $\left(n^{-1}S_{m}^{(1)}(\vectr{\beta},t),n^{-1}S_{l}^{(1)}(\vectr{\beta},t),n^{-1}S^{(0)}(\vectr{\beta},t)\right)^T$. That is,
	\begin{eqnarray}
	&& \frac{1}{n}\int_0^\tau\left[\left\{\frac{n^{-1}\vectr{S}^{(1)}(\vectr{\beta},t)}{n^{-1}S^{(0)}(\vectr{\beta},t)}\right\}^{\otimes2}_{ml} -	\left\{\frac{n^{-1}\vectr{S}_w^{(1)}(\vectr{\beta},t)}{n^{-1}S_w^{(0)}(\vectr{\beta},t)}\right\}^{\otimes2}_{ml}\right]dN_.(t)  \nonumber \\
	&& \hspace{0.4cm} = \frac{1}{n}\int_0^\tau\left[\frac{\xi_t}{\omega_t^2}\left\{n^{-1}S_{wm}^{(1)}(\vectr{\beta},t) - n^{-1}S_m^{(1)}(\vectr{\beta},t)\right\} \right. \nonumber \\
	&& \hspace{0.7cm} \left. + \frac{\eta_t}{\omega_t^2}\left\{n^{-1}S_{wl}^{(1)}(\vectr{\beta},t) - n^{-1}S_l^{(1)}(\vectr{\beta},t)\right\} \right. \nonumber \\
	&& \hspace{0.7cm}\left. - \frac{\xi_t\eta_t}{\omega_t^3}\left\{n^{-1}S_w^{(0)}(\vectr{\beta},t) - n^{-1}S^{(0)}(\vectr{\beta},t)\right\} \right]dN_.(t) \, , \nonumber 
	\end{eqnarray}
	where $(\eta_t$, $\xi_t$, $\omega_t)^T$ is on the line segment connecting $\left(n^{-1}S_{wm}^{(1)}(\vectr{\beta},t), n^{-1}S_{wl}^{(1)}(\vectr{\beta},t), \right. $ $\left. n^{-1}S_w^{(0)}(\vectr{\beta},t)\right)^T$ with $\left(n^{-1}S_m^{(1)}(\vectr{\beta},t), n^{-1}S_l^{(1)}(\vectr{\beta},t), n^{-1}S^{(0)}(\vectr{\beta},t)\right)^T$. Based on the continuous mapping theorem, Lemma 1, A.3 and A.7, Eq.(\ref{lemma3}) follows.
\end{proof}

We are now in position to prove Eq.(\ref{Theorem1:part1}) of Theorem 1.
\begin{proof}[Proof of Theorem 1(1)]
	First, let us introduce some additional notation as follows,
	
	$$S_{wkl}^{(1)}(\vectr{\beta},t) = \sum_{i=1}^{n} w_i e^{\vectr{\beta}^T\vectr{X}_i}Y_i(t)X_{ik}X_{il}R_i \, ,$$ and
	$$S_{wklm}^{(1)}(\vectr{\beta},t) = \sum_{i=1}^{n} w_i e^{\vectr{\beta}^T\vectr{X}_i}Y_i(t)X_{ik}X_{il}X_{im}R_i \, .$$
	
	Denote $l^{*'}_k(\vectr{\beta})$ as the derivative of
	$l^*(\vectr{\beta})$ with respect to $\beta_k$, and $\tilde{\matrx{\mathcal{I}}}_k$ as the $k$'th row of the matrix $\tilde{\matrx{\mathcal{I}}}$. A first order Taylor expansion for $n^{-1}l^{*'}_k(\tilde{\vectr{\beta}})$ about $\hat{\vectr{\beta}}_{PL}$ is
	\begin{equation} \label{Taylor_deriv_l*m_bhat}
	0=\frac{1}{n}l^{*'}_k(\tilde{\vectr{\beta}})=\frac{1}{n}l^{*'}_k(\hat{\vectr{\beta}}_{PL}) + \tilde{\matrx{\mathcal{I}}}_k^T(\hat{\vectr{\beta}}_{PL})(\tilde{\vectr{\beta}} - \hat{\vectr{\beta}}_{PL}) + \frac{1}{n}\mbox{Res}_k
	\end{equation}
	and 
	\[
	\frac{1}{n}\mbox{Res}_k= (\tilde{\vectr{\beta}} - \hat{\vectr{\beta}}_{PL})^T\frac{\partial \tilde{\matrx{\mathcal{I}}}_k(\vectr{\xi})}{\partial\vectr{\beta}^T}(\tilde{\vectr{\beta}} - \hat{\vectr{\beta}}_{PL}) \, ,
	\]
	where $\vectr{\xi}$ is on the line segment between $\tilde{\vectr{\beta}}$ and $\hat{\vectr{\beta}}_{PL}$, $k=1,\ldots,r$. Examining a general element of $\partial\tilde{\matrx{\mathcal{I}}}_k(\vectr{\beta}) / \partial\vectr{\beta}^T$, we get
	\begin{eqnarray}
	\frac{1}{n}\frac{\partial^2 l^{*'}_k(\vectr{\beta})}{\partial \vectr{\beta}_l \partial \vectr{\beta}_k} &=& 
	\frac{1}{n}\int_0^{\tau}\frac{1}{S_w^{(0)3}(\vectr{\beta},t)}\left[S_{wklm}^{(1)}(\vectr{\beta},t)S_w^{(0)2}(\vectr{\beta},t) + \left\{S_{wkl}^{(1)}(\vectr{\beta},t)S_{wm}^{(1)}(\vectr{\beta},t) \right. \right. \nonumber \\ 
	&& \left. \left. + S_{wkm}^{(1)}(\vectr{\beta},t)S_{wl}^{(1)}(\vectr{\beta},t) + S_{wlm}^{(1)}(\vectr{\beta},t)S_{wk}^{(1)}(\vectr{\beta},t)\right\}S_w^{(0)}(\vectr{\beta},t) \right. \nonumber \\ 
	&& \left. -2S_{wk}^{(1)}(\vectr{\beta},t)S_{wl}^{(1)}(\vectr{\beta},t)S_{wm}^{(1)}(\vectr{\beta},t)  \right]dN_.(t) \, . \nonumber
	\end{eqnarray}
	
	Now it will be shown that $\partial\tilde{\matrx{\mathcal{I}}}_k(\vectr{\xi}) / \partial\vectr{\beta}^T$ is bounded in conditional probability.
	Let us note that based on Lemma 1, the continuous mapping theorem and A.7, the subsample-based pseudo-score function $n^{-1}\partial l^*(\vectr{\beta})/\partial \vectr{\beta}^T$ converges in conditional probability to the full sample score function $n^{-1}\partial l(\vectr{\beta})\partial \vectr{\beta}^T$. Since $\mathcal{B}$ is compact and the function $l(\vectr{\beta})$ is continuous and convex, $\hat{\vectr{\beta}}_{PL}$ is its unique global maximizer, then based on \citet[Theorem 5.9, page 46]{vaart_1998} we know that 
	\begin{equation} \label{condconsis}
		\|\tilde{\vectr{\beta}} - \hat{\vectr{\beta}}_{PL}\|_2 = o_{P|\mathcal{D}_n}(1).
	\end{equation} 
	This result establishes that $\tilde{\vectr{\beta}}$ is indeed consistent for $\hat{\vectr{\beta}}_{PL}$ in the conditional space, but does not teach us about its rate of convergence. Due to continuity, and since $\vectr{\xi}$ is on the line segment between $\tilde{\vectr{\beta}}$ and $\hat{\vectr{\beta}}_{PL}$,  it follows that $\|\vectr{\xi} - \hat{\vectr{\beta}}_{PL}\|_2 = o_{P|\mathcal{D}_n}(1)$, hence $\vectr{\xi} = O_{P|\mathcal{D}_n}(1)$. Based on this observation, A.3 and A.8, one can verify that the functions $n^{-1}\vectr{S}^{(k)}_w(\vectr{\xi},t)$, $k=0,1,2$, are all bounded in conditional probability, being the mean of $q$ conditionally independent and bounded addends, plus the sum of $n_e$ bounded constants divided by $n$. Namely, let us write $$n^{-1}\vectr{S}^{(0)}_w(\vectr{\xi},t) = \frac{1}{q}\sum_{i=1}^q\frac{e^{\vectr{\xi}^TX^*_i}Y_i^*(t)}{np_i^*} + \frac{1}{n}\sum_{i \in \mathcal{E}}e^{\vectr{\xi}^TX_i}Y_i(t) \, ,$$ where the ``*" sign denotes sampling with replacement, and similarly the same presentation holds for $n^{-1}\vectr{S}^{(k)}_w(\vectr{\xi},t)$, $k=1,2$, $n^{-1}S_{wkl}^{(1)}(\vectr{\xi},t)$ and $n^{-1}S_{wklm}^{(1)}(\vectr{\xi},t)$. Therefore, and also based on A.7, we find that $\partial\tilde{\matrx{\mathcal{I}}}_k(\vectr{\xi}) / \partial\vectr{\beta}^T = O_{P|\mathcal{D}_n}(1)$, and as a result, for all $k=1,\ldots,r$,
	\begin{equation} \label{Res is Onorm}
	\frac{1}{n}\mbox{Res}_k = O_{P|\mathcal{D}_n}\left(\lVert\tilde{\vectr{\beta}} - \hat{\vectr{\beta}}_{PL}\rVert_2^2\right) \, .
	\end{equation}
	From Eq.'s (\ref{Taylor_deriv_l*m_bhat})--(\ref{Res is Onorm}) and A.9 we obtain
	\begin{equation} \label{beta tilde - beta PL}
	\tilde{\vectr{\beta}} - \hat{\vectr{\beta}}_{PL} = -\tilde{\matrx{\mathcal{I}}}^{-1}(\hat{\vectr{\beta}}_{PL})\left\{ \frac{1}{n}\frac{\partial l^*(\hat{\vectr{\beta}}_{PL})}{\partial \vectr{\beta}^T}  + O_{P|\mathcal{D}_n}\left(\lVert\tilde{\vectr{\beta}} - \hat{\vectr{\beta}}_{PL}\rVert_2^2\right) \right\} \, .
	\end{equation}
	Since matrix inversion is a continuous operation, then due to Lemma 3 and the continuous mapping theorem, $\tilde{\matrx{\mathcal{I}}}^{-1}(\hat{\vectr{\beta}}_{PL})-{\matrx{\mathcal{I}}}^{-1}(\hat{\vectr{\beta}}_{PL})=o_{P|\mathcal{D}_n}(1)$, which yields $\tilde{\matrx{\mathcal{I}}}^{-1}(\hat{\vectr{\beta}}_{PL})=O_{P|\mathcal{D}_n}(1)$. Therefore, combining Lemmas 1--2, Eq.(\ref{condconsis}) and Eq.(\ref{beta tilde - beta PL}), it holds that
	\begin{equation}
	\tilde{\vectr{\beta}} - \hat{\vectr{\beta}}_{PL} = O_{P|\mathcal{D}_n}(q^{-1/2}) + o_{P|\mathcal{D}_n}\left( \lVert \tilde{\vectr{\beta}} - \hat{\vectr{\beta}}_{PL}\rVert_2\right) \, ,\nonumber
	\end{equation}
	hence, 
	\begin{equation} \label{theorem1part1final}
	\tilde{\vectr{\beta}} - \hat{\vectr{\beta}}_{PL} = O_{P|\mathcal{D}_n}(q^{-1/2}) \, ,
	\end{equation}
	which completes the proof of Eq.(\ref{Theorem1:part1}) in Theorem 1.
\end{proof}

\begin{proof}[Proof of Theorem 1(2)]
	For proving Eq.(\ref{Theorem1:part2}), let us write for any $t\in(0,\tau)$
	\[
	\hat{\Lambda}_0(t,\tilde{\vectr{\beta}}) - \hat{\Lambda}_0(t,\hat{\vectr{\beta}}_{PL}) =
	\frac{1}{n}\int_{0}^t\left\{\frac{1}{n^{-1}S^{(0)}(\tilde{\vectr{\beta}},u)} - \frac{1}{n^{-1}S^{(0)}(\hat{\vectr{\beta}}_{PL},u)} \right\}dN_.(u) \, .
	\]
	Using a Taylor expansion about $\hat{\vectr{\beta}}_{PL}$ yields
	\begin{equation} \label{Theorem1.2.2}
	\hat{\Lambda}_0(t,\tilde{\vectr{\beta}}) - \hat{\Lambda}_0(t,\hat{\vectr{\beta}}_{PL}) = (\tilde{\vectr{\beta}} - \hat{\vectr{\beta}}_{PL})\frac{1}{n}\int_{0}^t\frac{S^{(1)}(\vectr{\xi},u)}{n^{-1}S^{(0)2}(\vectr{\xi},u)} dN_.(u) \, ,
	\end{equation}
	where $\vectr{\xi}$ is on the line segment between $\tilde{\vectr{\beta}}$ and  $\hat{\vectr{\beta}}_{PL}$. Based on Assumptions A.3, A.7, and Eq.(\ref{theorem1part1final}), Eq.(\ref{Theorem1.2.2}) is a product of a term which is $O_{P|\mathcal{D}_n}(q^{-1/2})$ and a term bounded in conditional probability, thus proving Eq.(\ref{Theorem1:part2}) and completing Theorem 1.
\end{proof}

We now turn to stating and proving Theorem 2, which concerns the asymptotic properties of the subsample-based estimators in the unconditional space.
\begin{theorem}
	Given that A.1--A.10 hold, then as $q \rightarrow \infty$ and $n\rightarrow \infty$,  
	\begin{equation}
	\sqrt{n}\mathbb{V}_{\tilde{\vectr{\beta}}}(\vectr{p},\vectr{\beta}^o)^{-1/2}(\tilde{\vectr{\beta}} - \vectr{\beta}^o) \xrightarrow{D} N(0,I) \label{Theorem2:part1}
	\end{equation}
	where $I$ is the identity matrix. And, for all $t\in[0,\tau]$,
	\begin{equation}
	\sqrt{n}\mathbb{V}_{\hat{\Lambda}_0(t,\tilde{\vectr{\beta}})}(\vectr{p},\vectr{\beta}^o,t)^{-1/2}\{\hat{\Lambda}_0(t,\tilde{\vectr{\beta}}) - \Lambda_0^o(t)\} \xrightarrow{D} N(0,1) \label{Theorem2:part2}
	\end{equation}
\end{theorem}
where $$\mathbb{V}_{\tilde{\vectr{\beta}}}(\vectr{p},\vectr{\beta}) = {\matrx{\mathcal{I}}}^{-1}(\vectr{\beta})+\frac{n}{q}{\matrx{\mathcal{I}}}^{-1}(\vectr{\beta})\matrx{\upvarphi}(\vectr{p},\vectr{\beta}){\matrx{\mathcal{I}}}^{-1}(\vectr{\beta}) \, ,$$ 
$$\matrx{\upvarphi}(\vectr{p},\vectr{\beta})=\frac{1}{n^2}\left\{\sum_{i\in\mathcal{C}}\frac{\vectr{a}_i(\vectr{\beta})\vectr{a}_i(\vectr{\beta})^T}{p_i} - \sum_{i,j\in\mathcal{C}}\vectr{a}_i(\vectr{\beta})\vectr{a}_j(\vectr{\beta})^T\right\}  \, ,$$
$$	\vectr{a}_i(\vectr{\beta})=\int_{0}^\tau \left\{ \vectr{X}_i - \frac{\vectr{S}^{(1)}(\vectr{\beta},t)}{S^{(0)}(\vectr{\beta},t)}\right\}\frac{Y_i(t)e^{\vectr{\beta}^T\vectr{X}_i}}{S^{(0)}(\vectr{\beta},t)}dN_.(t) \, ,$$
$$\mathbb{V}_{\hat{\Lambda}_0(t,\tilde{\vectr{\beta}})}(\vectr{p},\vectr{\beta},t) = \int_0^t\frac{dN_.(u)}{n^{-1}S^{(0)2}(\vectr{\beta},u)} + \frac{n}{q}\vectr{H}^T(\vectr{\beta},t) \mathbb{V}_{\tilde{\vectr{\beta}}}(\vectr{p},\vectr{\beta})\vectr{H}(\vectr{\beta},t) \, , $$
and,
$$\vectr{H}(\vectr{\beta},t) =  \int_0^t\frac{\vectr{S}^{(1)}(\vectr{\beta},u)}{S^{(0)2}(\vectr{\beta},u)}dN_.(u) \, .$$

First, let us show that $\tilde{\vectr{\beta}}$ is consistent to $\vectr{\beta}^o$.
	\begin{lemma} Given that A.1--A.9 hold, then as $q \rightarrow \infty$ and $n\rightarrow \infty$,
		\begin{equation} \label{Lemma5:1}
		\lim_{n,q\rightarrow \infty}\Pr(\Vert \tilde{\vectr{\beta}} - \vectr{\beta}^o \Vert_2 > \epsilon) = 0  \, \, , \forall \epsilon>0 \, .
		\end{equation}
	\end{lemma}
	\begin{proof}[Proof of Lemma 4]
		In Theorem 1 it was established that for all $\epsilon>0$,  $$\lim_{n,q\rightarrow \infty}\Pr\left(\Vert\tilde{\vectr{\beta}} - \hat{\vectr{\beta}}_{PL}\Vert_2 \ge \epsilon |\mathcal{D}_n \right)=0 \, .$$
		In the unconditional probability space, 
		$\Pr\left(\Vert\tilde{\vectr{\beta}} - \hat{\vectr{\beta}}_{PL}\Vert_2 \ge \epsilon |\mathcal{D}_n \right)$ is itself a random variable, let us denote it as $\pi_{n,q}$. It then follows that
		$$\Pr\left(\lim_{n,q\rightarrow \infty}\pi_{n,q} =0 \right)=1 \, ,$$
		meaning that $\pi_{n,q} \xrightarrow[n,q\rightarrow\infty]{a.s} 0 \, .$ Now we have that for all $\epsilon >0$,
		\begin{equation} \label{Theorem2:uncond:betahat}
		\lim_{n,q\rightarrow \infty}\Pr\left(\Vert \tilde{\vectr{\beta}} - \hat{\vectr{\beta}}_{PL} \Vert_2 > \epsilon\right) = \lim_{n,q\rightarrow \infty}E(\pi_{n,q}) = E(\lim_{n,q\rightarrow \infty}\pi_{n,q}) = 0 \, ,
		\end{equation}
		where the interchange of expectation and limit is allowed due to the dominated convergence theorem, since $\pi_{n,q}$ is trivially bounded by $1$. We proceed and write
		\begin{eqnarray}
		\Pr\left(\Vert\tilde{\vectr{\beta}} - \vectr{\beta}^o \Vert_2 \ge \epsilon \right) &=& \Pr\left(\Vert\tilde{\vectr{\beta}} -\hat{\vectr{\beta}}_{PL} +\hat{\vectr{\beta}}_{PL} - \vectr{\beta}^o \Vert_2 \ge \epsilon \right) \nonumber \\
		&\le& \Pr\left(\Vert\tilde{\vectr{\beta}} -\hat{\vectr{\beta}}_{PL}\Vert_2 +\Vert\hat{\vectr{\beta}}_{PL} - \vectr{\beta}^o \Vert_2 \ge \epsilon \right) \nonumber \\ 
		&\le& \Pr\left(\{\Vert\tilde{\vectr{\beta}} -\hat{\vectr{\beta}}_{PL}\Vert_2 \ge \epsilon/2\}\cup \{\Vert\hat{\vectr{\beta}}_{PL} -\vectr{\beta}^o\Vert_2 \ge \epsilon/2 \} \right) \nonumber \\
		&\le& \Pr\left(\Vert\tilde{\vectr{\beta}} -\hat{\vectr{\beta}}_{PL}\Vert_2 \ge \epsilon/2\right) + \Pr\left(\Vert\hat{\vectr{\beta}}_{PL} -\vectr{\beta}^o\Vert_2 \ge \epsilon/2\right) \, . \nonumber
		\end{eqnarray}
		Taking limits on both sides,
		\begin{eqnarray}
		&& \lim_{n,q\rightarrow \infty}\Pr\left(\Vert\tilde{\vectr{\beta}} - \vectr{\beta}^o \Vert_2 \ge \epsilon \right) \nonumber \\ 
		&& \hspace{0.7cm} \le \lim_{n,q\rightarrow \infty}\Pr\left(\Vert\tilde{\vectr{\beta}} -\hat{\vectr{\beta}}_{PL}\Vert_2 \ge \epsilon/2\right) 
		+ 	 \lim_{n,q\rightarrow \infty}\Pr\left(\Vert\hat{\vectr{\beta}}_{PL} -\vectr{\beta}^o\Vert_2 \ge \epsilon/2\right) = 0 \, , \nonumber
		\end{eqnarray}
		where the first addend on the right is $0$ due to Eq.(\ref{Theorem2:uncond:betahat}), and the second addend is $0$ based on known results for Cox regression and A.1--A.7. By that we have shown Eq.(\ref{Lemma5:1}). 
	\end{proof}

\begin{proof}[Proof of Theorem 2]  
	Similarly to Eq.(\ref{beta tilde - beta PL}), and based on Lemma 4, let us use a Taylor expansion for the subsample-based pseudo-score function evaluated at $\tilde{\vectr{\beta}}$, about $\vectr{\beta}^o$, and get
	\begin{equation} \label{beta tilde - beta0}
	\tilde{\vectr{\beta}} - \vectr{\beta}^o = -\tilde{\matrx{\mathcal{I}}}^{-1}(\vectr{\beta}^o)\left\{ \frac{1}{n}\frac{\partial l^*(\vectr{\beta}^o)}{\partial \vectr{\beta}^T}  + o_{P}\left(\lVert\tilde{\vectr{\beta}} - \vectr{\beta}^o\rVert_2\right) \right\} \, .
	\end{equation}
	Similarly to \cite{samuelsen1997psudolikelihood}, we decompose the subsample-based pseudo-score function into two separate components,
	\begin{equation} \label{Theorem2:partialscore}
	\frac{1}{n}\frac{\partial l^*(\vectr{\beta})}{\partial \vectr{\beta}^T} = \frac{1}{n}\frac{\partial l(\vectr{\beta})}{\partial \vectr{\beta}^T} + \frac{1}{n}\sum_{i\in\mathcal{C}}(1-w_iR_i)\breve{\vectr{a}}_i(\vectr{\beta}) \, ,
	\end{equation}
	where 
	\[
	\breve{\vectr{a}}_i(\vectr{\beta})= \int_{0}^\tau \left\{ \vectr{X}_i - \frac{\vectr{S}^{(1)}(\vectr{\beta},t)}{S^{(0)}(\vectr{\beta},t)}\right\}\frac{Y_i(t)e^{\vectr{\beta}^T\vectr{X}_i}}{S_w^{(0)}(\vectr{\beta},t)}dN_.(t) \, .
	\]
Verifying Eq.~(\ref{Theorem2:partialscore}),
	\begin{eqnarray}
		\sum_{i\in\mathcal{C}}(1-w_iR_i)\breve{\vectr{a}}_i(\vectr{\beta}) &=& 
		\sum_{i=1}^n(1-w_iR_i)\breve{\vectr{a}}_i(\vectr{\beta})  \nonumber \\
		&=& \sum_{i=1}^{n}\Delta_i\sum_{j=1}^n \left\{\vectr{X}_j - \frac{\vectr{S}^{(1)}(\vectr{\beta},T_i)}{S^{(0)}(\vectr{\beta},T_i)}\right\}\frac{Y_j(T_i)e^{\vectr{\beta}^{T}\vectr{X}_j}}{S_w^{(0)}(\vectr{\beta},T_i)} \nonumber \\
		&& - \sum_{i=1}^{n}\Delta_i\sum_{j=1}^n  w_jR_j\left\{\vectr{X}_j - \frac{\vectr{S}^{(1)}(\vectr{\beta},T_i)}{S^{(0)}(\vectr{\beta},T_i)}\right	\}\frac{Y_j(T_i)e^{\vectr{\beta}^{T}\vectr{X}_j}}{S_w^{(0)}(\vectr{\beta},T_i)} \nonumber
	\end{eqnarray}
	Let us proceed by exchanging the summation order,
	\begin{eqnarray}
		\sum_{i\in\mathcal{C}}(1-w_iR_i)\breve{\vectr{a}}_i(\vectr{\beta})
		&=& \sum_{i=1}^n\Delta_i\left\{\frac{\vectr{S}^{(1)}(\vectr{\beta},T_i)}{S_w^{(0)}(\vectr{\beta},T_i)} - \frac{\vectr{S}^{(1)}(\vectr{\beta},T_i)}{S^{(0)}(\vectr{\beta},T_i)}\frac{S^{(0)}(\vectr{\beta},T_i)}{S_w^{(0)}(\vectr{\beta},T_i)}\right\} \nonumber \\
		&& - \sum_{i=1}^n\Delta_i\left\{\frac{\vectr{S}_w^{(1)}(\vectr{\beta},T_i)}{S_w^{(0)}(\vectr{\beta},T_i)} - \frac{\vectr{S}^{(1)}(\vectr{\beta},T_i)}{S^{(0)}(\vectr{\beta},T_i)}\frac{S_w^{(0)}(\vectr{\beta},T_i)}{S_w^{(0)}(\vectr{\beta},T_i)}\right\} \nonumber \\
		&=& \sum_{i=1}^n\Delta_i\left\{\frac{\vectr{S}^{(1)}(\vectr{\beta},T_i)}{S^{(0)}(\vectr{\beta},T_i)} - \frac{\vectr{S}_w^{(1)}(\vectr{\beta},T_i)}{S_w^{(0)}(\vectr{\beta},T_i)} \right\} \nonumber \\
		&=& \frac{\partial l^*(\vectr{\beta})}{\partial \vectr{\beta}^T} - \frac{\partial l(\vectr{\beta})}{\partial \vectr{\beta}^T} \nonumber \, .
	\end{eqnarray}
	So, based on Eq's.(\ref{beta tilde - beta0})--(\ref{Theorem2:partialscore}) we can write
	\begin{equation} \label{Theorem2:beta_tilde - beta0:decomp}
	\sqrt{n}(\tilde{\vectr{\beta}} - \vectr{\beta}^o) = -\tilde{\matrx{\mathcal{I}}}^{-1}(\vectr{\beta}^o)\left[\frac{1}{\sqrt{n}}\frac{\partial l(\vectr{\beta}^o)}{\partial \vectr{\beta}^T} + \frac{1}{\sqrt{n}}\sum_{i\in\mathcal{C}}(1-w_iR_i)\breve{\vectr{a}}_i(\vectr{\beta}^o)\right] +  o_P(\sqrt{n}\Vert \tilde{\vectr{\beta}} - \vectr{\beta}^o \Vert_2) \, .
	\end{equation}
	We will now study the two addends $n^{-1/2}\partial l(\vectr{\beta}^o)/\partial \vectr{\beta}^T$ and $n^{-1/2}\sum_{i\in\mathcal{C}}(1-w_iR_i)\breve{\vectr{a}}_i(\vectr{\beta}^o)$, and it will be established that they are asymptotically independent, and each one is asymptotically normal, hence we will be able to characterize the asymptotic distribution of $\sqrt{n}(\tilde{\vectr{\beta}} - \vectr{\beta}^o)$.
	Let us examine the first addend. From standard Cox regression results, assuming A.1--A.7 hold,  ${\left\{-\matrx{\mathcal{I}}(\vectr{\beta}^o)\right\}^{-1/2}}n^{-1/2}\partial l(\vectr{\beta}^o)/\partial \vectr{\beta}^T\xrightarrow{D} N(0,I)$.
Therefore,  
\begin{equation} \label{Theorem2:firstaddend:norm}
n^{-1/2}\partial l(\vectr{\beta}^o)/\partial \vectr{\beta}^T \xrightarrow{D} N(0,\matrx{\Sigma}(\vectr{\beta^o})) \, ,
\end{equation}
where $\matrx{\Sigma}(\vectr{\beta})$ is defined in A.5. 

Our goal now is to show that the second addend is asymptotically normal. It should be emphasized that since the observations are sampled based on their status at the end of follow-up time, the weights are not predictable, and standard martingale theory does not apply here, so instead we pursue a different direction.
	First, one can write
	\begin{equation}\label{eq:atilde}
	\frac{\sqrt{q}}{n} \sum_{i\in\mathcal{C}}(1-w_iR_i)\breve{\vectr{a}}_i(\vectr{\beta}) = \frac{\sqrt{q}}{n}\int_{0}^{\tau}\frac{\vectr{S}^{(1)}(\vectr{\beta},t)S_w^{(0)}(\vectr{\beta},t) - \vectr{S}_w^{(1)}(\vectr{\beta},t)S^{(0)}(\vectr{\beta},t)}{S^{(0)}(\vectr{\beta},t)S_w^{(0)}(\vectr{\beta},t)}dN_.(t) \, ,
	\end{equation}
	and
	\begin{equation}\label{eq:a}
	\frac{\sqrt{q}}{n} \sum_{i\in\mathcal{C}}(1-w_iR_i)\vectr{a}_i(\vectr{\beta}) = \frac{\sqrt{q}}{n}\int_{0}^{\tau}\frac{\vectr{S}^{(1)}(\vectr{\beta},t)S_w^{(0)}(\vectr{\beta},t) - \vectr{S}_w^{(1)}(\vectr{\beta},t)S^{(0)}(\vectr{\beta},t)}{\left\{S^{(0)}(\vectr{\beta},t)\right\}^2}dN_.(t) \, .
	\end{equation}
	By means of the functional delta method, conditionally on $\mathcal{D}_n$, it can be shown that Eq.(\ref{eq:atilde}) is asymptotically normally distributed, with mean 0 and covariance matrix identical to that of Eq.(\ref{eq:a}). For the sake of simpler presentation and derivation, we show the asymptotic normality of Eq.(\ref{eq:a}) instead and identify its covariance matrix, and it will follow that Eq.(\ref{eq:atilde}) and Eq.(\ref{eq:a}) are asymptotically equivalent. From the conditional space, we proceed to the unconditional space and establish the corresponding results there.
	Importantly, $\vectr{a}_i(\vectr{\beta}^o)$ is constant upon conditioning on $\mathcal{D}_n$, so $$\frac{\sqrt{q}}{n}\sum_{i\in\mathcal{C}}w_iR_i\vectr{a}_i(\vectr{\beta}) $$ can be alternatively expressed as a sum of $q$ iid observations in the conditional space, marked with a ``*" sign, as follows, $$\frac{\sqrt{q}}{n}\sum_{i=1}^qw_i^*\vectr{a}^*_i(\vectr{\beta}^o) = \frac{1}{\sqrt{q}}\sum_{i=1}^q\frac{\vectr{a}^*_i(\vectr{\beta}^o)}{np_i^*} \equiv \frac{1}{\sqrt{q}}\sum_{i=1}^q\vectr{\gamma}_i(\vectr{p},\vectr{\beta}^o) \, .$$
	Since the distribution of $\vectr{\gamma}_i(\vectr{p},\vectr{\beta}^o)$ changes as $n$ increases, the Lindeberg-Feller \citep[proposition 2.27]{vaart_1998} condition should be established, as it covers the settings of triangular arrays.
	First, denote $\matrx{\upvarphi}(\vectr{p},\vectr{\beta}) \equiv Var(\vectr{\gamma}(\vectr{p},\vectr{\beta})|\mathcal{D}_n)$, and let us show that $\matrx{\upvarphi}(\vectr{p},\vectr{\beta}^o)$ is conditionally bounded.
	\begin{eqnarray}
	\matrx{\upvarphi}(\vectr{p},\vectr{\beta}^o) &=& E\left(\vectr{\gamma}(\vectr{p},\vectr{\beta}^o)\vectr{\gamma}^T(\vectr{p},\vectr{\beta}^o)|\mathcal{D}_n\right) - E\left(\vectr{\gamma}(\vectr{p},\vectr{\beta}^o)|\mathcal{D}_n\right)E\left(\vectr{\gamma}(\vectr{p},\vectr{\beta}^o)|\mathcal{D}_n\right)^T  \nonumber \\
	& =& \frac{1}{n^2}\left\{\sum_{i\in\mathcal{C}}\frac{\vectr{a}_i(\vectr{\beta}^o)\vectr{a}_i(\vectr{\beta}^o)^T}{p_i} - \sum_{i,j\in\mathcal{C}}\vectr{a}_i(\vectr{\beta}^o)\vectr{a}_j(\vectr{\beta}^o)^T\right\}  =  O_{|\mathcal{D}_n}(1) \, , \nonumber
	\end{eqnarray}
	where the last equality is due to {A.2--A.3, A.7--A.8}. It is interesting to note, that the sum of $\vectr{a}_i(\hat{\vectr{\beta}}_{PL})$ over all observations is equal to $\vectr{0}$, but in our case the sum is not $\vectr{0}$ because only the censored observations appear in the summation, and because the $\vectr{a}_i$'s are evaluated at the true parameter value $\vectr{\beta^o}$.
	Now, for every $\epsilon > 0$, and for some $\delta>0$,
	\begin{eqnarray}
	&& \sum_{j=1}^qE\left\{\Vert q^{-1/2}\vectr{\gamma}_i(\vectr{p},\vectr{\beta}^o)\Vert_2^2I\left(\Vert q^{-1/2}\vectr{\gamma}_i(\vectr{p},\vectr{\beta}^o)\Vert_2 > \epsilon\right)|\mathcal{D}_n \right\} \nonumber \\
	&& \hspace{0.4cm} \le \frac{1}{q^{1+\delta/2}\epsilon^\delta}\sum_{i=1}^qE\left\{\left\lVert \frac{\vectr{a}^*_i(\vectr{\beta}^o)}{np^*_i}\right\rVert_2^{2+\delta} \bigg|\mathcal{D}_n\right\} \nonumber \\
	&& \hspace{0.4cm} = \frac{1}{q^{\delta/2}\epsilon^\delta n^{2+\delta}}\sum_{i\in\mathcal{C}}\frac{\Vert \vectr{a}_i(\vectr{\beta}^o)\Vert_2^{2+\delta}}{p_i^{1+\delta}} = o_{|\mathcal{D}_n}(1) \, , \nonumber
	\end{eqnarray}
	where the first inequality is due to \citet[pg.21]{vaart_1998}. By that, since $E(1-w_iR_i|\mathcal{D}_n)=0$, it holds that $n^{-1}\sqrt{q}\matrx{\upvarphi}(\vectr{p},\vectr{\beta}^o)^{-1/2}\sum_{i\in\mathcal{C}}(1-w_iR_i)\breve{\vectr{a}}_i(\vectr{\beta}^o)$ converges conditionally on $\mathcal{D}_n$ to a standard multivariate normal distribution.
	Put differently, for all $\vectr{u}\in\mathbb{R}^r$,
	\begin{equation} \label{Theorem2:gamma:normal:1}
	P\left\{n^{-1}\sqrt{q}\matrx{\upvarphi}(\vectr{p},\vectr{\beta}^o)^{-1/2}\sum_{i\in\mathcal{C}}(1-w_iR_i)\breve{\vectr{a}}_i(\vectr{\beta}^o)  \le \vectr{u} \bigg|\mathcal{D}_n\right\} \xrightarrow{|\mathcal{D}_n} \Phi(\vectr{u}) \, ,
	\end{equation}
	where $\Phi$ is the standard multivariate normal cumulative distribution function. Since the conditional probability is a random variable in the unconditional space, then due to  Eq.(\ref{Theorem2:gamma:normal:1}) it converges almost surely to $\Phi(\vectr{u})$. Being additionally bounded, then due to the dominated convergence theorem, it follows that 
	\begin{equation} \label{Theorem2:gamma:normal:2}
	P\left\{n^{-1}\sqrt{q}\matrx{\upvarphi}(\vectr{p},\vectr{\beta}^o)^{-1/2}\sum_{i \in \mathcal{C}}(1-w_iR_i)\breve{\vectr{a}}_i(\vectr{\beta}^o)  \le \vectr{u} \right\} \rightarrow \Phi(\vectr{u}) \, .
	\end{equation}
	Suppose that  $\matrx{\upvarphi}(\vectr{p},\vectr{\beta}^o) \xrightarrow{P} \matrx{\Psi}(\vectr{p},\vectr{\beta}^o)$, where $ \matrx{\Psi}(\vectr{p},\vectr{\beta}^o)$ is a positive-definite matrix, then based on the results above, it will follow that $n^{-1/2}\sum_{i\in\mathcal{C}}(1-w_iR_i)\breve{\vectr{a}}_i(\vectr{\beta}^o) \xrightarrow{D} N(0,\alpha \matrx{\Psi}(\vectr{p},\vectr{\beta}^o))$, where $\alpha$ is the limiting constant of $n/q$, see Assumption A.10.

	Furthermore, the two addends are asymptotically independent, as seen by 
	\begin{eqnarray}
		&&\lim_{n,q\rightarrow \infty} \Pr\left(\frac{1}{\sqrt{n}}\frac{\partial l(\vectr{\beta}^o)}{\partial \vectr{\beta}^T} \le \vectr{u},  \frac{1}{\sqrt{n}}\sum_{i\in\mathcal{C}}(1-w_iR_i)\breve{\vectr{a}}_i(\vectr{\beta}^o) \le \vectr{v}\right) \nonumber \\
		&=&\lim_{n,q\rightarrow \infty} E\left(I\left\{\frac{1}{\sqrt{n}}\frac{\partial l(\vectr{\beta}^o)}{\partial \vectr{\beta}^T} \le \vectr{u}  \right\}\Pr\left\{ \frac{1}{\sqrt{n}}\sum_{i\in\mathcal{C}}(1-w_iR_i)\breve{\vectr{a}}_i(\vectr{\beta}^o) \le \vectr{v} \bigg| \mathcal{D}_n \right\}   \right) \nonumber \\
		&=& E\left(\lim_{n,q\rightarrow \infty}I\left\{\frac{1}{\sqrt{n}}\frac{\partial l(\vectr{\beta}^o)}{\partial \vectr{\beta}^T} \le \vectr{u}  \right\}\lim_{n,q\rightarrow \infty}\Pr\left\{ \frac{1}{\sqrt{n}}\sum_{i\in\mathcal{C}}(1-w_iR_i)\breve{\vectr{a}}_i(\vectr{\beta}^o) \le \vectr{v} \bigg| \mathcal{D}_n \right\}   \right) \nonumber \\
		&=& 
		 E\left(\lim_{n,q\rightarrow \infty}I\left\{\frac{1}{\sqrt{n}}\frac{\partial l(\vectr{\beta}^o)}{\partial \vectr{\beta}^T} \le \vectr{u}\right\} \Phi\left(\alpha^{-1/2}\matrx{\Psi}(\vectr{p},\vectr{\beta}^o)^{-1/2}\vectr{v}\right)   \right) \nonumber \\
		&=& \lim_{n,q\rightarrow \infty}E\left( I\left\{\frac{1}{\sqrt{n}}\frac{\partial l(\vectr{\beta}^o)}{\partial \vectr{\beta}^T} \le \vectr{u}  \right\} \right)\Phi\left(\alpha^{-1/2}\matrx{\Psi}(\vectr{p},\vectr{\beta}^o)^{-1/2}\vectr{v}\right) \nonumber \\
		&=& \lim_{n,q\rightarrow \infty}\Pr \left( \frac{1}{\sqrt{n}}\frac{\partial l(\vectr{\beta}^o)}{\partial \vectr{\beta}^T} \le \vectr{u}   \right)\Phi\left(\alpha^{-1/2}\matrx{\Psi}(\vectr{p},\vectr{\beta}^o)^{-1/2}\vectr{v}\right) \nonumber \\
		&=& \Phi(\matrx{\Sigma}(\vectr{\beta}^o)^{-1/2}\vectr{u}) \Phi\left(\alpha^{-1/2}\matrx{\Psi}(\vectr{p},\vectr{\beta}^o)^{-1/2}\vectr{v}\right)) \, , \nonumber
	\end{eqnarray}
	where we have used the dominated convergence theorem.

Lastly, based on Lemma 3 and the dominated convergence theorem, one can show that $-\tilde{\matrx{\mathcal{I}}}(\vectr{\beta}^o)$ is consistent to $\matrx{\Sigma}(\vectr{\beta}^o)$, see A.5, similarly to how Lemma 4 was proved. Therefore, when the sum of these two addends is multiplied from the left by  $-\tilde{\matrx{\mathcal{I}}}^{-1}(\vectr{\beta}^o)$, it will follow by Slutsky's theorem that the sum is still normal, with mean 0 and a covariance matrix asymptotically equivalent to $\mathbb{V}_{\tilde{\vectr{\beta}}}(\vectr{p},\vectr{\beta}^o)$ in Eq.(\ref{Theorem2:part1}).
 Therefore, based on the asymptotic independence, and by combining Eq.'s (\ref{Theorem2:beta_tilde - beta0:decomp})--(\ref{Theorem2:firstaddend:norm}) and Eq.(\ref{Theorem2:gamma:normal:2}), we arrive at Eq.(\ref{Theorem2:part1}). The two variance components correspond to two orthogonal sources of variance, namely the variance of the original full-data Cox regression estimates, and the additional variance generated by the subsampling procedure. 

Based on A.1--A.10 and Eq.(\ref{Theorem2:part1}), the same steps as in \citet[pg. 300]{fleming2011counting}, can be taken in order to prove Eq.(\ref{Theorem2:part2}).
\end{proof}
\subsection{Optimal Sampling Probabilities}
There is more than one way to define ``optimal", and one should choose a criterion to base optimality upon. In the field of optimal design of experiments, some well-known criteria are the A, D and E optimal designs \citep{pukelsheim2006optimal}, standing for ``Average", ``Determinant" and ``Eigen", respectively, which correspond to the minimization of the trace/determinant/maximal eigenvalue of the estimated regression coefficients covariance matrix. We choose in this paper to work with the A-optimal design criterion due to its analytical convenience and because it is equivalent to minimizing the asymptotical MSE of the estimated coefficients. Namely, the sampling probability vector which minimizes $Tr(\mathbb{V}_{\tilde{\vectr{\beta}}}(\vectr{p},\vectr{\beta}^o))$ is derived, where $Tr$ is the trace operator. 
\begin{theorem}
	The $A$-optimal sampling probabilities vector $\vectr{p}^A$ is of the form \\
	\begin{equation} \label{optA}
	p^A_m= \frac{\Vert \matrx{\mathcal{I}}^{-1}(\vectr{\beta}^o)\vectr{a}_{m}(\vectr{\beta}^o)\Vert_2}{\sum_{i\in\mathcal{C}}\Vert \matrx{\mathcal{I}}^{-1}(\vectr{\beta}^o)\vectr{a}_{i}(\vectr{\beta}^o)\Vert_2} \mbox{  for all} \, \,  m \in \mathcal{C} \, .
	\end{equation}
\end{theorem}
\begin{proof}[Proof of Theorem 3]
	\begin{eqnarray}
	Tr\left(\mathbb{V}_{\tilde{\vectr{\beta}}}(\vectr{p},\vectr{\beta}^o)\right) = Tr\left\{\matrx{\mathcal{I}}^{-1}(\vectr{\beta}^o)\matrx{\upvarphi}(\vectr{p},\vectr{\beta}^o)\matrx{\mathcal{I}}^{-1}(\vectr{\beta}^o)\right\} + d \, ,\nonumber 
	\end{eqnarray}
	where $d$ is a constant not involving the vector $\vectr{p}$. Proceeding with the first term,
	\begin{eqnarray}
	Tr\left\{\matrx{\mathcal{I}}^{-1}(\vectr{\beta}^o)\matrx{\upvarphi}(\vectr{p},\vectr{\beta}^o)\matrx{\mathcal{I}}^{-1}(\vectr{\beta}^o)\right\} &=& Tr\left[\frac{1}{n^2}\matrx{\mathcal{I}}^{-1}(\vectr{\beta}^o)\left\{\sum_{i\in\mathcal{C}}\frac{1}{p_i}\vectr{a}_i(\vectr{\beta}^o)\vectr{a}_i^T(\vectr{\beta}^o) \right. \right. \nonumber \\
	&& - \left. \left. \sum_{ i,j\in\mathcal{C}}\vectr{a}_i(\vectr{\beta}^o)\vectr{a}_j^T(\vectr{\beta}^o) \right\}\matrx{\mathcal{I}}^{-1}(\vectr{\beta}^o) \right] \, . \nonumber 
	\end{eqnarray}
	By omitting the part which does not involve $\vectr{p}$, 
	\begin{eqnarray}
	Tr\left\{\frac{1}{n^2}\sum_{i\in\mathcal{C}}\frac{1}{p_i}\matrx{\mathcal{I}}^{-1}(\vectr{\beta}^o)
	\vectr{a}_i(\vectr{\beta}^o)\vectr{a}_i^T(\vectr{\beta}^o)\matrx{\mathcal{I}}^{-1}(\vectr{\beta}^o)\right\} 
	&=& \frac{1}{n^2}\sum_{i\in\mathcal{C}}\frac{1}{p_i}Tr\left\{
	\vectr{a}_i^T(\vectr{\beta}^o)\matrx{\mathcal{I}}^{-2}(\vectr{\beta}^o)\vectr{a}_i(\vectr{\beta}^o)\right\} \nonumber \\
	&=& \frac{1}{n^2}\sum_{i\in\mathcal{C}}\frac{1}{p_i}\Vert \matrx{\mathcal{I}}^{-1}(\vectr{\beta}^o)\vectr{a}_{i}(\vectr{\beta}^o)\Vert^2_2 \nonumber \, .
	\end{eqnarray}
	Removing the factor of $n^{-2}$ does not alter the optimization solution, so let us now define the Lagrangian function, with multiplier $\gamma$, as
	$$g(\vectr{p}) = \sum_{i\in\mathcal{C}}\frac{1}{p_i}\Vert \matrx{\mathcal{I}}^{-1}(\vectr{\beta}^o)\vectr{a}_{i}(\vectr{\beta}^o)\Vert^2_2  + \gamma(1-\sum_{i\in\mathcal{C}}p_i) \, ,$$
	Differentiating with respect to $p_m, m\in\mathcal{C}$ and setting the derivative to $0$ we get
	$$\frac{\partial g(\vectr{p})}{\partial p_m} = -\frac{\Vert \matrx{\mathcal{I}}^{-1}(\vectr{\beta}^o)\vectr{a}_{m}(\vectr{\beta}^o)\Vert^2_2}{p_m^2} - \gamma \equiv 0 \, ,$$
	hence $$p_m = \frac{\Vert \matrx{\mathcal{I}}^{-1}(\vectr{\beta}^o)\vectr{a}_{m}(\vectr{\beta}^o)\Vert_2}{\sqrt{-\gamma}}  \, .$$
	Since all probabilities sum up to 1, it follows that
	$$\sqrt{-\gamma} = \sum_{i\in\mathcal{C}}\Vert \matrx{\mathcal{I}}^{-1}(\vectr{\beta}^o)\vectr{a}_{i}(\vectr{\beta}^o)\Vert_2 \, ,$$ which yields Eq.(\ref{optA}).
\end{proof}
These expressions are not readily usable, as they require the unknown vector $\vectr{\beta}^o$. Interestingly, had we plugged the PL estimator $\hat{\vectr{\beta}}_{PL}$ in place of $\vectr{\beta}^o$, it would turn out that these optimal probabilities are in fact proportional to the norm of the commonly used approximation for each observation's ``Dfbeta" statistic. The Dfbeta is defined as $\hat{\vectr{\beta}}_{PL}  - \hat{\vectr{\beta}}_{PL(i)}$, where $\hat{\vectr{\beta}}_{PL(i)}$ is the vector of estimated coefficients obtained by removing observation $i$ from the sample \citep[pg. 391]{klein2006survival}. The Dfbeta's are commonly used to detect influential observations, hence providing insight into which observations are assigned a higher sampling probability with this method. However, $\hat{\vectr{\beta}}_{PL}$ is also unknown, as it entails running Cox regression on the full sample, which is what we are trying to circumvent, and in practice a two-step estimator, in the spirit of \cite{wang2018optimal} will be used, as discussed in section 2.4.

We now turn to discuss a different, yet related, optimality criterion, motivated by the Loewner ordering \citep{pukelsheim2006optimal}, which enjoys a computational advantage. The Loewner ordering states that for two positive definite matrices $\matrx{M_1}$ and $\matrx{M_2}$, $\matrx{M_1}\le \matrx{M_2}$ if and only if $\matrx{M_2} - \matrx{M_1}$ is a nonnegative definite matrix. The part in $\mathbb{V}_{\tilde{\vectr{\beta}}}(\vectr{p},\vectr{\beta}^o)$ which involves $\vectr{p}$ is composed of 3 matrices ${\matrx{\mathcal{I}}}^{-1}(\vectr{\beta}^o)\matrx{\upvarphi}(\vectr{p},\vectr{\beta}^o) {\matrx{\mathcal{I}}}^{-1}(\vectr{\beta}^o)$ where $\matrx{\upvarphi}$ is the only one among them which actually contains $\vectr{p}$. It follows that for two potential sampling probability vectors $\vectr{p}^{(1)},\vectr{p}^{(2)}$,  $\mathbb{V}_{\tilde{\vectr{\beta}}}(\vectr{p}^{(1)},\vectr{\beta}^o) \le \mathbb{V}_{\tilde{\vectr{\beta}}}(\vectr{p}^{(2)},\vectr{\beta}^o)$ if and only if 
$\matrx{\upvarphi}(\vectr{p}^{(1)},\vectr{\beta}^o) \le \matrx{\upvarphi}(\vectr{p}^{(2)},\vectr{\beta}^o)$. This idea suggests that one can try to minimize $Tr\left(\matrx{\upvarphi}(\vectr{p},\vectr{\beta}^o)\right)$ instead, which is equivalent to the L-optimal criterion \citep{pukelsheim2006optimal}, standing for ``Linear" and aims at the minimization of the trace of the covariance matrix of a linearly transformed vector of estimated regression coefficients. In our case, minimizing the trace of $\matrx{\upvarphi}(\vectr{p},\vectr{\beta}^o)$ is equivalent to minimizing the asymptotical mean squared error (MSE) of $\matrx{\mathcal{I}}(\vectr{\beta}^o)\tilde{\vectr{\beta}}$.	

\begin{theorem}
	The L-optimal sampling probabilities vector $\vectr{p}^L$ is of the form \\
	\begin{equation} \label{optL}
	p^L_m = \frac{\Vert \vectr{a}_{m}(\vectr{\beta}^o)\Vert_2}{\sum_{i\in\mathcal{C}}\Vert \vectr{a}_{i}(\vectr{\beta}^o)\Vert_2} \mbox{  for all} \, \,  m \in \mathcal{C} \, .
	\end{equation}
\end{theorem}
The proof of Theorem 4 is straightforward and similar to the steps of Theorem 3, and thus is omitted. Again, using $\hat{\vectr{\beta}}_{PL}$ instead of $\vectr{\beta}^o$, one finds that each observation's sampling probability is in fact proportional to the norm of its vector of score residuals \citep[pg 385 Eq. 11.6.1]{therneau1990martingale,klein2006survival}. The score residual is another useful quantity for identification of influential observations, and is analogous to leverage scores in linear regression. Therefore, the ``L" in L-optimality can be loaded in our case with a second interpretation, namely ``Leverage-optimality". Roughly speaking, large values for $\Vert\vectr{a}_i(\vectr{\beta}^o)\Vert_2$ usually involve some of the following situations:
\begin{itemize}
	\item Extreme covariate values.
	\item The observation is at risk until a relatively late time.
	\item The linear predictor $\vectr{\beta}^{oT}\vectr{X}_i$ is relatively large.
\end{itemize}
Since $\vectr{\beta}^o$ is unknown, we employ the two-step estimator to be discussed in the next section.

The A-optimal probabilities take into account the variances and correlation between the covariates, through the information matrix $\matrx{\mathcal{I}}$, in contrast to the L-optimal criterion which treats each covariate on its own and is agnostic to their variances and correlation. It is thus expected that the A-optimality criterion outperform L-optimality under unequal variances and when covariates are strongly correlated. In the numerical studies in Section we show that the efficiency loss of L-opt compared to A-opt tends to be relatively small.

On the premise that one is particularly interested only in a subset of the covariates, while the rest are retained in the model just to control for confounders, the optimal sampling probabilities could be adjusted in order to target the subset of interest. Denote $\mathcal{S}$ the subset of interest, then 

\begin{equation} \label{optAsubset}
	p^{A(\mathcal{S})}_m= \frac{\Vert \matrx{\mathcal{I}}^{-1}_{[\mathcal{S}]}(\vectr{\beta}^o)\vectr{a}_{m}(\vectr{\beta}^o)\Vert_2}{\sum_{i\in\mathcal{C}}\Vert \matrx{\mathcal{I}}_{[\mathcal{S}]}^{-1}(\vectr{\beta}^o)\vectr{a}_{i}(\vectr{\beta}^o)\Vert_2} \mbox{  for all} \, \,  m \in \mathcal{C} \, ,
\end{equation}
and,
\begin{equation} \label{optLsubset}
		p^{L(\mathcal{S})}_m = \frac{\Vert \vectr{a}_{[\mathcal{S}]m}(\vectr{\beta}^o)\Vert_2}{\sum_{i\in\mathcal{C}}\Vert \vectr{a}_{[\mathcal{S}]i}(\vectr{\beta}^o)\Vert_2} \mbox{  for all} \, \,  m \in \mathcal{C} \, ,
\end{equation}
where $\matrx{\mathcal{I}}^{-1}_{[\mathcal{S}]}$ is the submatrix of  $\matrx{\mathcal{I}}^{-1}$, containing the rows corresponding to the covariates in $\mathcal{S}$, and $\vectr{a}_{[\mathcal{S}]}$ is the subvector of $\vectr{a}$ containing the entries corresponding to the covariates in $\mathcal{S}$. An analysis of this form, targeting one covariate at a time, is illustrated in Section 4.2.2. 

Finally, observations that were censored prior to the first observed failure time, have a sampling probability of 0, according to both optimal criteria. It is efficient, as these contribute no information at all to the original full-data PL.

\subsection{Two-Step Procedure - The Proposed Approach}
The results obtained in Eqs.(\ref{optA})--(\ref{optLsubset}) are impractical as they contain the unknown $\vectr{\beta}^o$. Instead, a two-step procedure is implemented, which can be described as follows. 
\begin{itemize}
\item[] {\bf Step 1:} 
Sample $q_0$ observations uniformly from $\mathcal{C}$, join them with the observed failure times to form $\mathcal{Q}_{pilot}$. Run a weighted Cox regression on $\mathcal{Q}_{pilot}$ to obtain an initial estimator $\tilde{\vectr{\beta}}_{U}$, and use it to derive approximated optimal sampling probabilities using one of Eq.(\ref{optA})--(\ref{optLsubset}).  
\item[] {\bf Step 2:} Sample another $q$ observations from $\mathcal{C}$ using the probabilities computed at Step 1. Combine these observations with the observed failure times to form $\mathcal{Q}$, and rerun weighted Cox regression on $\mathcal{Q}$ to obtain the two-step estimator $\tilde{\vectr{\beta}}_{TS}$.	
\end{itemize}

Theorem 5 is the analog of Theorem 2, and establishes asymptotic properties of $\tilde{\vectr{\beta}}_{TS}$ and  $\hat{\Lambda}_{0}(t,\tilde{\vectr{\beta}}_{TS})$, in the unconditional space.
\begin{theorem}
	Under Assumptions A.1--A.10, as $q \rightarrow \infty$ and $n\rightarrow \infty$, the following holds,
	
	\begin{equation} \label{Theorem5:eq1}
		\sqrt{n}\mathbb{V}_{\tilde{\vectr{\beta}}}(\vectr{p}^{opt},\vectr{\beta}^o)^{-1/2}(\tilde{\vectr{\beta}}_{TS} - \vectr{\beta}^o) \xrightarrow{D} N(0,I) \, ,
	\end{equation}
	and, for all $t\in[0,\tau]$,
	\begin{equation} \label{Theorem5:eq2}
		\sqrt{n}\mathbb{V}_{\hat{\Lambda}_0(t,\tilde{\vectr{\beta}})}(\vectr{p}^{opt},\vectr{\beta}^o,t)^{-1/2}\{\hat{\Lambda}_0(t,\tilde{\vectr{\beta}}_{TS}) - \Lambda_0^o(t)\} \xrightarrow{D} N(0,1) \, ,
	\end{equation}
\end{theorem}
where $\vectr{p}^{opt}$ is either $\vectr{p}^L$ or $\vectr{p}^A$, depending on the chosen optimality criterion.
\begin{proof}[Proof of Theorem 5]
	There are two key points essential for proving Eq.'s (\ref{Theorem5:eq1})--(\ref{Theorem5:eq2}). The first, is that in the conditional space, if $\tilde{\vectr{\beta}}_{U}$ is also conditioned upon, the sampling probabilities become deterministic, and we return to the settings of Theorem 2. The second, is that the consistency and normality results which were derived for $\tilde{\vectr{\beta}}$, apply for any vector of deterministic sampling probabilities that satisfies A.8. Now, for all $\vectr{u}\in\mathbb{R}^r$,
	\begin{eqnarray}
	&& \Pr\left\{\mathbb{V}_{\tilde{\vectr{\beta}}}^{-1/2}(\vectr{p}^{opt},\vectr{\beta}^o)\left(\sqrt{n}(\tilde{\vectr{\beta}}_{TS} - \vectr{\beta}^o) \right)\le \vectr{u}  \right\} \nonumber \\
	&& = E\left[P\left\{\mathbb{V}_{\tilde{\vectr{\beta}}}^{-1/2}(\vectr{p}^{opt},\vectr{\beta}^o)\left(\sqrt{n}(\tilde{\vectr{\beta}}_{TS} - \vectr{\beta}^o) \right)\le \vectr{u} \big| \mathcal{D}_n,\tilde{\vectr{\beta}}_{U} \right\}\right] \, . \nonumber
	\end{eqnarray}
	Based on the points mentioned above and Theorem 2, it follows that
	$$\Pr\left\{\mathbb{V}_{\tilde{\vectr{\beta}}}^{-1/2}(\vectr{p}^{opt},\vectr{\beta}^o)\left(\sqrt{n}(\tilde{\vectr{\beta}}_{TS} - \vectr{\beta}^o) \right)\le \vectr{u} \big| \mathcal{D}_n,\tilde{\vectr{\beta}}_{U} \right\} \xrightarrow{|\mathcal{D}_n,\tilde{\vectr{\beta}}_{U}} \Phi(\vectr{u}) \, , $$
	then, because the conditional probability is a bounded random variable in the unconditional space, which converges almost surely to $\Phi(\vectr{u})$, Eq.(\ref{Theorem5:eq1}) is implied by the dominated convergence theorem. The same arguments hold for proving Eq.(\ref{Theorem5:eq2}), and we skip the proof.\\
\end{proof}

Lastly, let us briefly address the computational complexity involved in computing $\tilde{\vectr{\beta}}_{TS}$. Since we are including all observed failure times in the analysis, and since it is assumed that the number of observed failure times constitutes a fixed (small) proportion of the overall sample size, we cannot use big-O notation in order to express the expected reduction in computation time. In the asymptotic big-O tools, which are commonly used to measure algorithm complexity, it holds that $O(\alpha n) = O(n)$, for any fixed $\alpha$, no matter how small. Therefore, whatever the big-O complexity of the original algorithm is, by taking a fixed proportion of the sample, the resultant big-O complexity is of the same order. However, we demonstrate by simulations and real data analysis that in practice our method results in a substantial computation time reduction.

\subsection{Variance Estimation}
Based on Eq.(\ref{Theorem5:eq1}), a natural estimator for the covariance matrix of $\tilde{\vectr{\beta}}_{TS}$ is $$n^{-1}\mathbb{V}_{\tilde{\vectr{\beta}}}(\vectr{p}^{opt},\tilde{\vectr{\beta}}_{TS}) = n^{-1}{\matrx{\mathcal{I}}}^{-1}(\tilde{\vectr{\beta}}_{TS})+q^{-1}{\matrx{\mathcal{I}}}^{-1}(\tilde{\vectr{\beta}}_{TS})\matrx{\upvarphi}(\vectr{p}^{opt},\tilde{\vectr{\beta}}_{TS}){\matrx{\mathcal{I}}}^{-1}(\tilde{\vectr{\beta}}_{TS}) \, .$$
However, calculation of $\matrx{\mathcal{I}}^{-1}(\tilde{\vectr{\beta}}_{TS})$ and $\matrx{\upvarphi}(\vectr{p}^{opt},\tilde{\vectr{\beta}}_{TS})$ involves the full data, and may be avoided by replacing these matrices with their subsampling-based counterparts, $\tilde{\matrx{\mathcal{I}}}^{-1}(\tilde{\vectr{\beta}}_{TS})$ and $\tilde{\matrx{\upvarphi}}(\vectr{p}^{opt},\tilde{\vectr{\beta}}_{TS})$, where
$$\tilde{\matrx{\upvarphi}}(\vectr{p}^{opt},\tilde{\vectr{\beta}}_{TS}) = \frac{1}{n^2}\left\{\frac{1}{q}\sum_{i=1}^q\frac{\tilde{\vectr{a}}_i(\tilde{\vectr{\beta}}_{TS})\tilde{\vectr{a}}_i(\tilde{\vectr{\beta}}_{TS})^T}{p_i^{*2}}-\frac{1}{q^2}\sum_{i=1}^q\frac{\tilde{\vectr{a}}_i(\tilde{\vectr{\beta}}_{TS})}{p_i^*}\left(\sum_{i=1}^q\frac{\tilde{\vectr{a}}_i(\tilde{\vectr{\beta}}_{TS})}{p_i^*}\right)^T\right\} \, ,$$
and
$$\tilde{\vectr{a}}_i(\vectr{\beta}) = \int_{0}^\tau \left\{ \vectr{X}_i - \frac{\vectr{S}_w^{(1)}(\vectr{\beta},t)}{S_w^{(0)}(\vectr{\beta},t)}\right\}\frac{Y_i(t)e^{\vectr{\beta}^T\vectr{X}_i}}{S_w^{(0)}(\vectr{\beta},t)}dN_.(t) \, .$$
This subsampling-based variance matrix can be denoted as $\tilde{\mathbb{V}}_{\tilde{\vectr{\beta}}}(\vectr{p}^{opt},\tilde{\vectr{\beta}}_{TS})$.
The variance estimator for $\hat{\Lambda}_{0}(t,\tilde{\vectr{\beta}}_{TS})$  is simply $n^{-1}\mathbb{V}_{\hat{\Lambda}_0(t,\tilde{\vectr{\beta}})}(\vectr{p}^{opt},\tilde{\vectr{\beta}}_{TS},t)$. 
Section 4 shows that our proposed variance estimators approximate the covariance matrices very well.

\section{Cox Model Refinements}
Our methods are easily extended to more intricate analyses, illustrating its potential utility. The previously derived asymptotic results hold for each of the following cases under some suitable adjusments for the list of assumptions, as discussed below.

\subsection{Delayed Entry (Left Truncation)}
For right censored and delayed entry data, the standard riskset correction approach is implemented. Namely, suppose observation $i$ has a delayed entry time $L_i\ge0$, then the modified at-risk process is given by $Y_i(t)=I(L_i\le t\le T_i)$. Observations whose entry and censoring times both occurred between two successive observed failure times, contribute no information to the PL. The value of $\vectr{a}_i(\vectr{\beta})$ for these observations is 0 and therefore they are assigned a sampling probability of 0 according to both L-opt and A-opt. Its implication is that our methods remove these observations from the dataset, and do not allow their inclusion in the subsample. Under these modifications and A.6, all previous asymptotic results hold.
\subsection{Time-Dependent Covariates}
For using the R {\tt coxph} function with time-dependent covariates, the data should be organized in ``pseudo-observations'' form, as guided by \cite{therneau2017using}. Each observation may be broken into several distinct pseudo-observations, such that each has its ``entrance'' and ``exit'' times, creating non-overlapping intervals, that together reconstruct the original time interval. Each pseudo-observation has fixed covariate values, and by treating them as if they were independent observations, the resultant PL is the same as that of the original data. This elegant computational trick, although providing a convenient tool for analyzing data with time-dependent covariates, also suffers from a computational downside by inflating the number of ``observations'' inserted to the optimization routine. When the number of observations is large to begin with, the number of pseudo-observations may be daunting. 

We propose two approaches to use our subsampling methods to cope with the computation challenge, with the first approach being the recommendeded one. \\
\underline{\textbf{Approach 1}}:
This approach simply amounts to sampling from the censored pseudo-observations, such that only the most informative pieces of information are kept from the original observations. Since regarding all pseudo-observations as independent results in the original full-data PL, our derived optimality results extend to sampling from the pseudo-observations. Similarly to the discussion in the delayed entry subsection, it may be the case now that pseudo-observations are assigned sampling probability of 0. The implication is that some parts of an observation's follow-up trajectory, may contribute no information to the estimation procedure. \\ 
\underline{\textbf{Approach 2}}:
When splitting the data into pseudo-observations, each one of them is assigned an ID number, corresponding to their originating observation. In case one finds themselves uncomfortable using just fractions of observations, another possibility is to compute $\vectr{a}(\vectr{\beta})$ for all pseudo-observations, and then add them by ID, thus reassembling all $\vectr{a}_i(\vectr{\beta})$'s of the original censored observations. This alternative has three drawbacks, the first is that sampling an observation results in sampling a block of pseudo-observations, and it is hard to predict what the size of the subsampled data will be, as observations vary in the number of their corresponding pseudo-observations. The second, is that for the same computational costs, more informative pseudo-observations could have been selected. The third, is that some additional costs are involved in reconstructing the original $\vectr{a}_i(\vectr{\beta})$, incurred by the ID matching. 

Since an efficient approximation to the full-data PL estimator is sought, we opt for approach 1, as it provides a better approximation, and the idea of using fractions of observations is conceptually new. 

\subsection{Stratified Analysis}
In a stratified analysis, each observation's riskset consists of its respective stratum. Therefore, in order to compute sampling probabilities, one can derive all censored observations' $\vectr{a}_i(\vectr{\beta})$'s in each stratum separately. If the A-opt option is used, the information matrices should be derived for all strata, and then summed, to get the information matrix of the full data. Having all $\vectr{a}_i(\vectr{\beta})$'s and the information matrix, the sampling probabilities can be worked out, and the two-step subsampling procedure can be used. In order for the asymptotic results to hold for this case, Assumption A.1 should be extended to the cumulative hazard functions of all strata.

\subsection{Time-Dependent Coefficients}
In a time-dependent coefficient analysis, it is customary to specify some function to model the time-dependent part of a covariate's effect. For instance, it may be assumed that the coefficient of a given covariate is $\beta t$ or $\beta \ln t$. The {\tt coxph} function supports such an analysis using a user defined function, as explained in \cite{therneau2017using}, and during the execution process, the data is transformed into time-dependent form. Specifically, each observation is divided into time intervals between all successive observed failure times to which it was at risk. This transformation heavily inflates the number of pseudo-observations, such that even an innocent-looking dataset of 10,000 observations with about 1,000 observed failure times, is transformed into over 4 million pseudo-observations. Our method can be used here to pick a small and informative subset of pseudo-observations to reduce the computational burden. For the asymptotic results to hold, it is assumed that $\beta(t) = \eta f(t)$, where $f(t)$ is a uniformly bounded function of $t$ in the interval $(0,\tau)$, and $\eta$ replaces $\beta$ in all the assumptions where the latter is involved.      

\section{Numerical Results}
The simulation study as well as usecase 2 in the UKB analysis, to be described below, were performed on a Dell XPS 15 9500, with an Intel processor i7-10750H CPU. The analysis of usecase 1 in the UKB analysis was performed on an AWS instance r5.8xlarge. {\tt R} code for the data analysis and reported simulations is available
at Github site: \\ {\tt https://github.com/nirkeret/subsampling}.

\subsection{Simulation Study}
In order to evaluate the performance of our proposed methods, compared to the full-data PL estimator, and to the NCC and classic CC (uniform subsampling from censored), we carried out a simulation study. Several data characteristics were examined in our simulations: \\ $\bullet$ Time-fixed covariates with different dependency structures, with and without delayed entry. \\
$\bullet$ Time-dependent covariates, affecting both censoring and event times, which are conditionally independent.

For studying these scenarios under the different methods, 500 samples were generated, each of size $n=15,000$, using a vector of coefficients $\vectr{\beta}^o$ of size $r=6$, $\vectr{\beta}^o=0.1(3,-5,1,-1,1,-3)^T$. For the time-independent covariates simulations, the censoring times were generated from an exponential distribution with rate 0.2, independently of failure times. The instantaneous baseline hazard rates were set to be $$\lambda_0(t) = 0.001I(t<6) + c_{\lambda_0} I(t\ge6)  \, ,$$ 
where $c_{\lambda_0}$ was different for each setting ``A", ``B" or ``C", as described below.  The distribution of the covariates $\matrx{X}$ also differed between all three settings, as follows.
\begin{itemize}
	\item [\textbf{A}.] $X_i \sim Unif(0,4)$, for all $i=1,
	\ldots,6$, and are independent. $c_{\lambda_0}=0.05$ without delayed entry and $c_{\lambda_0}=0.015$ with delayed entry. This is the simple setting of equal variances and no correlation between the covariates.
	\item[\textbf{B.}] $X_i \sim Unif(0,u_i)$, $u_i = 1,6,2,2,1,6$, for $i=1,\ldots,6$, respectively, and are independent. $c_{\lambda_0}=0.15$ without delayed entry, and $c_{\lambda_0}=0.05$ with delayed entry. This is the setting of unequal variances, but no correlation between the covariates.
	\item[\textbf{C.}] $X_1,X_2,X_3$ are independently sampled from $U(0,4)$. $X_4 = 0.5X_1+0.5X_2 + \varepsilon_1$. $X_5=X_1 + \varepsilon_2$. $X_6 = X_1  + \varepsilon_3$, where $\varepsilon_1 \sim N(0,0.1)$, $\varepsilon_2 \sim N(0,1)$, $\varepsilon_3 \sim N(1,1.5)$ and the $\varepsilon$'s are independent.  $c_{\lambda_0} = 0.05$ without delayed entry, and $c_{\lambda_0} = 0.025$ with delayed entry. This is the setting of correlated covariates with (mildly) unequal variances. The strongest pairwise Pearson's correlation is about $0.75$.
\end{itemize}
These settings were each replicated with and without delayed entry, and for different numbers of sampled censored observations. For the delayed entry settings, entry times were sampled uniformly from $0$ to the $3$'rd quartile of the observed times, independently for each observation, and $n=15,000$ observations were then randomly sampled only out of those satisfying $T>R$. 

For the time-dependent covariate, we replicated all previous settings, and added a covariate that mimics the number of CRC screening tests that individuals may experience throughout their life. The times between two consecutive tests were sampled from a $Unif(3,12)$ distribution, and each observation may undergo up to 4 tests. At each time, the covariate holds the cumulative number of tests up to that moment. The time-dependent covariate's coefficient was set to $\beta^o_{dep}=0.15$, and in setting B, $c_{\lambda_0}=0.05$. The censoring times are sampled similarly to the failure times, with a vector of coefficients $\beta^C=(0.15,-0.1,0.15,-0.1,0.15,-0.1)^T$, $\beta^C_{dep}=0.1$, and $\lambda^C_0(t) = 0.2I(t<6) + 0.15 I(t\ge6)$. Conditionally on the covariates, the censoring and failure times are independent.

For matching controls to cases in NCC, we used the {\tt ccwc} function in the {\tt Epi R} package \citep{Epi}. For the classic NCC (``NCC-c''), a stratified analysis was performed with {\tt coxph}, whereas for Samuelsen's improved NCC (``NCC-S'') \citep{samuelsen1997psudolikelihood}, the {\tt R} package {\tt MultipleNCC} was used. 

The methods were compared based on three metrics. The root mean squared error (RMSE) for the vector of coefficients was derived for each method, averaged over the 500 samples. The RMSE was calculated with respect to both $\vectr{\beta}^o$ and $\hat{\vectr{\beta}}_{PL}$, the latter serving as the gold standard. The RMSE of an estimator $\hat{\vectr{\beta}}$ with respect to $\vectr{\beta}^o$ is defined as $$\frac{1}{500}\sum_{j=1}^{500}\sqrt{\sum_{i=1}^r\left(\hat{\beta}_i^{(j)} - \beta^o_i\right)^2} \, ,$$ and with respect to $\hat{\vectr{\beta}}_{PL}$, it is defined as $$\frac{1}{500}\sum_{j=1}^{500}\sqrt{\sum_{i=1}^r\left(\hat{\beta}_i^{(j)} - \hat{\beta}_{PLi}^{(j)}\right)^2} \, ,$$ where the superscript $(j)$ indicates the estimator computed from the $j$'th sample. The average runtime (in seconds) and the relative RMSE (RR) are also presented, where RR is defined as the ratio of each method's RMSE to that of $\hat{\vectr{\beta}}_{PL}$'s, with respect to the true parameter $\vectr{\beta}^o$.

With regards to RMSE and RR, Tables \ref{tab_sim01}--\ref{tab_sim03} show that the L-opt and A-opt methods tend to perform similarly, with some advantage to A-opt under setting ``C''. This advantage seems to shrink as the number of sampled observations increases. Additionally, our methods outperform the NCC-c, NCC-S and the uniform methods under all settings. With regards to runtime, L-opt is slightly faster than A-opt, and both are considerably faster than the full-data PL and the NCC methods, but as expected, are slower than the uniform method, which does not require computation of sampling probabilities. The conclusion is that L-opt and A-opt may offer a computational advantage over the full-data PL, without sacrificing a lot of efficiency, in contrast to the uniform sampling which may potentially incur a high efficiency loss. 

The simulation results of the time-dependent covariates setting are summarized in Table \ref{tab_sim_td_03}. It should be kept in mind that sampling pseudo-observations (Approach 1), marked with ``ps'' in the tables, and Approach 2 (``L-opt'', ``A-opt'') are not readily comparable. In Approach 1, $q$ reflects the number of pseudo-observations, or rows in the dataset, included in the analysis, whereas in Approach 2 $q$ reflects the number of observations, such that each may translate into several rows. Despite Approach 2 using overall more pseudo-observations than Approach 1, the latter still has an advantage in setting C, in terms of RMSE and RR. The explanation is that Approach 1 selects the most informative pseudo-observations, while Approach 2 selects an observation as a whole, which is not optimal. For illustration, suppose that the integral from time 0 to some time $t$ in $\vectr{a}(\vectr{\beta})$ is negative, but from time $t$ to time $\tau$ it is positive. Overall, $\vectr{a}(\vectr{\beta})$ may be small, due to the two parts canceling out each other, even though each part may be informative on its own, and could have been selected using Approach 1.
 
\begin{figure} 
	\centering
	\makebox{\includegraphics[width=90mm]{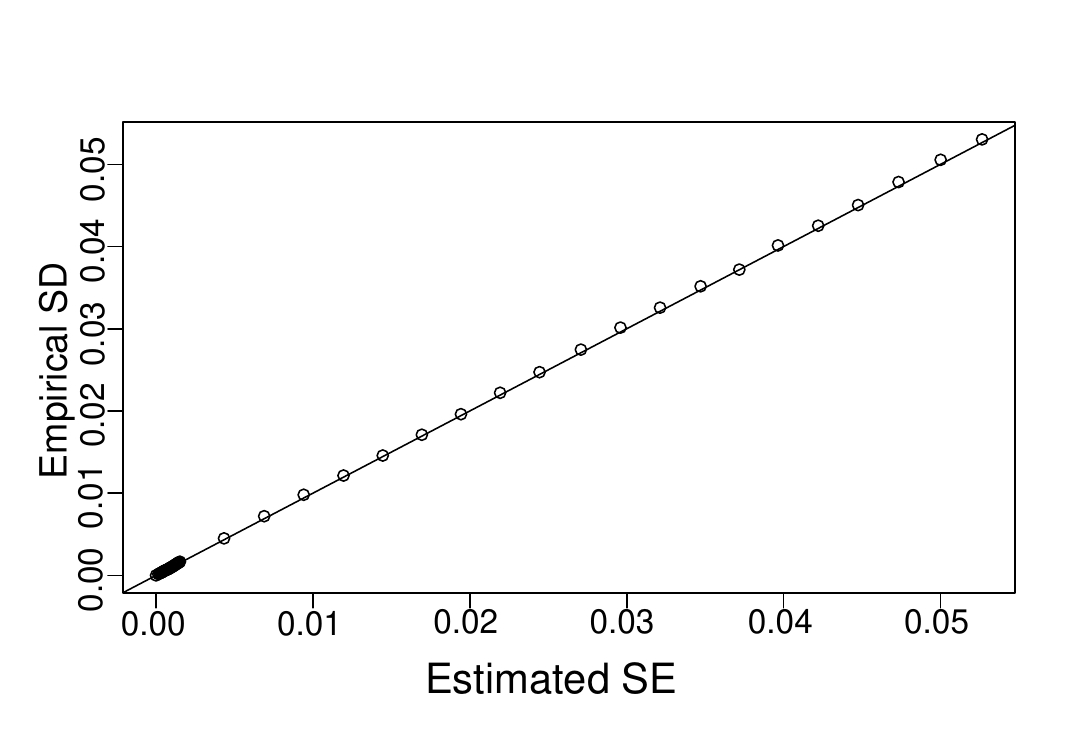}}
	\caption{\label{fig:haz:var} Simulation results of right-censored data without delayed entry, under setting ``C", for the A-opt method, with three censored observations per observed failure time: estimated SE and empirical SD for the cumulative hazard function at various time points.}
\end{figure}

For demonstrating the finite sample performance of our proposed variance estimators, we estimated the covariance matrices of the coefficients, under setting ``C'', without delayed entry, and with three censored observations per observed failure time, for all 500 samples, and averaged them. Table \ref{tab_SE} shows the empirical SD and average estimated SE of each estimated coefficient, as well as the average Frobenius distance between the empirical and estimated covariance matrices. For instance, the average Forbenius distance for L-opt is 	

$$\frac{1}{500}\sum_{j=1}^{500}\| n^{-1}\tilde{\mathbb{V}}_{\tilde{\vectr{\beta}}}(\vectr{p}^{L(j)},\tilde{\vectr{\beta}}^{(j)}_{TS}) - \mbox{Cov}(\tilde{\vectr{\beta}}_{TS}) \|_F \, ,$$
where $\mbox{Cov}(\tilde{\vectr{\beta}}_{TS})$ is the empirical covariance matrix computed from the 500 vectors of estimated regression coefficients.
For the cumulative hazard function variance estimator, we took a grid of time points, ranging from 0 to 10, with jumps of size 0.2, and calculated the empirical SD and estimated SE at these points. In Figure \ref{fig:haz:var} the empirical SD at each time point is plotted against its estimated counterpart. Both Table \ref{tab_SE} and Figure \ref{fig:haz:var} show excellent agreement between the estimated and empirical variances. 
Additionally, the Wald-based 95\% confidence interval  coverage rate (CR) is presented for the time-fixed and the time-dependent covariates simulations under Approach 1. Since Approach 2 to the time-dependent covariates involves a cumbersome reconstruction of the original $\vectr{a}(\vectr{\beta})$ vectors, the computation of the variance matrix was troublesome, and as we do not recommend this approach anyway, we skipped it. The CRs were calculated with the built-in variance estimators in the {\tt R} functions for the full-data PL and the two NCC methods, and by using the estimated variance formula as described in Section 2.5, for the subsampling methods. It is evident that the variance estimation and the normalily results are accurate, except perhaps for the uniform subsampling when q was equal to the number of failures.

\subsection{UK Biobank Data }

The following analyses include the well-established environmental CRC risk factors: body mass index (BMI), smoking status (yes/no), history of CRC in family (yes/no), physical activity (yes/no), sex (female/male) , alcohol consumption (3 levels: ``non or occasional",``light frequent drinker", ``very frequent drinker"), education (5 levels: ``prefer not to answer", ``lower than highschool", ``highschool", ``higher vocational education", ``collage or university graduate"), drug use (3 levels: ``none", ``Aspirin", ``Ibuprofen"), post-menopausal hormones (yes/no); as well as 72 SNPs that have been linked to CRC by GWAS \citep{jeon2018determining}. The SNPs were standardized to have mean zero and unit variance.

In the following sections, we describe two use cases of the CRC UKB data, for which a subsampling approach can be of great practical value. The first one, is the main motivation of this paper, namely a CRC risk-prediction tool, with a time-dependent coefficient. The secondary example is a GWAS SNP marginal analysis.

\subsubsection{Time-Dependent Coefficients}
\begin{figure}
\begin{tabular}{cc}
		\includegraphics[width=65mm]{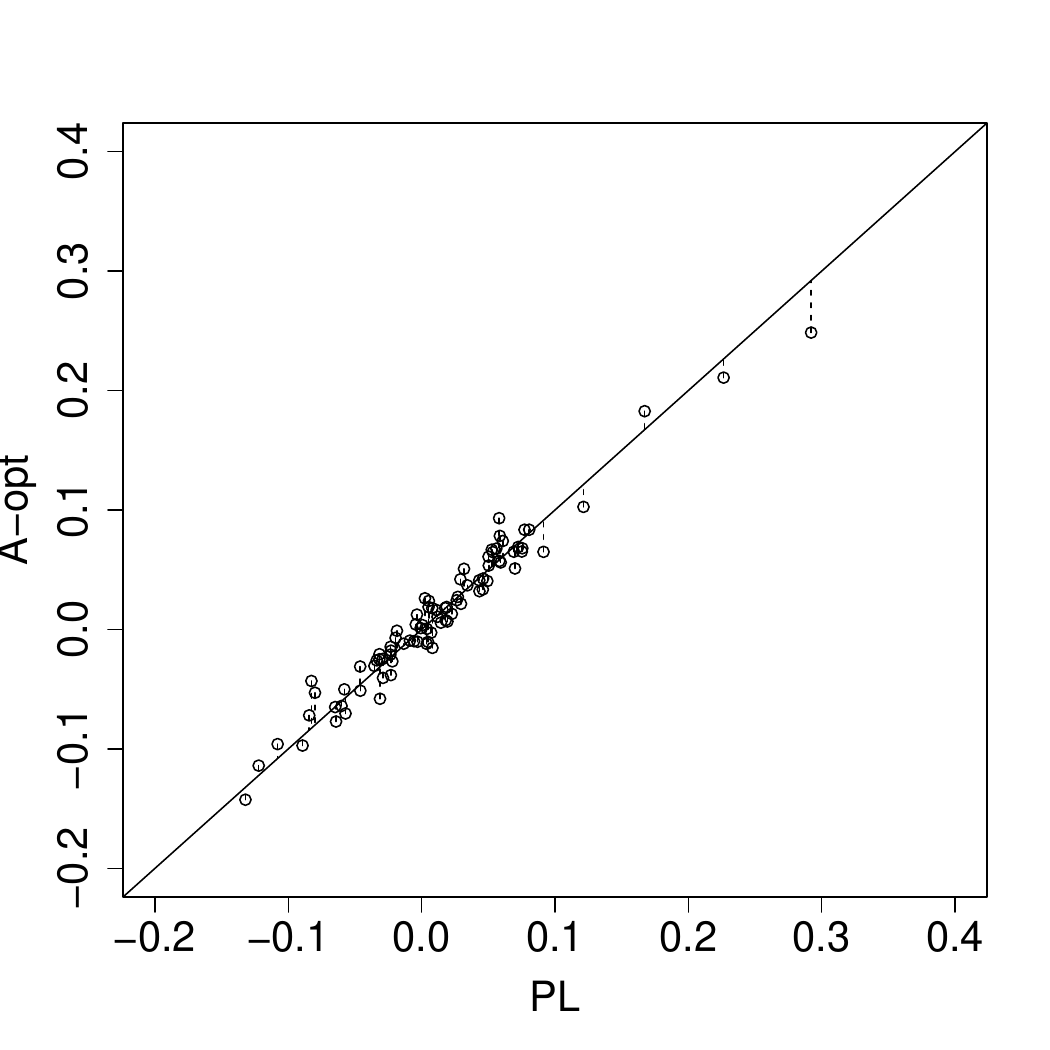} &   \includegraphics[width=65mm]{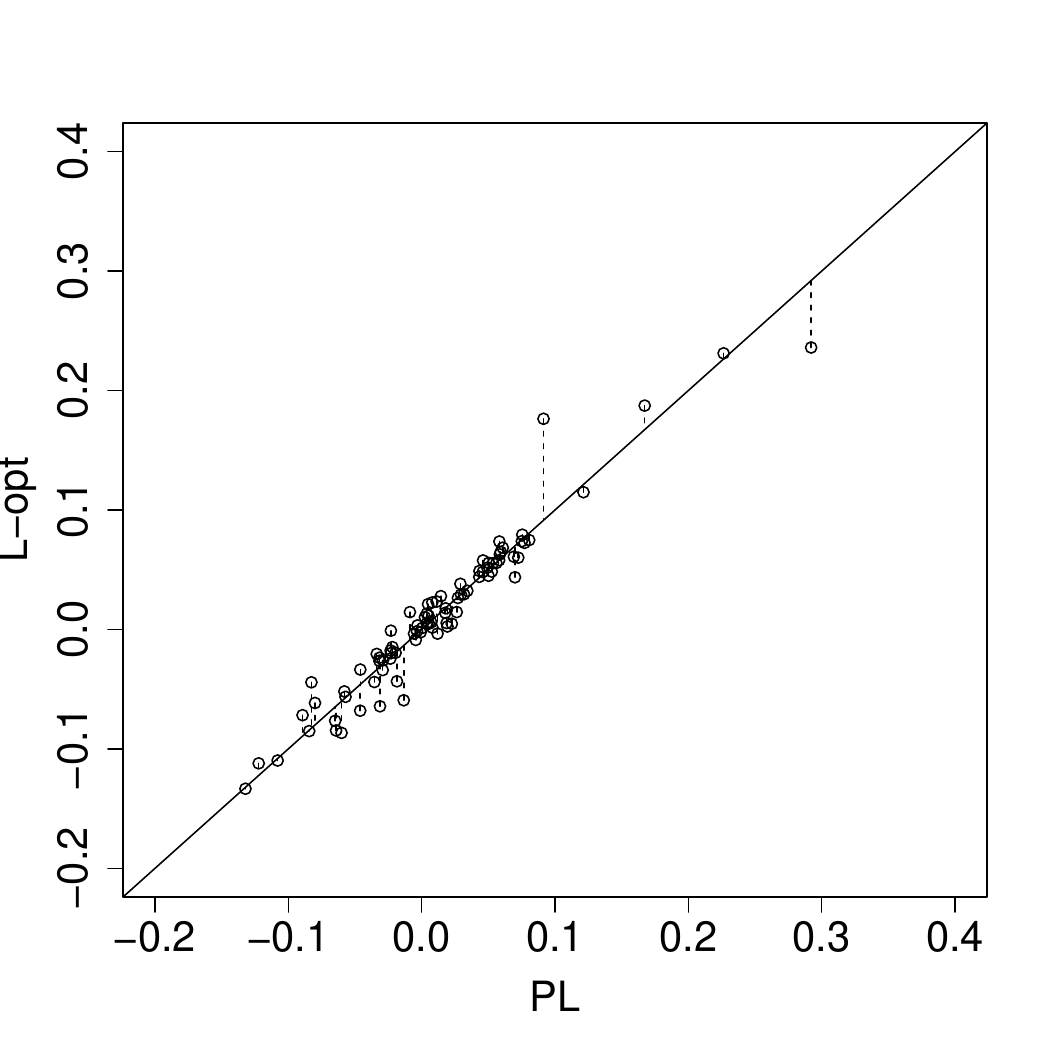} \\
		\includegraphics[width=65mm]{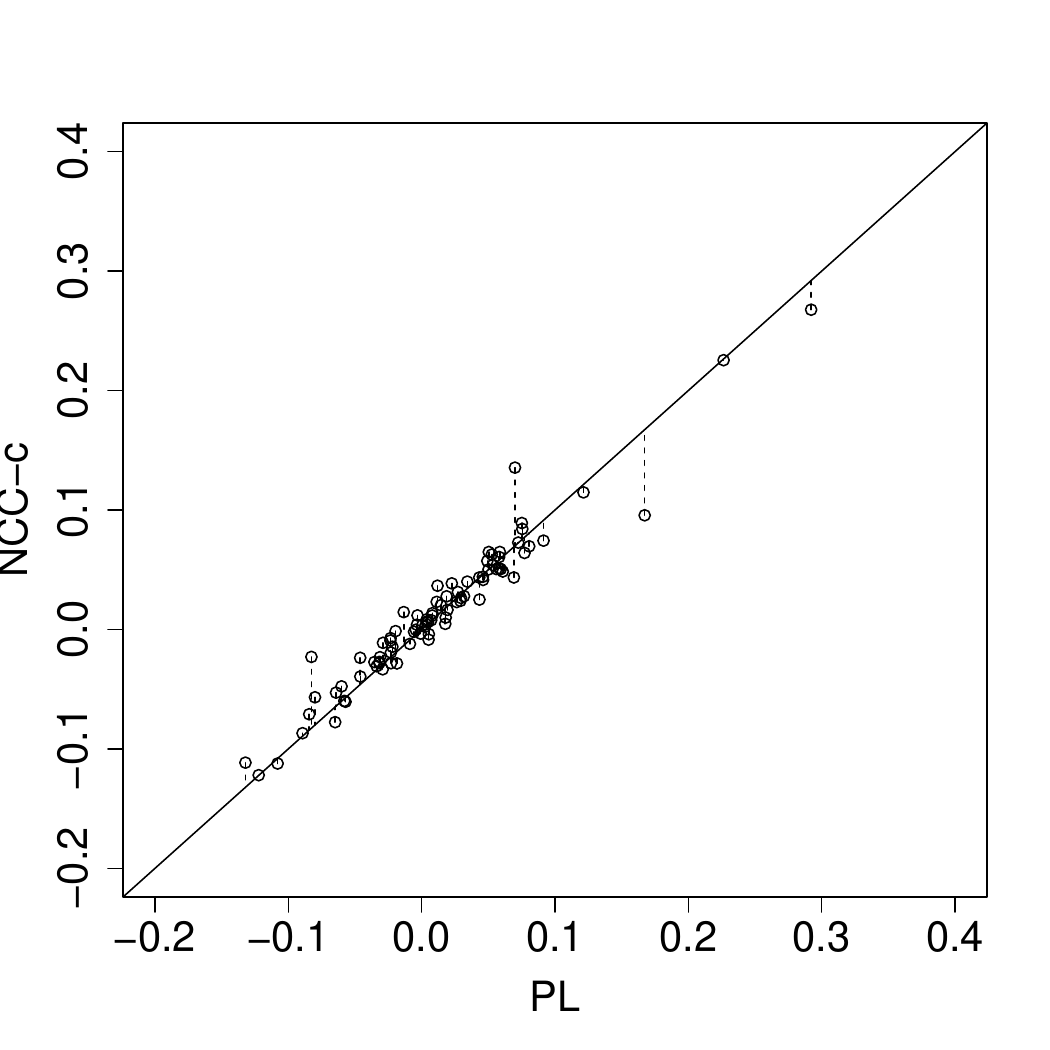} &   \includegraphics[width=65mm]{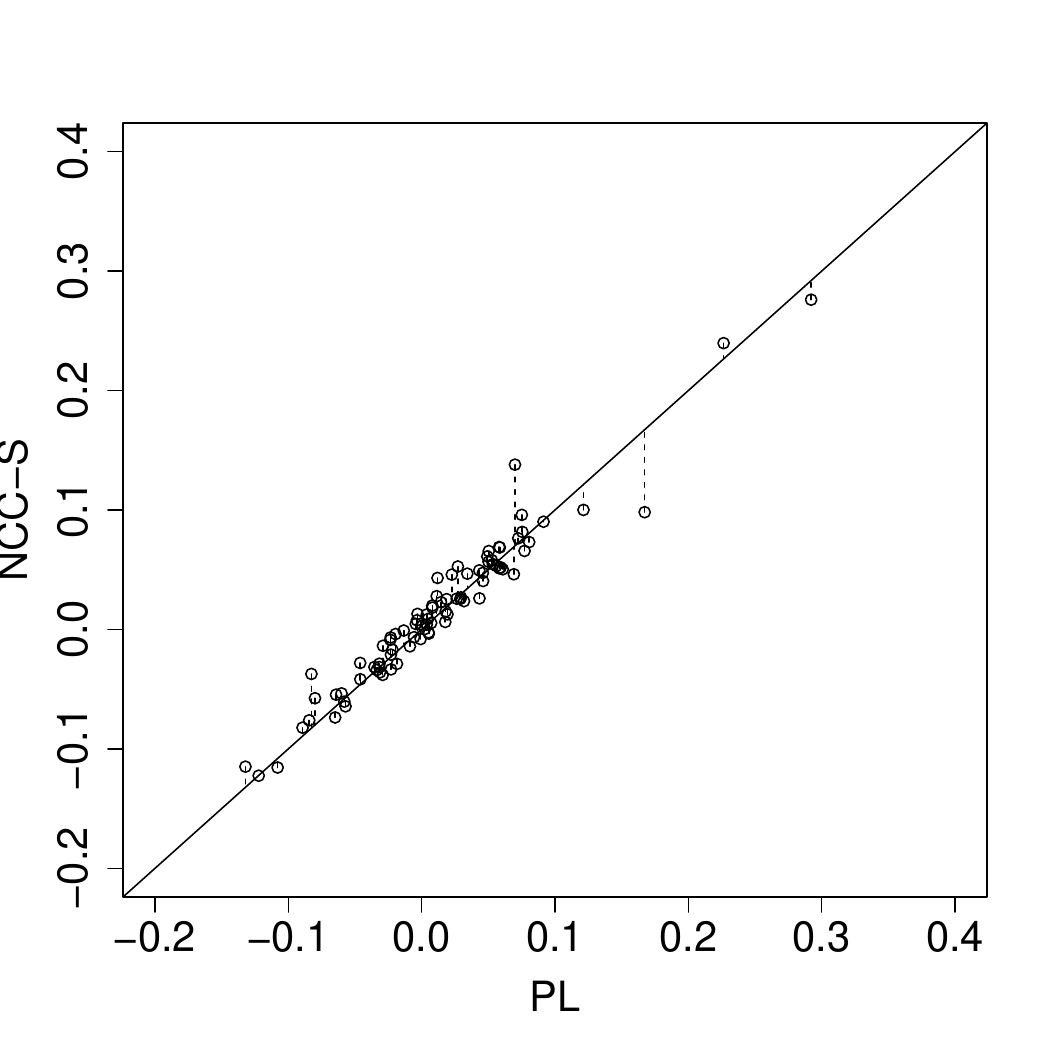} \\
		\multicolumn{2}{c}{\includegraphics[width=65mm]{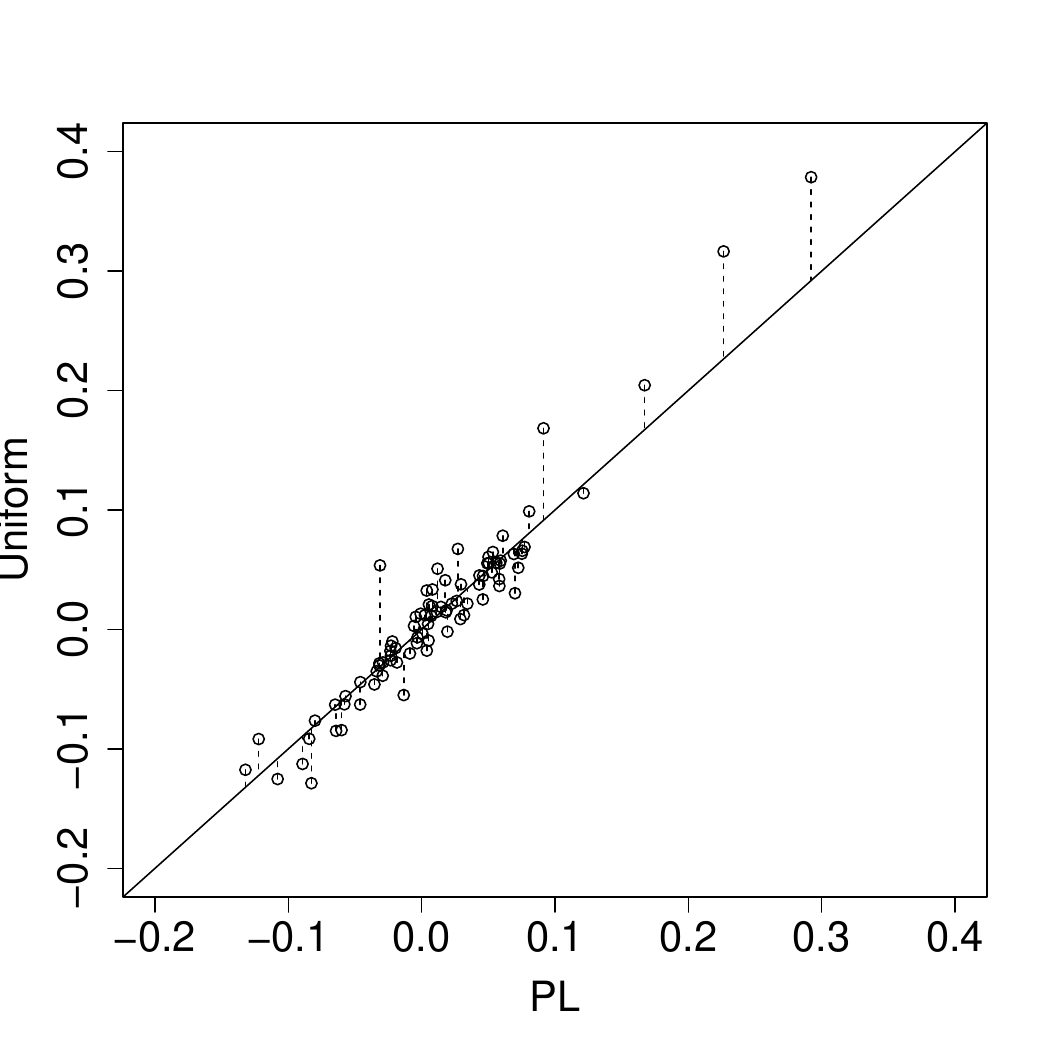} }\\
	\end{tabular}
	\caption{\label{fig:UKB:beta} UKB CRC data analysis: comparison between the estimated regression coefficients of each method (L-opt, A-opt, NCC-c, NCC-S, uniform) against the PL-based estimator.}
\end{figure}

Our goal in this section is to build a prediction model for CRC, based on the UKB data, using Cox regression. The model includes all 72 identified SNPs, environmental risk factors, and the interactions between sex and physical activity, BMI, smoking status and alcohol consumption, so there are overall 91 regression coefficients. The full data analysis takes 93.4 seconds, while the runtimes for the uniform, L-opt, A-opt, NCC-c and NCC-S, with 4 sampled censored observations per observed failure time, were 1.7, 4.4, 9.9, 41.6 and 1765.5 seconds, respectively. The results of these analyses are provided in Tables \ref{tab_ukb_reg_metrics}--\ref{tab_ukb_reg_all} and Figure \ref{fig:UKB:beta}. Table  \ref{tab_ukb_reg_metrics} demonstrates that the L-opt and A-opt methods perform substantially better than the other methods, considering both the runtime and the distance to the full-data PL estimator, measured by the RMSE (for the coefficients) and the Frobenius norm (for the covariance matrix). However, since 93.4 seconds is not too prohibitive, the subsampling methods are not direly needed. 

\begin{figure}
	\centering
	\includegraphics[width=150mm,]{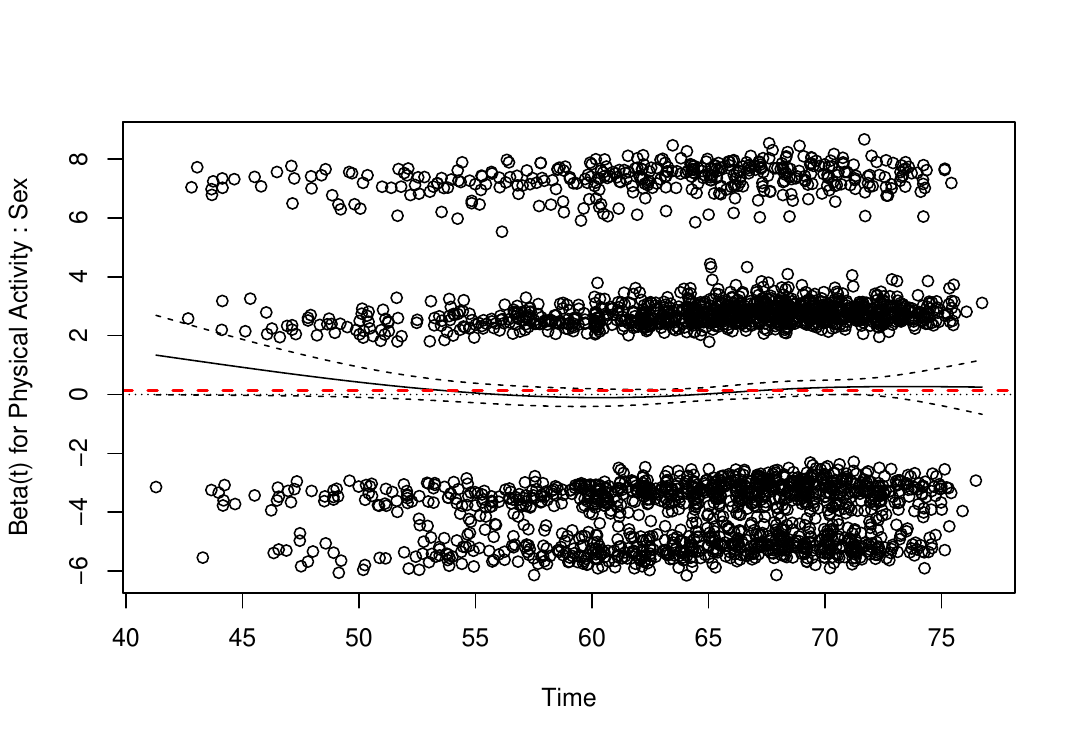}
	\caption{\label{fig:UKB:td:activity_sex} UKB CRC data analysis: Diagnostic graph for evaluating necessity of a time-dependent coefficient for the interaction term of physical activity and sex. The red dashed line shows the time-fixed coefficient for physical activity : sex.  }
\end{figure}

\begin{figure}
	\centering
	\begin{tabular}{cc}
		\includegraphics[width=65mm]{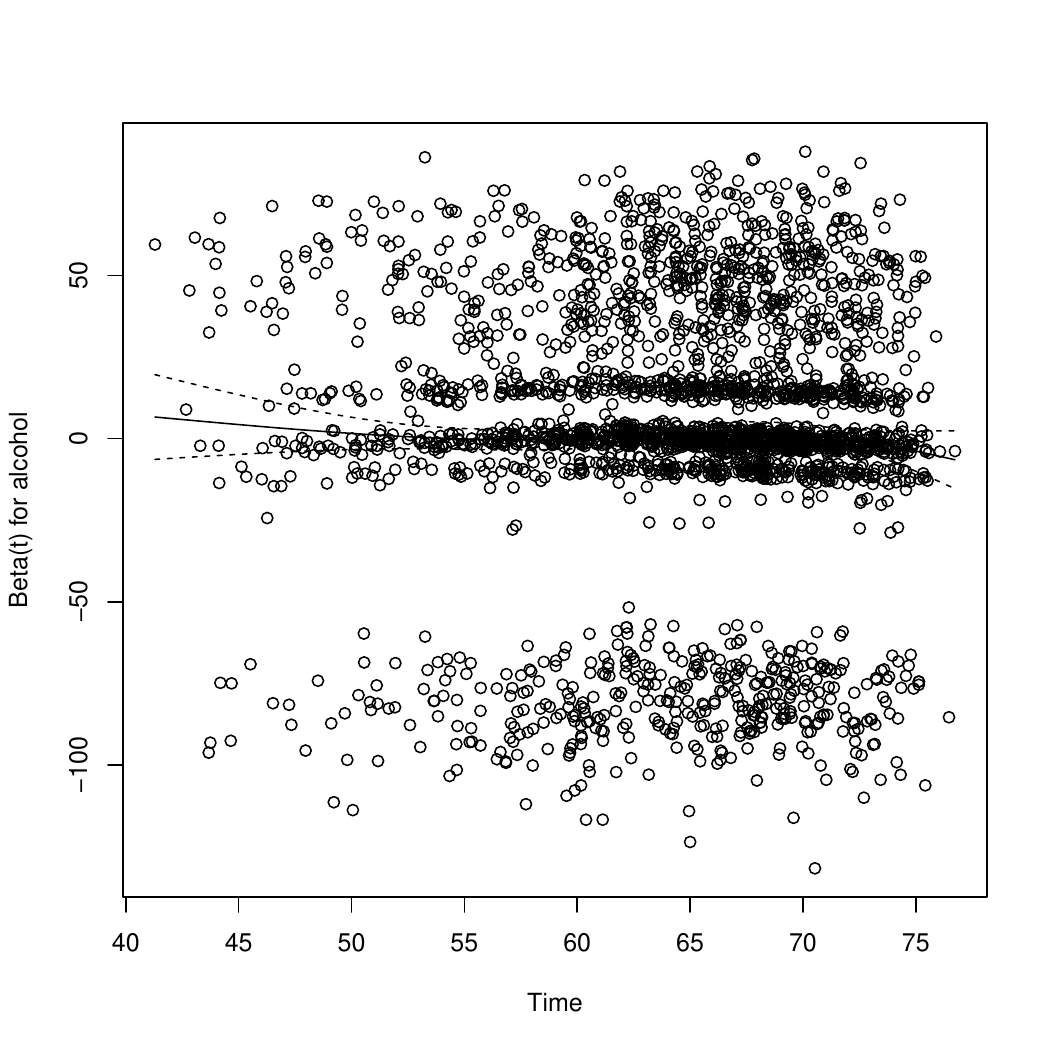} &   \includegraphics[width=65mm]{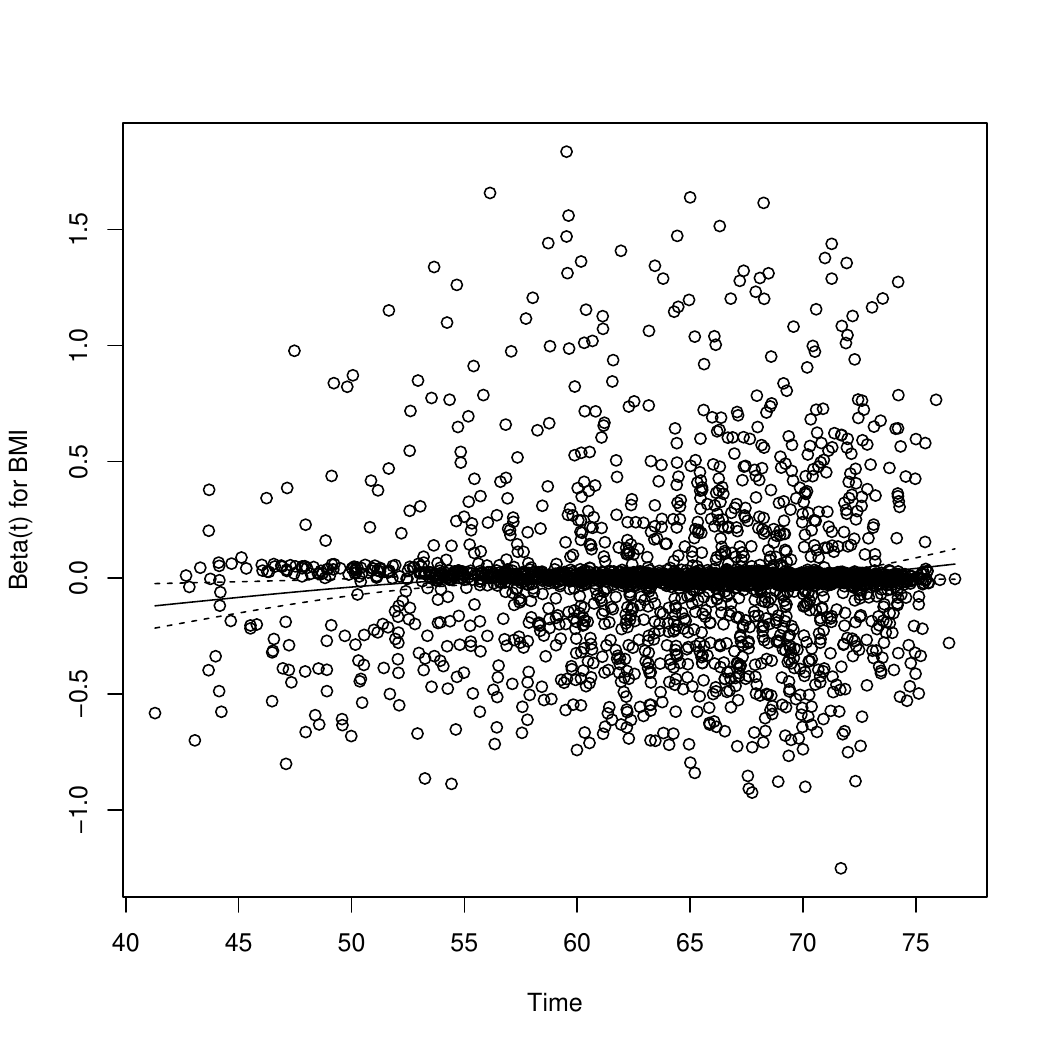} \\  \includegraphics[width=65mm]{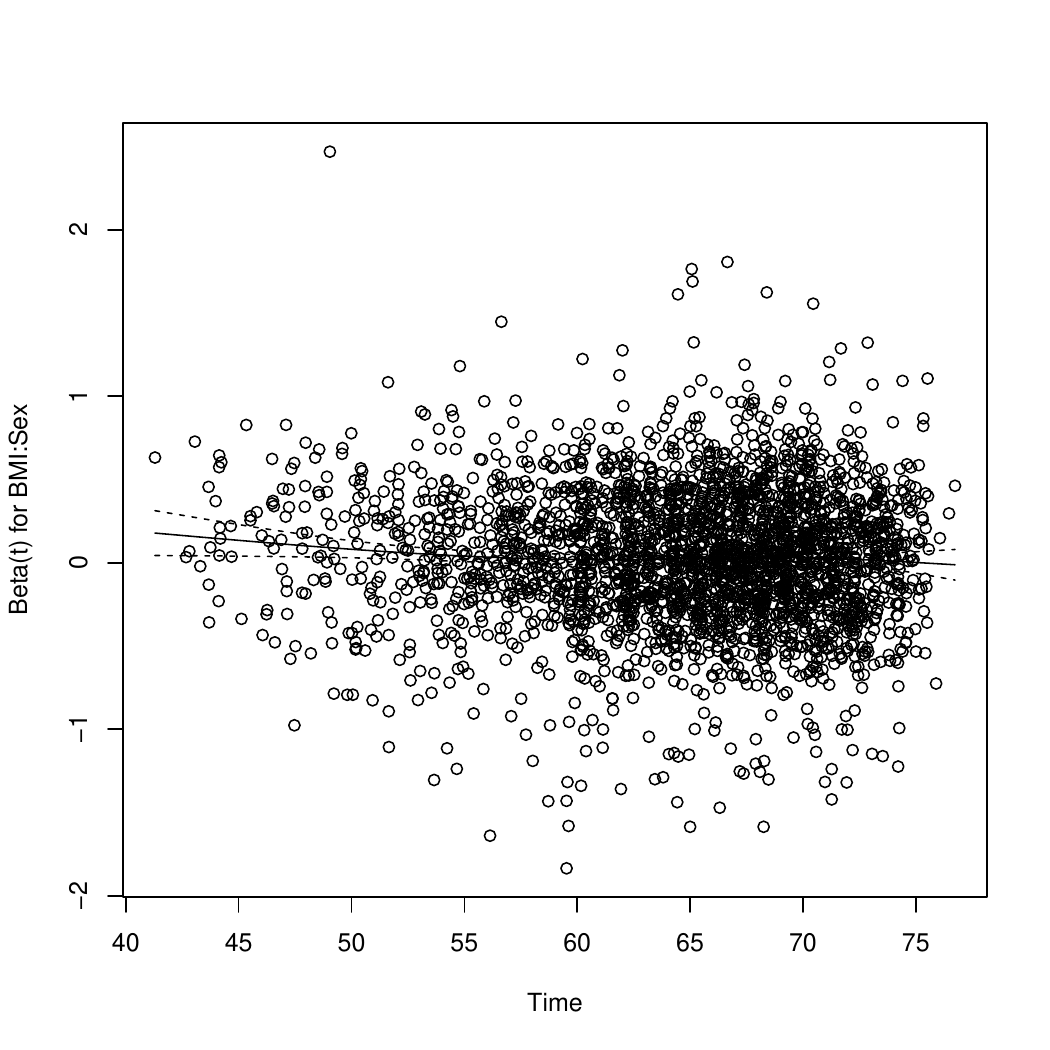} & \includegraphics[width=65mm]{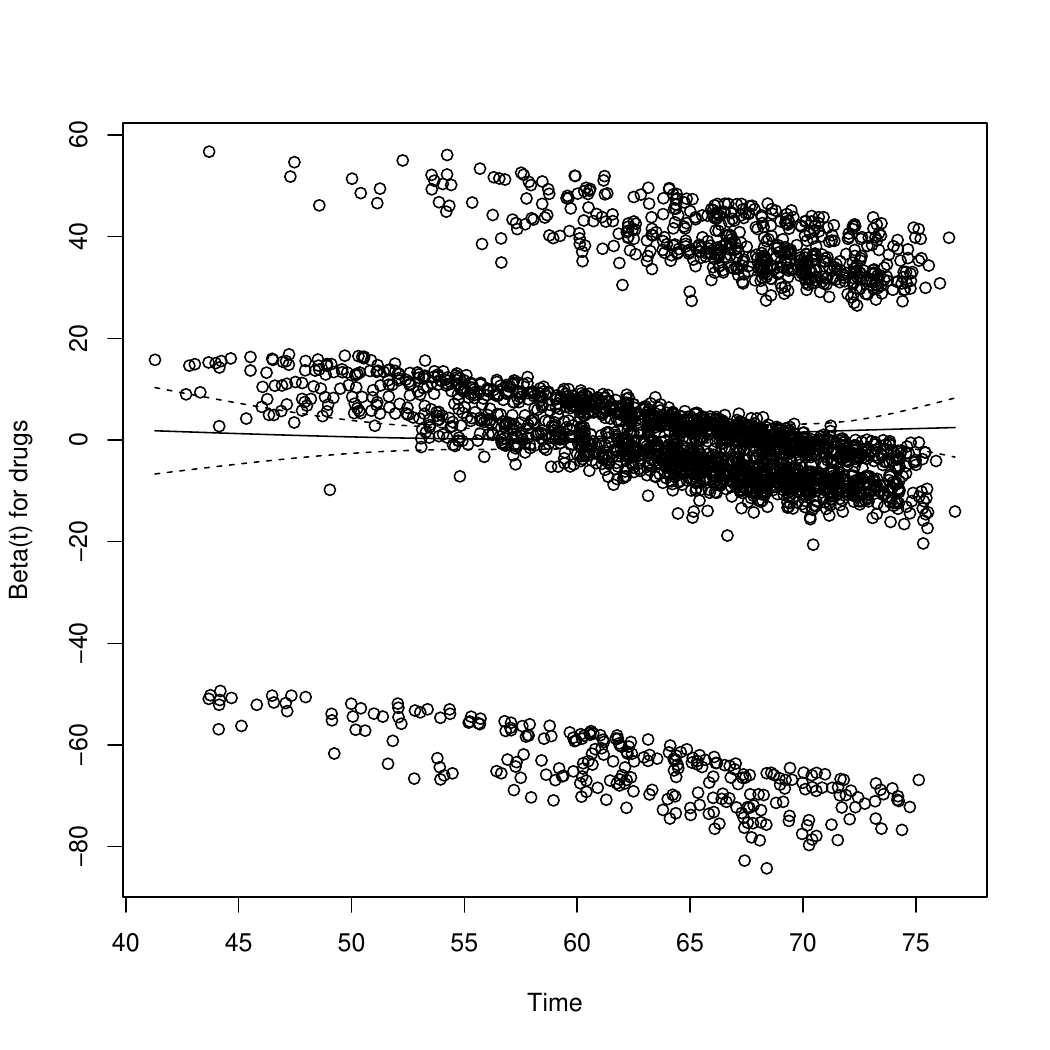} \\
		\includegraphics[width=65mm]{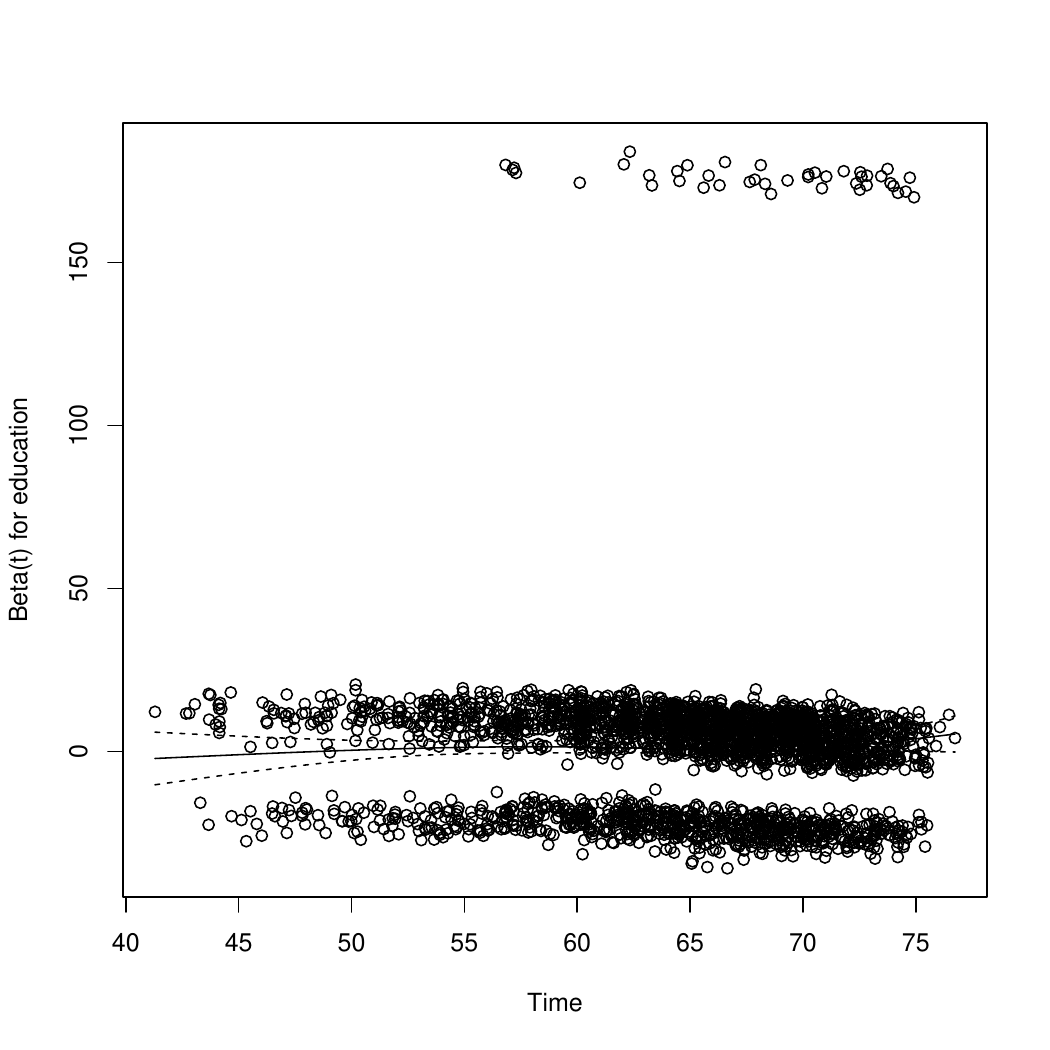}  & 	\includegraphics[width=65mm]{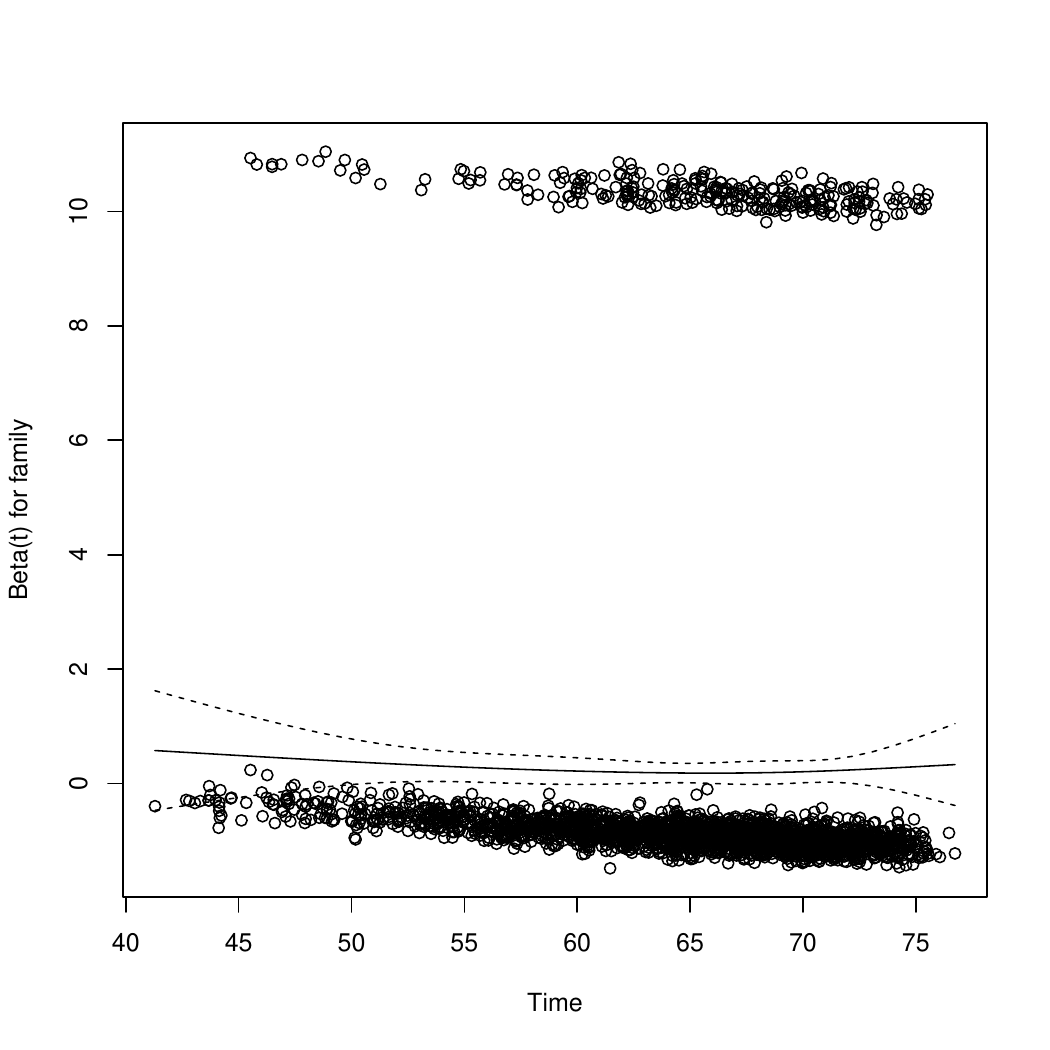} \\
	\end{tabular}
	\caption{\label{fig:UKB:td:graphs} UKB CRC data analysis: Graphical tests for evaluating the necessity of time-dependent coefficients, as produced by the {\tt cox.zph} function - part 1.} 
\end{figure}

\begin{figure}
	\centering
	\begin{tabular}{cc}
		\includegraphics[width=65mm]{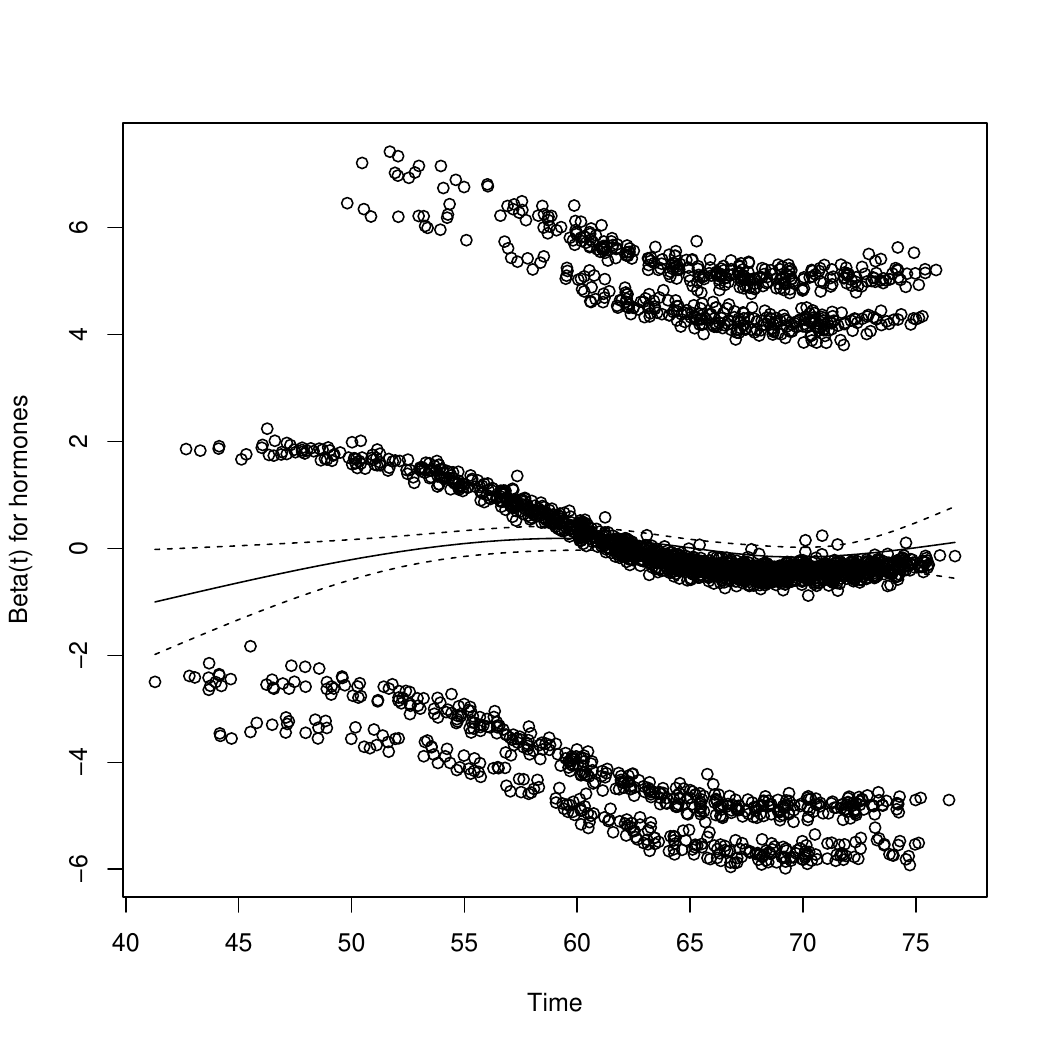} &   \includegraphics[width=65mm]{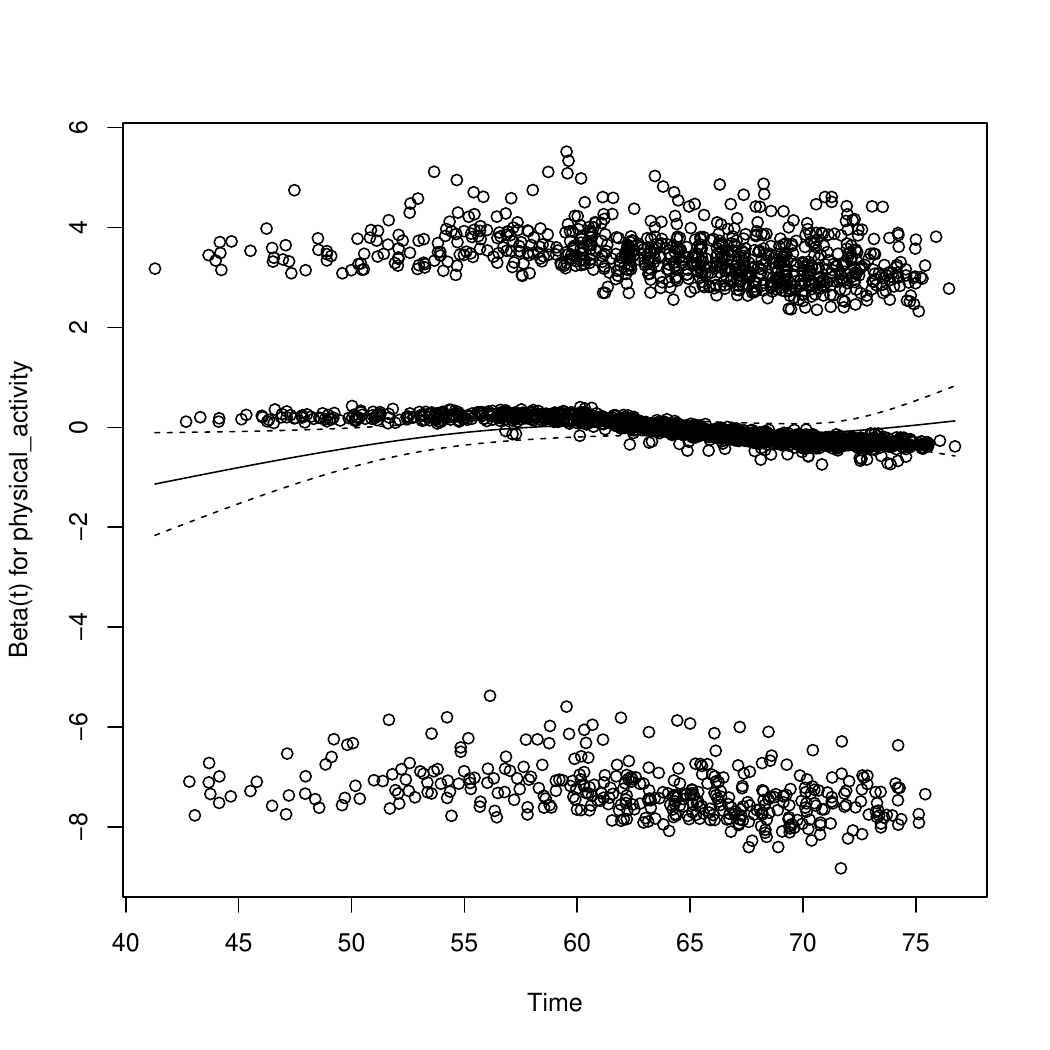} \\  \includegraphics[width=65mm]{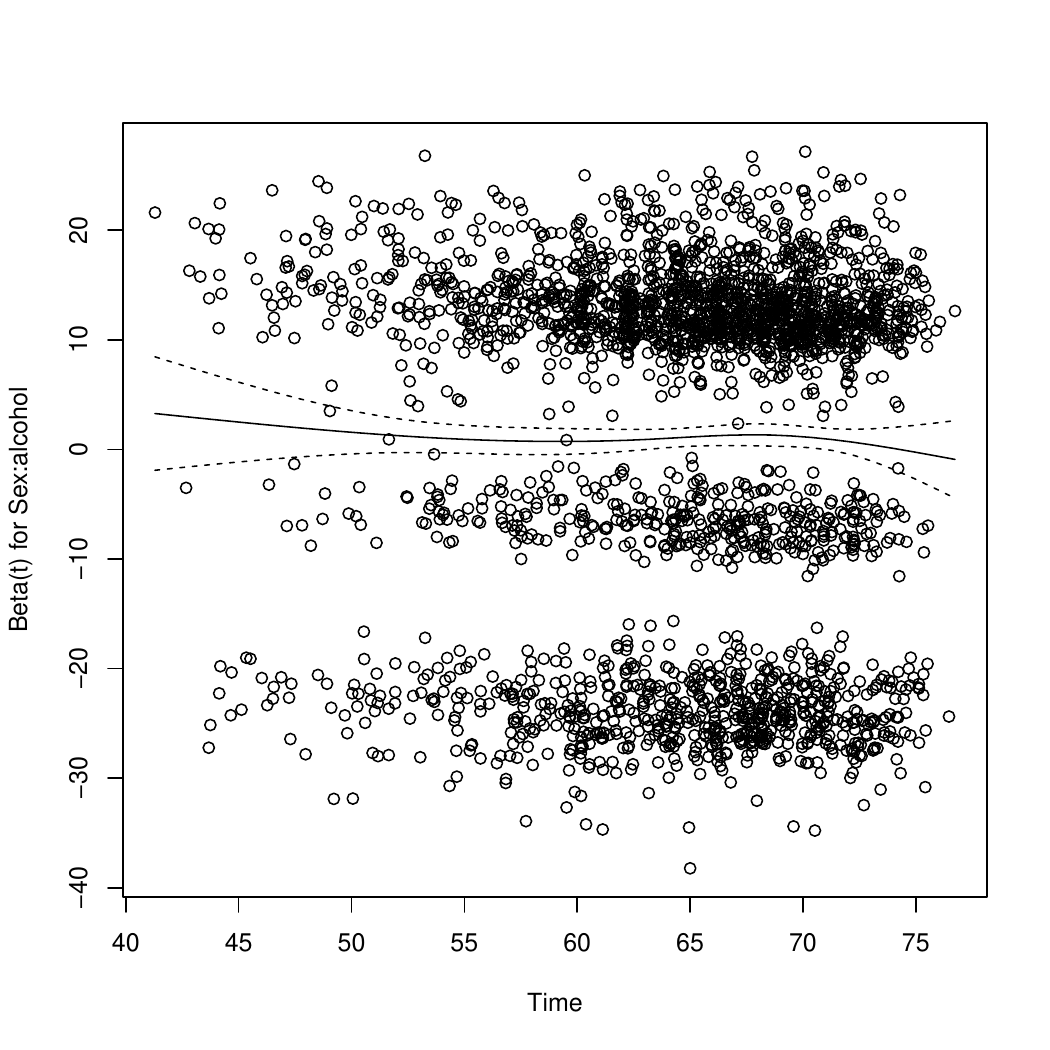} & \includegraphics[width=65mm]{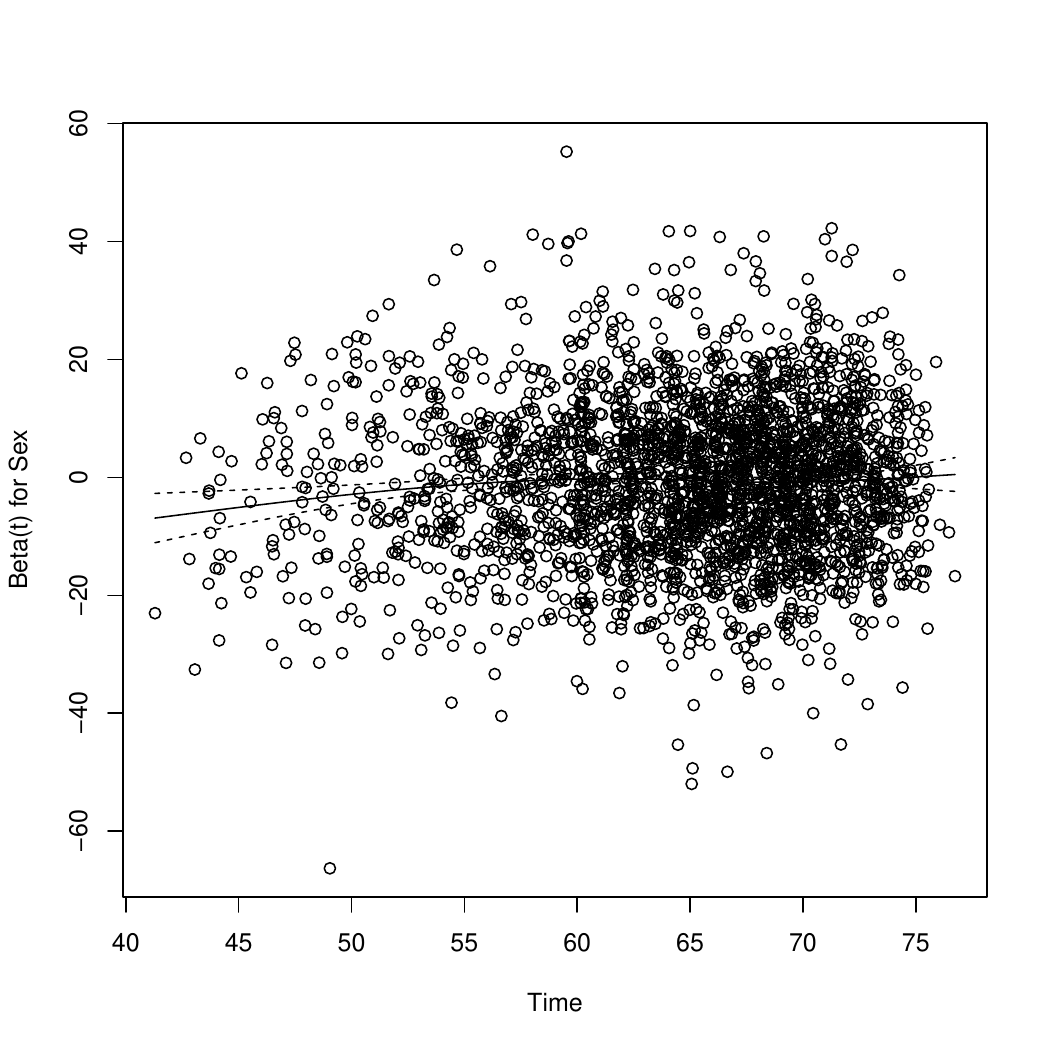} \\
		\includegraphics[width=65mm]{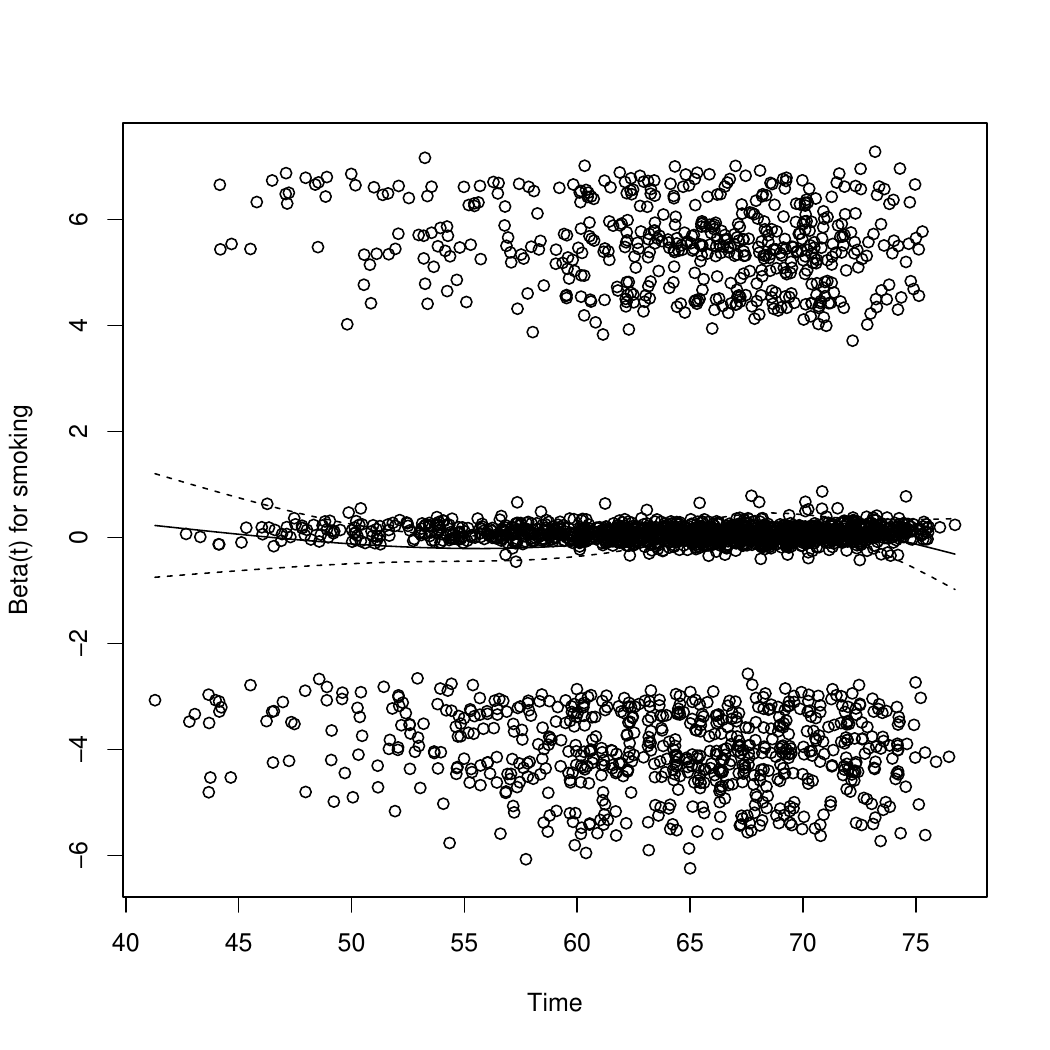}  & 	\includegraphics[width=65mm]{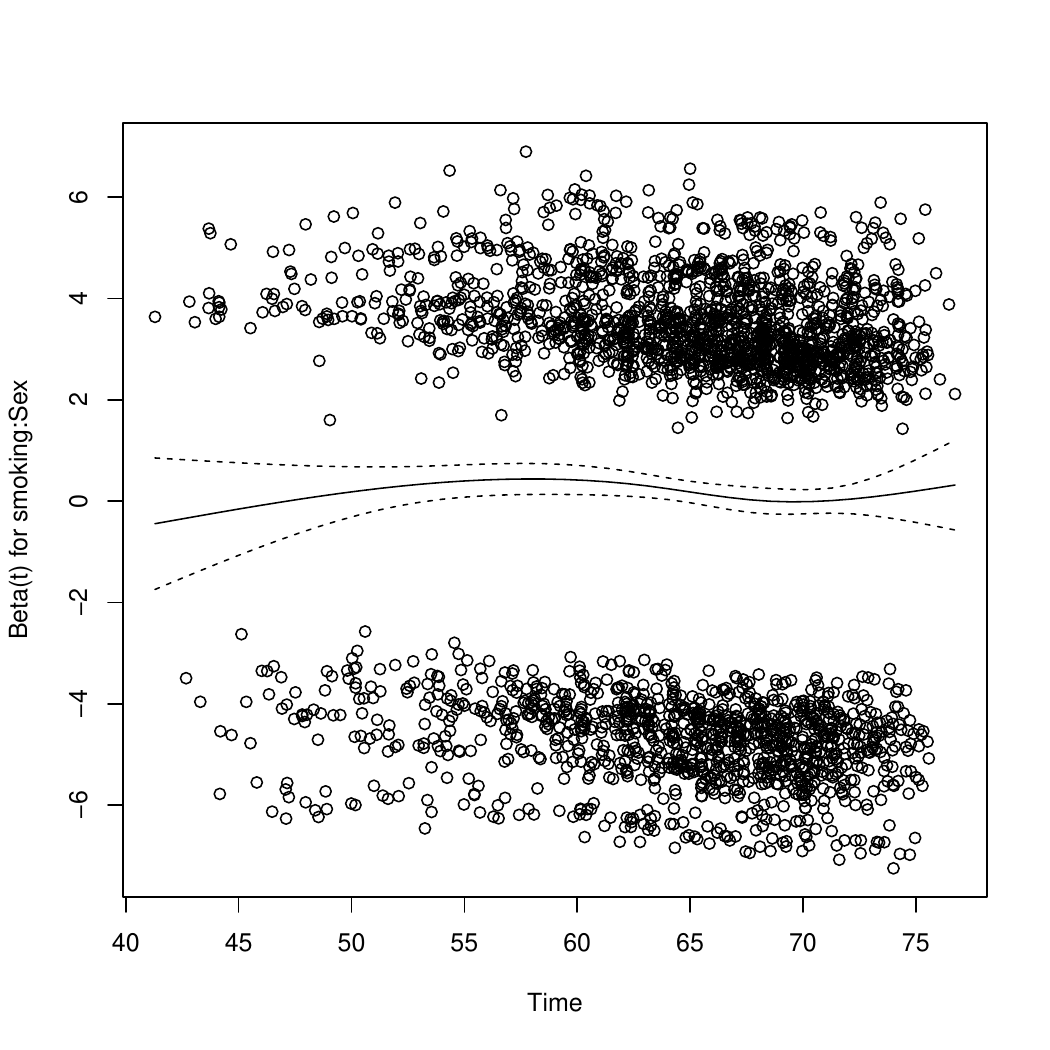} \\
	\end{tabular}
	\caption{\label{fig:UKB:td:graphs2} UKB CRC data analysis: Graphical tests for evaluating the necessity of time-dependent coefficients, as produced by the {\tt cox.zph} function - part 2.}
\end{figure}

A critical assumption for the Cox model is the proportional hazards. The useful function {\tt cox.zph} in the {\tt R} Survival package provides Schoenfeld-residuals-based graphical and statistical tests for evaluating the necessity of time-dependent coefficients. The output of {\tt cox.zph} indicates that the proportionality assumption for the interaction sex*exercise is potentially violated (p-value = 0.0068). Figure \ref{fig:UKB:td:activity_sex} suggests that there appears to be a stronger effect for early onset of CRC, associated with this interaction term. The graphical and statistical tests for the other covariates are provided in \ref{fig:UKB:td:graphs}--\ref{fig:UKB:td:graphs2} and Table \ref{tab:ukb:td:tests}. We thus fit a time-dependent effect $\beta \ln(t)$ for the sex*exercise interaction term, where the $\ln$ function is chosen to shrink the time scale.
For fitting such a model, one could use the ``{\tt tt}'' argument in the {\tt coxph} function, or equivalently, transform a dataset into time-dependent form by means of the {\tt survSplit} function in the Survival package, with the sorted vector of distinct failure times as cut points. More details about these mechanisms are provided in \cite{therneau2017using}. The more distinct observed failure times there are, the bigger the time-dependent dataset becomes, so all observed failure times are rounded to their first decimal digit, producing a maximal rounding error of about 18 days, quite negligible for cancer research with long follow-up duration. The resultant UKB-based time-dependent dataset comprises about 33 million rows, with a median of 71 pseudo-observations per one original observation. Due to the long runtime required by the NCC methods, we only considered the subsampling approaches in this analysis with $q/n_e = 15, 30, 71$, whose results are provided in Tables \ref{tab_ukb_timedep_metrics} -- \ref{tab:ukb:td:all71}. The L-opt method offers a good balance of runtime and accuracy. It substantially reduces the runtime from the full-data PL's 25.7 minutes to approximately 6 minutes, when $q/n_e=71$, while maintaining an advantage over the uniform subsampling in terms of the RMSE and Frobenius norm metrics. When $q/n_e=15, 30$ the L-opt's Frobenius norm metric is about one third of the uniform's, and when $q/n_e=71$, it is about one half, and therefore we advocate using L-opt.

It is evident that as the number of sampled pseudo-observations increases, the different methods give better approximations to the full PL results, both with regards to the point estimates, and with regards to the estimated standard errors, so that inference on the covariate effects can be performed using the subsampled data, almost with no loss of efficiency.

As to the propotional-hazards assumption, upon examining the time-dependent coefficient under consideration, it turns out to be non-significant at the 0.05 level under all methods.

\subsubsection{SNP Marginal Analysis}

Another potentially interesting application for the subsampling approach, is the case of SNP marginal analysis. As mentioned, for example, in \cite{yang2012conditional}, SNPs are usually tested for associations with a trait based on a one-SNP-at-a-time model, while controlling for other clinical and environmental covariates \citep{bush2012genome}. Large electronic health records (EHR) data and biobanks promote discoveries of new genetic variants, thus advancing precision medicine research \citep{denny2018influence,hughey2019cox}. The Cox regression model was found to increase statistical power in the detection of phenotype-associated SNPs, over the traditionally used logistic regression model \citep{van2008cox,staley2017comparison,hughey2019cox}, hence it would be beneficial to alleviate its computational burden while losing minimal efficiency. The number of SNPs to be tested for an association with a trait of interest can range from thousands to millions \citep{abed2019comparing}, implying that a similar number of models need be fitted. In the UKB, there is a total number of about 96 million testable variants \citep{bycroft2018uk}, and about 5 millions of them are expected to be common SNPs. \cite{hughey2019cox} ran a Cox-regression-based GWAS analysis with 795,850 common SNPs, for 50 different phenotypes, and a sample size of almost 50,000. The overall runtime, pararellized on 36 cores, was 7.1 days. Additionally, the number of observed failure times for the different phenotypes ranged from 104 to 7972, hence our methods are well suited for these phenotypes, as the rare event assumption seems to hold. 

For the sake of demonstration, we conducted a marginal analysis for the UKB CRC data just with the 72 SNPs and the environmental risk factors described above. In an actual GWAS study, one could scan the approximately 5 million common SNPs in the UKB. We derived the optimal sampling probabilities targeting each SNP in its respective model. The results are provided in Tables \ref{tab:ukb:marg:metrics}--\ref{tab:ukb:marg:allSNP}. Table \ref{tab:ukb:marg:metrics} presents the RMSE of each method with regards to the full-data PL estimates, the $l_2$ distance between the variances of the different methods to that of the full-data PL, and the required runtimes. Although the analysis included environmental risk factors, Table \ref{tab:ukb:marg:allSNP} contains just the estimates of the 72 SNPs.  While the full-data PL required about 11 minutes, only about 1, 1.2 and 0.25 minutes were required by the L-opt, A-opt and uniform, respectively. In practice, one would also use multiple cores in order to parallelize over the different SNPs. Evidently, our methods are adequate for analyzing millions of SNPs, while maintaining statistical efficiency. Currently, more work is underway to further improve the running time, and then we intend to conduct a comprehensive GWAS analysis.

\section{Discussion and Future Research}
We have proposed efficient and fast subsampling-based estimators for the Cox proportional-hazards model, motivated by two optimality criteria, under the settings of rare events. These estimators help alleviate the computational burden in analyzing massive datasets, while sacrificing minor estimation efficiency. In this work all observed failure times were included, while the censored observations are downsampled using optimal sampling probabilities.

\subsection{Memory Considerations}
When it comes to memory constraints, if the dataset is too big to be loaded into the RAM, we suggest the following adjustment. Divide the dataset into $K$ mutually exclusive sub-datasets, such that the first sub-dataset  contains all the observed failure times, and the others contain random subsets of censored observations. Then, one can load into RAM the ``failure'' dataset, and loop through the other $K-1$ datasets, one at a time, and for each censored observation in the $k$'th sub-dataset, $k=2,\ldots, K$, compute $\|\tilde{\vectr{a}}_{(k)}(\tilde{\vectr{\beta}}_U)\|_2 $ or $\|\matrx{\tilde{\mathcal{I}}}^{-1}(\tilde{\vectr{\beta}}_U)\tilde{\vectr{a}}_{(k)}(\tilde{\vectr{\beta}}_U)\|_2$, such that $\tilde{\vectr{a}}_{(k)}(\tilde{\vectr{\beta}}_U)$ is computed based on the observations currently loaded into RAM, where each censored observation is given a weight of $K-1$. Since all the observed failures are always used, we expect that this procedure will yield a very good approximation to the quantities that would have been calculated based on all data together. $\tilde{\vectr{\beta}}_U$ and $\matrx{\tilde{\mathcal{I}}}^{-1}(\tilde{\vectr{\beta}}_U)$ can be computed by randomly loading $q$ lines from the censored observations into RAM. 
Eventually, a vector of size $n_c$ with the corresponding norm values is obtained, and can be saved in RAM, so the approximated optimal sampling probabilities can be derived. After determining the subsample according to the optimal probabilities, only the relevant lines of data should then be loaded to RAM and the weighted Cox regression can be run as before. This idea has not been explored in simulations, but we expect that the previously derived results will hold also for this adaptation.

When a time-dependent coefficient is included in the model, the dataset typically inflate to a very large extent, which may pose memory-related problems. A different solution for this case is to subsample the data by regarding the time-dependent coefficient as time-fixed, and then fitting a time-dependent coefficient model on the subsample. This way, the computational burden will be reduced substaintially, at the price that the sampling probabilities will no longer be optimally adapted to the time-dependent coefficient, but rather to its time-fixed version.

\subsection{Future Research}

There are a number of potential research directions which can be pursued following this work:

$\bullet$ One could extend these methods to the case of non-rare events, such that the observed failure times need also be downsampled.

$\bullet$ This work stands as a cornerstone for development of subsampling tools for other Cox-regression-based methods, such as (semi-)competing risks, frailty models, correlated data, recurrent events, multistate models and more. Existing methods for analyzing these models are usually more computationally intensive, and therefore our work is an important first step for reducing computational burden when dealing with more complex data and modeling situations. 

$\bullet$ Cox regression might be useful as a working model for computing sampling probabilities for other computationally intensive methods, where tailor-made analytic derivation is intractable. Particularly, it would make sense to consider PL and log-rank-based random forests \citep{ishwaran2008random}, gradient boosting \citep{binder2009boosting}, among others. 

On a related note, when the number of covariates is large, a regularized Cox model is often employed. One could use our method without regularization to obtain a subsample, and then use it to run a weighted regularized Cox model, by using, for instance, the {\tt glmnet} function in the eponymous R package \cite{simon2011regularization}. We expect that our method will perform well on this use-case, however its optimality cannot be guaranteed.

$\bullet$ Our method can be evaluated against other NCC and CC methods, on the premise of missing covariates. When some covariates are too costly to measure, one can apply our method on all fully observed data and obtain a subset of observations to undergo measuring. It is interesting to find under which scenarios an improvement upon the classically used NCC and CC can be achieved.


$\bullet$ In this work, $q$ is chosen according to the researchers' available computational resources. An interesting research objective, which will be addressed in a separate work, is quantifying the efficiency loss in terms of $q$. 

\section*{Acknowledgments}
The author MG gratefully acknowledges support from the U.S.-Israel Binational
Science Foundation (BSF, 2016126), and the Israel Science Foundation (ISF, 1067/17) in carrying out this work. We would like to thank Asaf Ben-Arie for assistance with the programming involved in this work and Li Hsu for fruitful discussions. This research has been conducted using the UK Biobank Resource, under project number 56885.

\newpage

\begin{table}
	\caption{\label{tab_sim01} Simulation results of right-censored data without and with delayed entry for scenarios A, B, C: the value of $q$ was set equal to the number of failures, hence its mean, $\bar{q}$ is reported (standard errors multiplied by $\sqrt{500}$ are reported).} 
	\centering
	\fbox{%
		\begin{tabular}{lclccccc}
			 & $\bar{q}$ & Method & RMSE $\vectr{\beta}^o$ & RR & RMSE $\hat{\vectr{\beta}}_{PL}$ & CR & Runtime in sec.\\ 
			\hline 
			\multicolumn{8}{l}{No delayed entry}\\
			A  & 585 (23) & PL & 0.086 (0.025) & 1.000 & 0.000 (0.000) & 0.954 & 0.086 (0.037) \\ 
			&  & L-opt & 0.113 (0.033) & 1.322 & 0.074 (0.021) & 0.947 & 0.032 (0.021) \\ 
			&  & A-opt & 0.112 (0.034) & 1.315 & 0.072 (0.022) & 0.952 & 0.025 (0.022) \\ 
			&  & Uniform & 0.238 (0.070) & 2.787 & 0.220 (0.065) & 0.913 & 0.012 (0.021) \\ 
			&  & NCC-c & 0.140 (0.045) & 1.635 & 0.112 (0.036) & 0.953 & 0.341 (0.073) \\ 
			&  & NCC-S & 0.127 (0.038) & 1.488 & 0.093 (0.027) & 0.953 & 0.519 (0.075) \\
\end{tabular}}
\end{table}

\begin{table}
	\caption{\label{tab_sim02} Simulation results of right-censored data without and with delayed entry for scenarios A, B, C:  the value of $q$ was set equal to the number of failures$\times 2$, hence its mean, $\bar{q}$, is reported (standard errors multiplied by $\sqrt{500}$ are reported).} 
	\centering
	\fbox{%
		\begin{tabular}{lclccccc}
			& $\bar{q}$ & Method & RMSE $\vectr{\beta}^o$ & RR & RMSE $\hat{\vectr{\beta}}_{PL}$ & CR & Runtime in sec.\\ 
			\hline 
			\multicolumn{8}{l}{No delayed entry}\\
			A  & 1169 (46) & PL & 0.086 (0.025) & 1.000 & 0.000 (0.000) & 0.954 & 0.080 (0.028) \\ 
			&  & L-opt & 0.100 (0.029) & 1.167 & 0.051 (0.015) & 0.950 & 0.033 (0.017) \\ 
			&  & A-opt & 0.097 (0.030) & 1.140 & 0.049 (0.015) & 0.954 & 0.028 (0.018) \\ 
			&  & Uniform & 0.178 (0.052) & 2.083 & 0.154 (0.046) & 0.939 & 0.013 (0.017) \\ 
			&  & NCC-c & 0.115 (0.034) & 1.350 & 0.079 (0.024) & 0.951 & 0.333 (0.054) \\ 
			&  & NCC-S & 0.105 (0.030) & 1.232 & 0.062 (0.019) & 0.951 & 0.532 (0.082) \\
\end{tabular}}
\end{table}

\begin{table}
	\caption{\label{tab_sim03} Simulation results of right-censored data without and with delayed entry for scenarios A, B, C:  the value of $q$ was set equal to the number of failures$\times 3$, hence its mean, $\bar{q}$, is reported (standard errors multiplied by $\sqrt{500}$ are reported).} 
	\centering
	\fbox{%
		\begin{tabular}{lclccccc}
			& $\bar{q}$ & Method & RMSE $\vectr{\beta}^o$ & RR & RMSE $\hat{\vectr{\beta}}_{PL}$ & CR & Runtime in sec.\\ 
			\hline 
			\multicolumn{8}{l}{No delayed entry}\\
			A   & 1754 (69) & PL & 0.086 (0.025) & 1.000 & 0.000 (0.000) & 0.954 & 0.077 (0.025) \\ 
			&  & L-opt & 0.094 (0.028) & 1.096 & 0.041 (0.013) & 0.950 & 0.045 (0.015) \\ 
			&  & A-opt & 0.095 (0.028) & 1.110 & 0.041 (0.012) & 0.954 & 0.034 (0.020) \\ 
			&  & Uniform & 0.153 (0.046) & 1.793 & 0.124 (0.037) & 0.940 & 0.015 (0.015) \\ 
			&  & NCC-c & 0.108 (0.033) & 1.262 & 0.063 (0.020) & 0.945 & 0.331 (0.049) \\ 
			&  & NCC-S & 0.099 (0.030) & 1.161 & 0.048 (0.014) & 0.950 & 0.522 (0.072) \\
\end{tabular}}
\end{table}

\begin{table}
	\caption{\label{tab_sim_td_03} Simulation results of right-censored data with time dependent covariates: the value of $q$ was set equal to the number of failures$\times 3$, hence the mean, $\bar{q}$, and standard deviation of $q$, SD($q$), are reported.} 
	\centering
	\fbox{%
		\begin{tabular}{lclccccc}
			& $\bar{q}$ & Method & RMSE $\vectr{\beta}^o$ & RR & RMSE $\hat{\vectr{\beta}}_{PL}$ & CR & Runtime in sec.\\ 
			\hline 
			\multicolumn{8}{l}{No delayed entry}\\
			A   & 1061 (54) & PL & 0.128 (0.036) & 1.000 & 0.000 (0.000) & 0.957 & 0.199 (0.042) \\ 
			&  & L-opt & 0.142 (0.039) & 1.106 & 0.062 (0.017) &    -- & 0.084 (0.028) \\ 
			&  & A-opt & 0.142 (0.039) & 1.106 & 0.061 (0.017) &    -- & 0.091 (0.028) \\ 
			&  & Uniform & 0.260 (0.075) & 2.026 & 0.223 (0.064) &    -- & 0.023 (0.015) \\ 
			&  & L-opt-ps & 0.143 (0.041) & 1.115 & 0.070 (0.019) & 0.959 & 0.039 (0.022) \\ 
			&  & A-opt-ps & 0.145 (0.039) & 1.128 & 0.068 (0.018) & 0.960 & 0.041 (0.024) \\ 
			B & 1322 (62) & PL & 0.259 (0.100) & 1.000 & 0.000 (0.000) & 0.954 & 0.294 (0.045) \\ 
			&  & L-opt & 0.291 (0.110) & 1.125 & 0.138 (0.051) &    -- & 0.113 (0.031) \\ 
			&  & A-opt & 0.288 (0.110) & 1.113 & 0.131 (0.050) &    -- & 0.124 (0.032) \\ 
			&  & Uniform & 0.441 (0.171) & 1.705 & 0.358 (0.139) &    -- & 0.033 (0.019) \\ 
			&  & L-opt-ps & 0.300 (0.115) & 1.160 & 0.147 (0.053) & 0.957 & 0.048 (0.019) \\ 
			&  & A-opt-ps & 0.288 (0.112) & 1.114 & 0.139 (0.050) & 0.952 & 0.052 (0.020) \\ 
			C & 1039 (56) & PL & 0.569 (0.391) & 1.000 & 0.000 (0.000) & 0.947 & 0.196 (0.032) \\ 
			&  & L-opt & 0.647 (0.457) & 1.138 & 0.301 (0.207) &    -- & 0.080 (0.025) \\ 
			&  & A-opt & 0.651 (0.455) & 1.144 & 0.302 (0.219) &    -- & 0.087 (0.028) \\ 
			&  & Uniform & 1.201 (0.845) & 2.113 & 0.980 (0.700) &    -- & 0.022 (0.017) \\ 
			&  & L-opt-ps & 0.647 (0.453) & 1.137 & 0.317 (0.223) & 0.949 & 0.034 (0.010) \\ 
			&  & A-opt-ps & 0.637 (0.435) & 1.120 & 0.279 (0.173) & 0.952 & 0.037 (0.016) \\ 
		\end{tabular}
	}
\end{table}

\begin{table}
	\centering
	\caption{\label{tab_SE} Simulation results of right-censored data without delayed entry, under setting ``C", with three censored observations per observed failure time: empirical SE (Emp.), estimated SEs based on the full sample (Full) and on the subsample (Sub) for each estimated regression coefficient. The Frobenius norm was used to calculate the distance between the empirical and estimated variance matrices.}
	\begin{tabular}{|c|ccc|ccc|ccc|}
		\hline
		&  & Uniform &  &  & L-opt &  &  & A-opt &  \\ \hline
		& Emp. & Full & Sub & Emp. & Full & Sub & Emp. & Full & Sub \\ 
$ \beta_1 $& 0.4621 & 0.4622 & 0.4370 & 0.2874 & 0.2787 & 0.2794 & 0.2728 & 0.2666 & 0.2667 \\ 
$ \beta_2 $& 0.4463 & 0.4493 & 0.4243 & 0.2820 & 0.2720 & 0.2727 & 0.2604 & 0.2597 & 0.2598 \\ 
$ \beta_3 $& 0.0793 & 0.0768 & 0.0728 & 0.0457 & 0.0459 & 0.0460 & 0.0498 & 0.0488 & 0.0488 \\ 
$ \beta_4 $& 0.8901 & 0.8852 & 0.8359 & 0.5513 & 0.5345 & 0.5357 & 0.5144 & 0.5089 & 0.5091 \\ 
$ \beta_5 $& 0.0832 & 0.0886 & 0.0838 & 0.0548 & 0.0529 & 0.0530 & 0.0563 & 0.0556 & 0.0556 \\ 
$ \beta_6 $& 0.0598 & 0.0607 & 0.0560 & 0.0372 & 0.0354 & 0.0354 & 0.0377 & 0.0378 & 0.0379 \\ \hline
		$ \Vert . \Vert_F $ & 0.0000 & 0.0112 & 0.1339 & 0.0000 & 0.0297 & 0.0277 & 0.0000 & 0.0108 & 0.0106 \\ \hline
	\end{tabular}
\end{table}

\begin{table}
	\caption{\label{tab_ukb_reg_metrics} UKB CRC data analysis: metrics for evaluating the different methods. The RMSE and Frobenius distance are with respect to the full-data PL estimator. 4 censored observations were sampled for each observed failure time.}
	\centering
	\fbox{%
\begin{tabular}{lccc}
	& RMSE & $ \Vert . \Vert_F $ & Time (seconds) \\ 
	\hline
	PL & 0.0000 & 0.0000 & 93.4000 \\ 
	L-opt & 0.1871 & 0.0369 & 4.4200 \\ 
	A-opt & 0.1738 & 0.0196 & 9.9000 \\ 
	Uniform & 0.3547 & 0.0664 & 1.7000 \\ 
	NCC-c & 0.2017 & 0.0374 & 41.6100 \\ 
	NCC-S & 0.2137 & 0.0609 & 1765.5000 \\ 
\end{tabular}}
\end{table}

\begin{table}[ht]
	\caption{\label{tab_ukb_reg_all} UKB CRC data analysis: estimated regression coefficients and standard errors of each method. 4 censored observations were sampled for each observed failure time.}
	\tiny
	\centering
	\fbox{%
	\begin{tabular}{lcccccccccccc}
		& PL & PL se & L-opt & L se & A-opt & A se & Unif & Unif se & NCC-c & NCC-c se & NCC-S & NCC-S se \\ 
		\hline
		BMI & 0.0005 & 0.0059 & 0.0000 & 0.0066 & 0.0024 & 0.0065 & -0.0034 & 0.0068 & 0.0037 & 0.0064 & 0.0032 & 0.0064 \\ 
		Smoking & 0.0700 & 0.0595 & 0.0387 & 0.0666 & 0.0496 & 0.0670 & 0.0303 & 0.0678 & 0.1355 & 0.0651 & 0.1380 & 0.0664 \\ 
		Family & 0.2265 & 0.0635 & 0.2389 & 0.0718 & 0.2656 & 0.0715 & 0.3165 & 0.0803 & 0.2254 & 0.0735 & 0.2397 & 0.0765 \\ 
		Activity & -0.0800 & 0.0624 & -0.0762 & 0.0697 & -0.0869 & 0.0698 & -0.0764 & 0.0717 & -0.0568 & 0.0685 & -0.0576 & 0.0703 \\ 
		Sex & -0.7535 & 0.2580 & -0.8085 & 0.2943 & -0.6431 & 0.2885 & -1.0261 & 0.3085 & -0.6301 & 0.2879 & -0.6260 & 0.2886 \\ 
		Alcohol2 & -0.0135 & 0.0743 & 0.0033 & 0.0836 & -0.0207 & 0.0832 & -0.0550 & 0.0856 & 0.0145 & 0.0819 & -0.0009 & 0.0834 \\ 
		Alcohol3 & -0.0827 & 0.0696 & -0.0801 & 0.0782 & -0.0535 & 0.0784 & -0.1287 & 0.0803 & -0.0230 & 0.0767 & -0.0373 & 0.0784 \\
\end{tabular}}
\end{table}

\begin{table}[ht]
	\caption{\label{tab:ukb:td:tests} UKB CRC data analysis: Statistical tests for evaluating the necessity of time-dependent coefficients, as produced by the {\tt cox.zph} function.}
	\centering
	\fbox{%
		\begin{tabular}{lccc}
			& Chisq & df & P-value \\ 
			\hline
			BMI & 2.5101 & 1.0000 & 0.1131 \\ 
			Smoking & 3.6420 & 1.0000 & 0.0563 \\ 
			Family & 0.3278 & 1.0000 & 0.5670 \\ 
			Activity & 1.9921 & 1.0000 & 0.1581 \\ 
			Sex & 6.6049 & 1.0000 & 0.0102 \\ 
			Alcohol & 0.1426 & 2.0000 & 0.9312 \\ 
			Education & 6.0541 & 4.0000 & 0.1951 \\ 
			Drugs & 1.8821 & 2.0000 & 0.3902 \\ 
			Hormones & 4.8073 & 1.0000 & 0.0283 \\ 
			BMI:Sex & 6.1406 & 1.0000 & 0.0132 \\ 
			Smoking:Sex & 3.6072 & 1.0000 & 0.0575 \\ 
			Sex:alcohol & 2.4712 & 2.0000 & 0.2907 \\ 
			Activity:Sex & 7.3212 & 1.0000 & 0.0068 \\ 
	\end{tabular}}
\end{table}

\begin{table}[ht]
	\caption{\label{tab_ukb_timedep_metrics} UKB CRC time-dependent coefficient data analysis:  metrics for evaluating the different methods. 15, 30 and 71 pseudo-observations were sampled per each observed failure time. The RMSE and Frobenius distance are with respect to the full-data PL estimator.}
	\centering
\fbox{%
	\begin{tabular}{lccc}
		& RMSE & $ \Vert . \Vert_F $ & Runtime in sec. \\
		\hline 
		PL & 0.0000 & 0.0000 & 1542.0000 \\ 
		\hline
		15 \\ 
		\hline
		L-opt & 0.1680 & 0.0087 & 342.5340 \\ 
		A-opt & 0.1069 & 0.0052 & 890.0220 \\ 
		Uniform & 0.2547 & 0.0220 & 164.9380 \\ 
		\hline
		30 \\
		\hline
		L-opt & 0.1067 & 0.0041 & 343.9890 \\ 
		A-opt & 0.0867 & 0.0021 & 892.4160 \\ 
		Uniform & 0.2529 & 0.0112 & 170.9510 \\ 
		\hline
		71 \\
		\hline
		L-opt & 0.0467 & 0.0022 & 358.7080 \\ 
		A-opt & 0.0513 & 0.0011 & 906.9360 \\ 
		Uniform & 0.0722 & 0.0045 & 194.4480 \\ 
	\end{tabular}}
\end{table}

\begin{table}[ht]
	\caption{\label{tab:ukb:td:all15} UKB CRC time-dependent coefficient data analysis: estimated regression coefficients and standard errors of each method. 15 pseudo-observations were sampled for each observed failure time.}
\tiny
\centering
\fbox{%
	\begin{tabular}{lcccccccc}
		& PL & PL se & L-opt & L-opt se & A-opt & A-opt se & Unif & Unif se \\ 
		\hline
		BMI & 0.0005 & 0.0059 & 0.0005 & 0.0061 & -0.0001 & 0.0061 & 0.0002 & 0.0062 \\ 
		Smoking & 0.0699 & 0.0595 & 0.0728 & 0.0616 & 0.0548 & 0.0617 & 0.0675 & 0.0624 \\ 
		Family & 0.2265 & 0.0635 & 0.2284 & 0.0659 & 0.2299 & 0.0659 & 0.2894 & 0.0689 \\ 
		Activity & -0.0863 & 0.0621 & -0.0721 & 0.0644 & -0.1038 & 0.0645 & -0.0821 & 0.0652 \\ 
		Sex & -0.7631 & 0.2578 & -0.8341 & 0.2672 & -0.8083 & 0.2663 & -0.8397 & 0.2771 \\ 
		Alcohol2 & -0.0136 & 0.0743 & -0.0159 & 0.0767 & -0.0468 & 0.0770 & -0.0234 & 0.0781 \\ 
		Alcohol3 & -0.0827 & 0.0696 & -0.0872 & 0.0719 & -0.0953 & 0.0721 & -0.0886 & 0.0730 \\
\end{tabular}}
\end{table}

\begin{table}[ht]
	\caption{\label{tab:ukb:td:all30} UKB CRC time-dependent coefficient data analysis: estimated regression coefficients and standard errors of each method. 30 pseudo-observations were sampled for each observed failure time.}
	\tiny
	\centering
	\fbox{%
		\begin{tabular}{lcccccccc}
		& PL & PL se & L-opt & L-opt se & A-opt & A-opt se & Unif & Unif se \\ 
		\hline
		BMI & 0.0005 & 0.0059 & 0.0014 & 0.0059 & 0.0008 & 0.0059 & -0.0018 & 0.0060 \\ 
		Smoking & 0.0699 & 0.0595 & 0.0525 & 0.0606 & 0.0704 & 0.0605 & 0.0739 & 0.0609 \\ 
		Family & 0.2265 & 0.0635 & 0.2167 & 0.0647 & 0.2284 & 0.0647 & 0.2479 & 0.0662 \\ 
		Activity & -0.0863 & 0.0621 & -0.0962 & 0.0633 & -0.1005 & 0.0633 & -0.0866 & 0.0637 \\ 
		Sex & -0.7631 & 0.2578 & -0.7162 & 0.2615 & -0.6907 & 0.2611 & -0.8819 & 0.2674 \\ 
		Alcohol & -0.0136 & 0.0743 & -0.0275 & 0.0755 & -0.0050 & 0.0756 & -0.0271 & 0.0762 \\ 
		Alcohol & -0.0827 & 0.0696 & -0.0878 & 0.0707 & -0.0843 & 0.0709 & -0.0945 & 0.0714 \\
\end{tabular}}
\end{table}

\begin{table}[ht]
	\caption{\label{tab:ukb:td:all71} UKB CRC time-dependent coefficient data analysis: estimated regression coefficients and standard errors of each method. 71 pseudo-observations were sampled for each observed failure time.}
	\tiny
	\centering
	\fbox{%
		\begin{tabular}{lcccccccc}
			& PL & PL se & L-opt & L-opt se & A-opt & A-opt se & Unif & Unif se \\ 
		\hline
		BMI & 0.0005 & 0.0059 & -0.0002 & 0.0059 & 0.0003 & 0.0059 & -0.0000 & 0.0059 \\ 
		Smoking & 0.0699 & 0.0595 & 0.0675 & 0.0599 & 0.0740 & 0.0600 & 0.0752 & 0.0601 \\ 
		Family & 0.2265 & 0.0635 & 0.2327 & 0.0640 & 0.2390 & 0.0640 & 0.2283 & 0.0647 \\ 
		Activity & -0.0863 & 0.0621 & -0.1010 & 0.0626 & -0.0938 & 0.0626 & -0.0873 & 0.0628 \\ 
		Sex & -0.7631 & 0.2578 & -0.7472 & 0.2605 & -0.7691 & 0.2595 & -0.8063 & 0.2620 \\ 
		Alcohol & -0.0136 & 0.0743 & -0.0139 & 0.0748 & 0.0059 & 0.0748 & -0.0177 & 0.0751 \\ 
		Alcohol & -0.0827 & 0.0696 & -0.0755 & 0.0701 & -0.0952 & 0.0702 & -0.0816 & 0.0703 \\
\end{tabular}}
\end{table}

\begin{table}[ht]
\caption{\label{tab:ukb:marg:metrics} UKB CRC marginal analysis: metrics for evaluating the different methods. The RMSE and variance $l_2$ distance are calculated with respect to the full-data PL estimator. 4 censored observations were sampled per each observed failure time.}
		\centering
		\fbox{%
		\begin{tabular}{lcccc}
		& RMSE &  $\| .\|_2$  & Runtime in sec.  \\ 
			\hline
		PL & 0.000 &  0.000 & 667 \\  
		L-opt & 0.063 &  $ 5\cdot10^{-4}$ & 60.7 \\ 
		A-opt & 0.066 &  $ 5\cdot10^{-4}$ & 80.1 \\ 
		Uniform & 0.097  & 0.0011  & 14.5 \\
		\end{tabular}}
	\end{table}

\clearpage

\begin{table}[ht]
	\caption{\label{tab:ukb:marg:allSNP} UKB CRC marginal analysis: estimated regression coefficients and standard errors of each method for all 72 SNPs. 4 censored observations were sampled for each observed failure time.}
	\centering
	\tiny
	\fbox{%
	\begin{tabular}{lcccccccc} 
		& PL & PL se & L-opt & L se & A-opt & A se & Unif & Unif se \\ 
		\hline
	SNP1 & -0.057 & 0.020 & -0.048 & 0.021 & -0.054 & 0.021 & -0.054 & 0.023 \\ 
	SNP2 & -0.061 & 0.019 & -0.071 & 0.021 & -0.056 & 0.021 & -0.060 & 0.022 \\ 
	SNP3 & 0.047 & 0.019 & 0.038 & 0.021 & 0.052 & 0.020 & 0.036 & 0.022 \\ 
	SNP4 & 0.005 & 0.019 & 0.012 & 0.021 & 0.009 & 0.020 & 0.007 & 0.022 \\ 
	SNP5 & 0.053 & 0.019 & 0.056 & 0.021 & 0.056 & 0.021 & 0.089 & 0.022 \\ 
	SNP6 & 0.003 & 0.019 & -0.004 & 0.021 & -0.002 & 0.021 & 0.001 & 0.022 \\ 
	SNP7 & 0.044 & 0.018 & 0.053 & 0.020 & 0.044 & 0.020 & 0.035 & 0.021 \\
\end{tabular}}
\end{table}

\bibliography{literature}

\end{document}